\documentclass[onefignum,onetabnum]{siamart171218}

\usepackage{lipsum}
\usepackage{amsfonts}
\usepackage{graphicx}
\usepackage{epstopdf}
\usepackage{algorithmic}
\usepackage{amsmath}
\usepackage{amscd}
\usepackage{amssymb}
\usepackage{gensymb}
\usepackage{enumerate}
\usepackage{url}
\usepackage{nicefrac}
\usepackage{mathrsfs}
\usepackage{graphicx}
\usepackage{xcolor}
\usepackage[outercaption]{sidecap}
\usepackage{amsfonts,epigraph}%
\usepackage{booktabs} 
\usepackage[all,cmtip]{xy}
\usepackage{tikz}
\usetikzlibrary{shapes,arrows}
\usepackage[utf8]{inputenc}
\usetikzlibrary{calc}
\usetikzlibrary{arrows.meta}

\definecolor{boxColor}{cmyk}{0.33,0.10,0,0}
\ifpdf
\DeclareGraphicsExtensions{.eps,.pdf,.png,.jpg}
\else
\DeclareGraphicsExtensions{.eps}
\fi

\newsiamremark{remark}{Remark}
\newsiamremark{hypothesis}{Hypothesis}
\crefname{hypothesis}{Hypothesis}{Hypotheses}
\newsiamthm{claim}{Claim}

\headers{Multiplicity of time scales in climate, matter, life, and economy}{B. Booss--Bavnbek, R. K. Pedersen, and U. R. Pedersen}

\title{Multiplicity of time scales in the modelling of climate, matter, life, and economy\footnote{\nolinebreak {Put on arXiv Jan}. 31, 2020.
		\funding{This work was supported by \textit{CIRCLES --- Roskilde University's Centre for Interdisciplinary Research and Education in Circular Economy and Sustainability}.}}}

\author{Bernhelm Booss--Bavnbek
	\footnote{Department of Science and Environment, Roskilde University, P.O. Box 260, DK-4000 Roskilde, Denmark, \email{booss@ruc.dk}, \url{https://orcid.org/0000-0002-8865-7298}, \email{rakrpe@ruc.dk},
		\url{https://orcid.org/0000-0001-5946-8220}, \email{urp@ruc.dk}, \url{https://orcid.org/0000-0003-2567-555X}.}
	\and Rasmus K. Pedersen
	\footnotemark[2]
	\and Ulf R. Pedersen
	\footnotemark[2]}

\usepackage{amsopn}

\newcommand{\dd}[1]{\frac{d}{d#1}}

\def\e{\varepsilon}

\def\la{\lambda}

\def\CC{{\mathbb C}}

\DeclareRobustCommand*{\nicefrac}[2]{\ifmmode\mathnicefrac{#1}
	{ #2}%
	\else\textnicefrac{#1}{#2}\fi}
\newcommand*{\textnicefrac}[2]{\check@mathfonts%
	\mbox{\raisebox{.5ex}{\fontsize\sf@size\z@\selectfont#1}\kern-.
		1em%
		/\kern-.1em\raisebox{- .25ex}{\fontsize\sf@size\z@\selectfont#2} }}
\newcommand*{\mathnicefrac}[2]{%
	\mathchoice
	{\m@fr@c{\scriptstyle}{#1}{#2}}
	{\m@fr@c{\scriptstyle}{#1}{#2}}
	{\m@fr@c{\scriptscriptstyle}{#1}{#2}}
	{\m@fr@c{\scriptscriptstyle}{#1}{#2}}}

\def\fequal#1{\stackrel{#1}{=}}

\def\sqm1{\sqrt{-1}}

\def\tand{\mbox{\ \rm  and }}

\def\x{\times}

\def\={\cong}
\def\>{\supset}
\def\<{\subset}
\def\ii{^{-1}}
\def\12{\frac{1}{2}}
\def\0{^{\circ}}

\newcommand{\md}{\operatorname{md}}

\def\fequal#1{\stackrel{#1}{=}}
\newcommand{\auindex}[1]{}

\ifpdf
\hypersetup{
  pdftitle={Time scales in climate, matter, life, and economy},
  pdfauthor={B. Booss--Bavnbek, R. K. Pedersen, and U. R. Pedersen}
}
\fi

\begin{document}

\maketitle

\begin{abstract}
 This topic review communicates working experiences regarding interaction of a multiplicity of processes. Our experiences come from climate change modelling, materials science, cell physiology and public health, and macroeconomic modelling. We look at the astonishing advances of recent years in broad-band temporal frequency sampling, multiscale modelling and fast large-scale numerical simulation of complex systems, but also the continuing uncertainty of many science-based results. 
 We describe and analyse properties that depend on the time scale of the measurement; structural instability; tipping points; thresholds; hysteresis; feedback mechanisms with runaways or stabilizations or delays. We point to grave disorientation in statistical sampling, the interpretation of observations and the design of control when neglecting the presence or emergence of multiple characteristic times.
 We explain what these working experiences can demonstrate for environmental research.
\end{abstract}



\section{Introduction}

It is well known that differences between characteristic time lengths provide 
difficulties in mathematical modelling, 
statistical sampling, and 
numerical simulation.
Through the last three decades, many of these difficulties have been overcome by astonishing advances in high and broad-band temporal frequency observation, in  multiscale modelling of complex systems, and in fast and large-scale scientific computing and numerical simulation.
These advances within the \textit{mathematical, scientific and design communities} are impressive, but we are worried that there may be a lack of emphasis on multiscale aspects in environmental research and correspondingly only a low awareness of differences between characteristic times in \textit{public response} to the challenges of climate, environment, public health, and economy. Disregard of the multiscale aspects of a problem can become misleading in analysing and forecasting trends, designing counteracting, mitigating and/or adaptation means and communicating threats and solutions, see our conclusions in  \Cref{s:conclusions}. 

\subsection{Scientification of politics and politicisation of sciences --- triumphs and pitfalls}

Since many years, quantitative measurements, scientific theories, mathematical models and numerical simulations of complex systems have inspired decision making in environmental management and many other segments of the real world. They have illustrated supposed consequences, confirmed or contradicted prejudices and occasionally enlarged the views of decision makers by devising new approaches and alternative means, procedures and design. Their \textit{heuristic} value was undisputed. Only in recent times, however, we witness a \textit{scientification} (“Verwissenschaftlichung”) of decisions regarding ecological and other complex systems, where huge collections of data, asserted theories, intricate mathematical models, extensive statistical regressions and time series analyses, and gigantic numerical simulations deliver often far reaching decisions when confronted with urgent needs and/or public pressure.  Scientific arguments lead to so-called imperative necessities.

This trend has become manifest in many different fields: in macroeconomic decisions (taxes and social security regulations etc.), in public health (vaccination, early diagnosis and prevention programs), in environmental administration (e.g., fishing quota and eutrophication restoration), and, perhaps most outspoken, in climate change mitigation and adaptation (e.g., Paris Accord). 

Understandably and to deplore as misleading, vested interest of population groups and common prejudices have raised doubts about the objectivity and credibility of the underlying models, simulations, predictions and prescriptions. They blame intricate mathematical models and large numerical simulations to be social constructs designed to hide a political agenda.

Equally understandable and perhaps to deplore as misleading and hardly sustainable for a campaign over years, proponents of science based provisions act on the authority of alleged non-criticisable science. The Paris Agreement of 30 Nov. - 12 Dec. 2015 gave one example when it linked greenhouse gas emissions quantitatively to global temperatures: ``hold the increase in the global average temperature to below 2 \degree C above pre-industrial levels by reducing emissions to 40 gigatonnes" \cite[Decision 1/CP.21, Article 17]{ParisAgreement:2015}. Clearly, the official recognition of 195 states of the anthropogenic heat forcing was a great step forward in taking the warnings of climate researchers seriously. One might, however, worry that supposing a clear quantifiable and definitive link between emissions and temperature suggests a much higher credibility of estimates than available data, established theory and computer simulations presently can guarantee for such forecasts. Worse, it may suggest an immense hidden scientific and technological power to trust upon for reversing these trends when mandatory. In reality, the excessive emissions are on the way of providing atmospheric and climate change researchers with huge amounts of new data in that real-time terrestrial experiment. 

The young Swedish climate activist \textsc{Greta Thunberg} may have raised similar worries inadvertently when she told world leaders at the opening of a United Nations conference \cite{GretaThunberg:2019}
``For more than 30 years, the science has been crystal clear... To have a 67 per cent chance of staying below a 1.5 degrees of global temperature rise, the best odds given by the IPCC, the world had 420 gigatons of CO$_2$ left to emit back on January 1, 2018." Excellent as a warning, but also misleading in supposing that climate change researchers would know much more than they do.

The \textit{triumphs} of the scientification of politics have led to a politicisation of sciences, where scientists are  flattered and cajoled to deliver figures and directions, when they rather should counter\textit{ naive beliefs} in chosen quantitative data, complex mathematical models and non-transparent leviathan numerical simulations, as well as to counter\textit{ poorly informed doubts} that deny basic theory and well-established evidence. 

Statistical sampling, mathematical models and numerical simulations of environmental, climate, health and economic change have obtained a status of high public attention. The corresponding politicisation of sciences  has its own \textit{pitfalls}: Geophysicists and meteorologists, e.g., have left their offices and laboratories to become communicators and, sometimes against their will, opinion leaders. Their models can be presented and perceived quite differently along three axes: suggestive vs. counter-intuitive; realistic vs. fancied; science-based vs. speculative. In their role as opinion leaders in focus of growing public attention, they can feel obliged to restrain their professional imagination and their scepticism as scientists and stick to broad convincing arguments, to widely accepted facts and to conservative, calming estimates.

While none of us are atmosphere physicists or active in other segments of climate modelling, in this paper we take the liberty of focusing on the counter-intuitive, of criticizing too simplistic reality perceptions and of opening up speculations that are based on simple transparent mathematical models (see \Cref{ss:toy-model}), and our working experiences with matter, life and economics 
(see \Cref{s:matter,s:life,s:society}) and reach beyond conservative estimates.

\subsection{Structure of this paper}

In \Cref{s:emergence-in-math}, we point to multiple time scales in climate modelling and give a toy model of the dynamic interaction of \textit{Homo Sapiens} with the \textit{Earth System}. We recall basic mathematical concepts to explain the eventual structural instability of the interaction of multiple processes with different characteristic time scales and the related challenges in the expert--public communication. 

In \Cref{s:matter}, we introduce a typical multiscale example of materials science and take a closer look at the multiple time scales arising in the computer simulation of liquid dynamics of viscous materials. To us, this case provides the most fundamental and most dramatic example of the emergence of a multiplicity of time scales. This case is intended to support the ongoing change in mind sets in environmental and energy management by pointing to the universality of the multiplicity of characteristic times and the commanding need to focus on them. 

In \Cref{s:life}, we summarize two multiscale examples of life sciences. We emphasize the challenges of statistical sampling in living tissue, where different processes become visible depending on the choice of time steps. We take a closer look at the modelling and computer simulation of the production of blood and the development of some blood cancers and of the biphasic insulin secretion of pancreatic beta--cells. For the decisive role of a meaningful choice of characteristic time scales in public health we refer to the literature on modelling of infectious diseases.

In \Cref{s:society}, we summarize typical multiscale examples of the macrodynamics of capitalism and take a closer look at empirical data of a long-term Kondratiev-Schumpeter wave and a mathematical model of embedded Keynesian short-term business cycles. As always in studies of social systems, the interpretation of observations may, or may not, depend on the application of presupposed theoretical perceptions or political goals.   

In \Cref{s:conclusions},  we summarize our findings and sketch a conclusion for scientific, communicative and political challenges.

In   \Cref{ss:communicating},  we address the public disregard even of the most elementary multiscale aspects, like the emergence of multiple time scales.

In each section, we distinguish between (i) \textit{localisable} multiscale problems where a dominant macroscale model must be supplemented by microscopic models only for localized defects; (ii) \textit{global} multiscale problems, when, e.g., repetitive small disturbances and feedback can become decisive for global stability or instability; (iii)  \textit{evolving} multiscale problems, where there is no multiplicity of models and scales to begin with and the multiplicity of time scales emerges as part of the dynamics. We provide \textit{simple} explicit mathematical models of intricate systems with multiple time scales, e.g., a simple system of coupled ordinary differential equations.
In that way we hope to make people familiar with the \textit{counter--intuitive} effects of having two or even more characteristic times.
\smallskip

\noindent{
	This paper is an extended version of a review presented at the international conference on \textit{Transforming for Sustainability}, UN City of Copenhagen, 28-29 November 2018, accessible at \url{http://dirac.ruc.dk/~urp/2018/timescales/oral/}.}

%
%
%
%
%
\section{Time scales and climate change modelling}\label{s:emergence-in-math}

\subsection{Mathematical toy models with two characteristic time scales}\label{ss:toy-model}

\subsubsection{The dynamic interaction of \textit{Homo Sapiens} with the \textit{Earth System}}\label{s:toy-models}
In \Cref{f:toy-model} we visualize the challenges by a three-compartmental toy model of the dynamic interaction of \textit{Homo Sapiens} with the \textit{Earth System}. 
Confronted with the serious tasks of transforming for sustainability, a technical discussion on multiple time scales may appear rather abstract. Who cares whether they exist or not, are universal or particular, inevitable or avertible?
The following toy model shall illustrate how much we may compromise when we are not aware of possible multiple time scales.

\tikzstyle{block} = [draw, fill=blue!20, rectangle, 
minimum height=3em, minimum width=6em]
\tikzstyle{sum} = [draw, fill=blue!20, circle, node distance=2cm]
\tikzstyle{input} = [coordinate]
\tikzstyle{output} = [coordinate]
\tikzstyle{pinstyle} = [pin edge={to-,thin,black}]

\begin{figure}[ht]
	\begin{tikzpicture}[auto, node distance=3cm,>=latex']
	\node [input, name=input] {};
	\node [sum, right of=input] (sum) {};
	\node [block, right of=sum] (controller) {$\begin{matrix}
		\text{General Public}
		\end{matrix}$};
	\node [block, right of=controller, pin={[pinstyle]above:Disturbances},
	node distance=4cm] (system) {$\begin{matrix}
		\text{Earth}\\
		\text{System}
		\end{matrix}$};
	\draw [->] (controller) -- node[name=u] {$\ u\ $} (system);
	\node [output, right of=system] (output) {};
	\node [block, below of=u] (measurements) {$\begin{matrix}
		\text{Scientific Community}
		\end{matrix}$};
	%
	\draw [draw,->] (input) -- node {$r$} (sum);
	\draw [->] (sum) -- node {$e$} (controller);
	\draw [->] (system) -- node [name=y] {$T$}(output);
	\draw [->] (y) |- (measurements);
	\draw [->] (measurements) -| node[pos=0.99] {$-$} 
	node [near end] {$y_m$} (sum);
	\end{tikzpicture}
	\caption{Compartment model of the Earth Climate System with the mean atmospheric temperature $T$ as output variable, subject to natural disturbances and anthropogenic forcing $u$, generated by a general public which is guided by enlightenment and politics $e$ that is formed in a struggle between vested interests $r$ and science based mathematical modelling and simulation $y_m$\/.}
	\label{f:toy-model}
\end{figure}
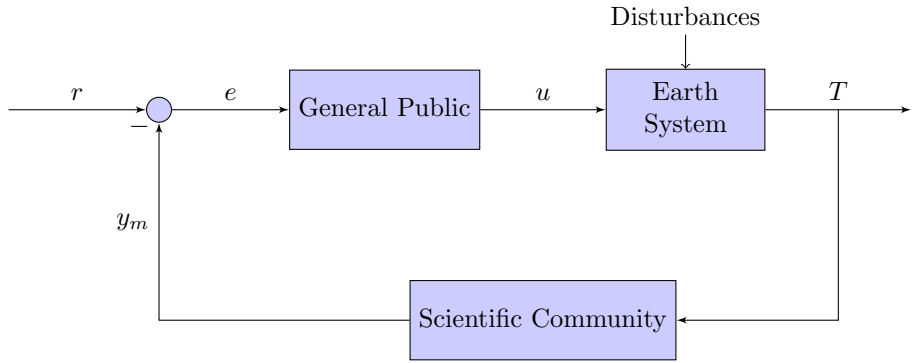

Roughly speaking, transforming to sustainability involves three compartments, the \textit{Earth System} and the two compartments of \textit{Homo Sapiens}, the \textit{General Public} and the \textit{Scientific Community}. Changes of the compartment \textit{Earth System} are subject both to human forcing $u$ by the control compartment \textit{General Public} and to natural \textit{Disturbances}, resulting in an output variable, here named $T$ like \textit{Temperature}. For simplicity we here assume that the state of the Earth System can be represented by a single variable ($T$). Output is measured and analysed by the third compartment, the \textit{Scientific Community}. The activity of the compartment \textit{General Public} is essentially determined by the arrow $e$, indicating \textit{Enlightenment} and \textit{Politics}. It originates from a node, where there is an ongoing struggle between spontaneous, ideologically or commercially magnified influences $r$ on the one side and the flow $y_m$ on the other side, made by science based mathematical modelling and interpretations of scientific measurements and public experiences.

\subsubsection{A scenario of different awareness competences between the general public and the scientific community}
For our toy model we imagine that the general public and the scientific community have quite different awareness competences regarding the output variable $T:=$  average annual Earth temperature. Typically, ``ordinary people" will feel dramatic changes of $T$, and the media and the education will often emphasize just that, in mathematical terms $T'= \dd {t}{T}$, the first derivative of $T$. One will be attentive to increasing rates of change, while decreasing rates of change will be perceived often as a relaxation even when $T'$ remains positive. 
On the contrary, in mathematics and sciences, we would perceive a continuing increase of the output variable $T$ as highly alarming even when the rates of increase should decrease. 

Reason for concern is that a variety of feedback loops (like the decreasing albedo effect and the increasing natural methane release by increasing temperature, see below) may force our output variable onto radically different trajectories. On the other side, we know, e.g., that the heat capacity of the atmosphere of a weight of 1 atmosphere [at] $\times$ area, equals approximately the heat capacity of 10 meters of water (roughly of the same weight) of the oceans  that have a mean depth of about 4000 meter. Therefore, there is room for much heat exchange between the atmosphere and the oceans. That heat exchange depends on the slow and not very well understood stirring in the oceans. So, we may argue that in the very long run the Earth System can handle the anthropogenic temperature forcing. 

However, how long is the ``very long run" and what about the other known feedback mechanisms? 
Recent measurements of the heat of the oceans have revealed that indeed about 93\% of the radiative energy imbalance (due to anthropogene emissions) accumulates in the oceans as increased ocean heat content (OHC), see \cite{ClimateChange:2019} and \cite{Cheng:2019} reporting on the measurements of the recently completed worldwide grid of 3000 deep water temperature gauges and the perhaps controversial \cite{Resplandy:2018} using ocean warming
outgassing of O$_2$ and CO$_2$, which can be isolated from the direct effects of anthropogenic emissions and CO$_2$ sinks, to independently estimate changes in OHC over time after 1991. 
That means, inter alia, that for now we see only a 7\% tip of the climate change problem on the Earth surface.

\subsubsection{Different kinds of model credibility and uncertainty}
It is obvious that inaccurate or inaccessible data is one of the predominant sources of uncertainty about the future path of the output variable and the effect of different ways and levels of transforming for sustainability. In mathematical modelling and simulation, that uncertainty is called \textit{aleatoric}. In principle, it could be diminished by higher investments in sensor networks and research, as emphasized by \textsc{J. Behrens} \cite[p. 286]{Behrens:2016} --- though hardly eliminated totally, as explained earlier in \Cref{ss:climate-change}. Contrary to that, the model uncertainty described above is hard to reduce, since there can be many different characteristic time scales, e.g., one for forcing by radiation on the every day / annual scale and the other one for the various feedback loops on the scale of decades or centuries or millennia. There are just different regimes, and it is unclear beforehand which regime is dominant and for how long. This kind of uncertainties is called \textit{epistemic}, since, as \textsc{Behrens} reminds us, ``model uncertainty is inherent in the process of understanding nature by simplifying it to natural laws." 

This is particularly true for climate change modelling, where \textsc{Popper}'s methodological demand of \textit{falsification} is impossible to satisfy: There are simply no experiments nor observations of the decisive atmospheric, terrestrial and oceanic processes and their interaction on the relevant scales. We only have mathematical models with parameters of quite different origin. Some parameters are (i) well-established world constants, immediately derived from physical first principles; some are (ii) measurable in a laboratory; but some are (iii) estimates from fitting somehow available data series to chosen and uncertain systems of equations. With (i) and (ii), the situation is typical for multiscale modelling and simulation. The deviation comes with (iii).

\begin{figure}[t]
	  \centering
	\rotatebox{90} {
		\includegraphics[scale=0.355]{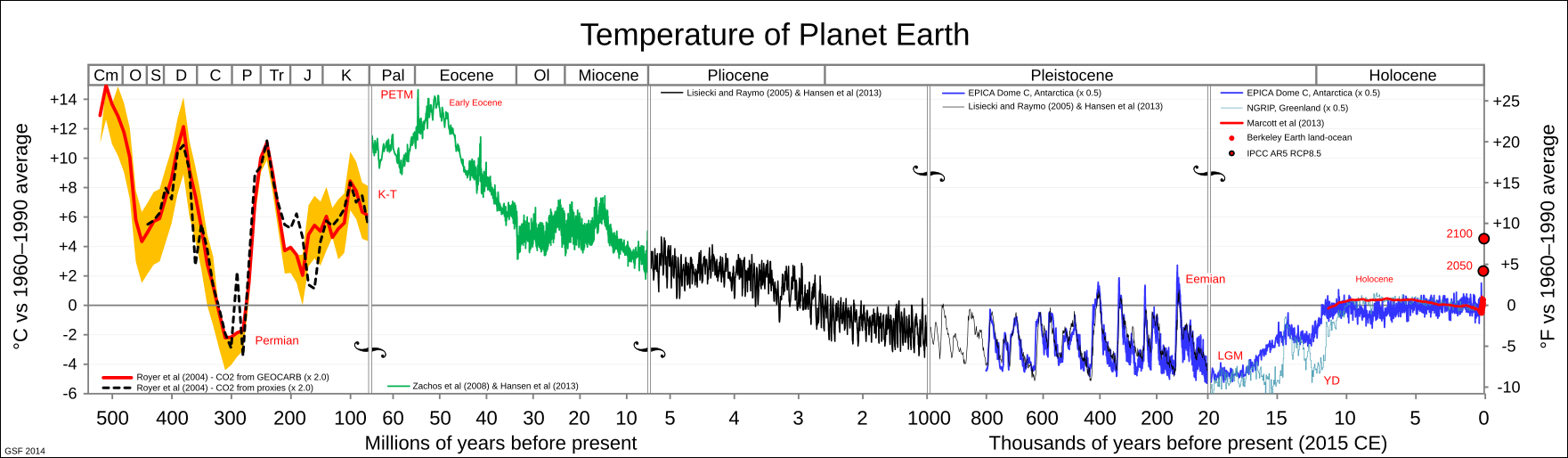}
	}
	\caption{Earth temperature with multiple time scales, adapted from \cite{Hansen-et-al:2013}, from \url{https://commons.wikimedia.org/wiki/File:All_palaeotemps.svg}, licensed under the Creative Commons Attribution-Share Alike 3.0 Unported license https://creativecommons.org/licenses/by-sa/3.0/}
	\label{f:earth-temperature}
\end{figure}

To give an example: In 1996, the U.S.A. and Russia reached an agreement (informal, but since then honored by all nuclear powers, except for India, Pakistan and North Korea) to halt \textit{all} nuclear test explosions. According to the testimony of the nuclear engineer \textsc{M.F. Horstemeyer} \cite[Section 1.3, pp. 4f]{Horstemeyer:2012}, they could do so because even the most abstract multiscale simulations of nuclear weapons and their equations and constants can be checked against previous large scale test observations, see also our comments below in \Cref{s:conclusions}. Astrophysicists like \textsc{C. Sagan} and geologists and geophysicists have argued for the use of abundant planetary and paleontology data for checking equations and constants of planetary climate change; until now, apparently, without substantial progress for Earth climate modelling. 

Of course, looking back at earlier periods of Earth history may help visualizing the probable outcome of continuing excessive Anthropocene greenhouse gas emissions in the years 2030, 2050 or 2100, as \textsc{K. Burke} and collaborators explain in \cite[Abstract]{Burke-et-al:2018}: ``Past Earth system states offer possible \textit{model systems} (our emphasis) for the warming world of the coming decades. These include the climate states of the Early Eocene (ca. 50 Ma), the Mid--Pliocene (3.3{\textendash}3.0 Ma), the Last Interglacial (129{\textendash}116 ka), the Mid--Holocene (6 ka), preindustrial (ca. 1850 CE), and the 20th century." See also our  \Cref{f:earth-temperature}. In spite of all advisable reservations against disseminating dystopian views in the scientific literature, it certainly is meritorious to draw parallels between past and future climates. We agree with \textsc{F. Lehner}, a project scientist at the US National Center for Atmospheric Research, however, in his comment \cite{Lehner:2018} to \textsc{Burke}'s study, that many uncertainties make it challenging to reconstruct and understand hydroclimate change, even over the last 1,000 years. As in our \Cref{tab:time-scales-climate-change},  to us the essential point is to be aware of the huge time scale difference between the supposed few years, be it 12, 30, or 100, to unleash mechanisms that can bring the Earth System back to mid--Pliocene temperatures, and the million years it may take the Earth System afterwards to get back to a climate similar to the present.

In the frame of this paper, our main mathematical conclusion is not necessarily to \textit{trust} the existing mathematical models of climate change in lack of better alternative models, but 
\begin{itemize}
	\item to stick to the established wisdom, that we know the basic atmospheric, terrestrial and oceanic processes sufficiently well separately, 
	\item to point to the presence of multiple time scales, and 
	\item to argue for a corresponding \textit{change of the mindsets} within the science community and the public.
\end{itemize}

\subsection{The emergence of two characteristic time scales --- a textbook example}\label{ss:emergence+public}
We summarize the communicative challenges of dealing with multiple characteristic time lengths of the Earth System (see \Cref{tab:time-scales-climate-change}), i.e.,
\begin{enumerate}
	\item dealing with situations that turn out as less frightening than predicted in the short time, 
	\item guarding against situations that will kick off a chain reaction that makes further temperature rises unstoppable for a possibly very, very long run (as in the clathrate gun hypothesis, see below \Cref{ss:climate-change}), and
	\item reversing the present imbalance between political intentions and actions, where real world data are ahead of real world intentions.
\end{enumerate}

\begin{figure}[ht]
	\includegraphics[scale=0.855]{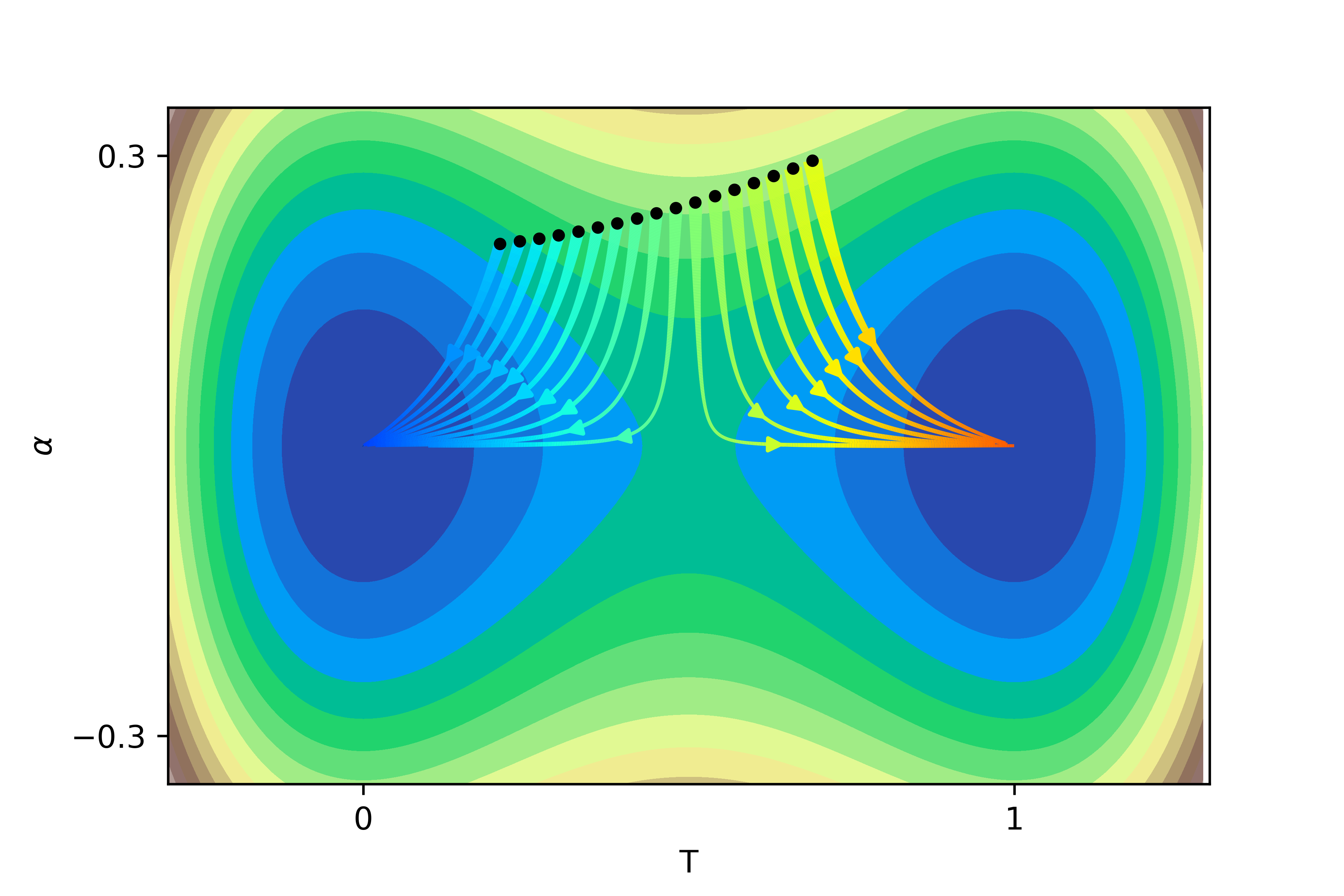}
	\caption{Dynamical system with the level curves of $V(T, \alpha) = T^2(T-1)^2+\alpha^2$\/, gradient lines of $(T',\alpha')=-\operatorname{grad} V(T,\alpha)$ and wells at $(0,0)$ and $(1,0)$ representing the relatively stable and relatively moderate present climate and a possibly triggered and also relatively stable future hot bed. See also \Cref{f:fourth-order-poly} with a simple Landau--Langevin model of that situation and our simulation on \url{http://dirac.ruc.dk/~urp/2018/timescales/oral/}}
	\label{f:reclining-eight}
\end{figure}

\subsubsection{A simple Landau--Langevin model of structural instability} \label{sss:landau-Langevin} 

A simple mathematical model in two dimensions, inspired by \textsc{Hirsch} and \textsc{Smale} \cite[p. 203f]{Hirsch-Smale:1974} and easily to demonstrate in laboratory with Landau-Langevin diffusion, 
illustrates aspects (1) and (2): In \Cref{f:reclining-eight}, we depict the present time Earth System of \Cref{f:toy-model} as a dynamical system in two variables, the global mean temperature $T$ (horizontal axis, \textit{abscissa}) and a universal control parameter $\alpha$ (vertical axis, \textit{ordinate}). We imagine that the system is subjected to a wide range of geophysical laws, here depicted by level curves of a single aggregate potential 
\[
V(T,\alpha)\ =\ T^2(T-1)^2+\alpha^2\,. 
\]
The level curve shaped like a reclining figure eight is $V^{-1}(\frac 1{16})$.
The drawn gradient lines represent typical trajectories (development paths in the time $t$) of the dynamical system $(T',\alpha')=-\operatorname{grad} V(T,\alpha)$ with
a saddle point at $(\frac 12,0)$ between two stable equilibria at $(0,0)$ and $(1,0)$.

Trajectories to the left of the threshold line $x = \frac 12$ tend toward $(0, 0)$
(as $t\to +\infty$) in agreement with our short time expectations, namely that it should be possible to return to a benign climate equilibrium (at $(0, 0)$) by transforming for sustainability. However, trajectories to the right tend toward $(1, 0)$ which may indicate a hot bed faraway from present temperatures, subjected to strong internal feedback mechanisms and out of the range of human control.

The worst thing happening 
in this simple model is that we become accustomed to the effective short range chances of optimistic human stewardship, advocated eloquently, e.g., by \textsc{H. Rosling} in \cite{Rosling:2018}, but failing to notice the proximity of the threshold and so foolishly glide over the threshold into a region where processes take over that work on much larger time scales and therefore will in the beginning not be easy to notice, but which will be irreversible for long Earth periods and harmful or even fatal for mankind.

\subsubsection{Supposed jumps between climate trajectories}\label{sss:climate-trajectories}
\begin{figure}[ht]
	\includegraphics[scale=0.53]{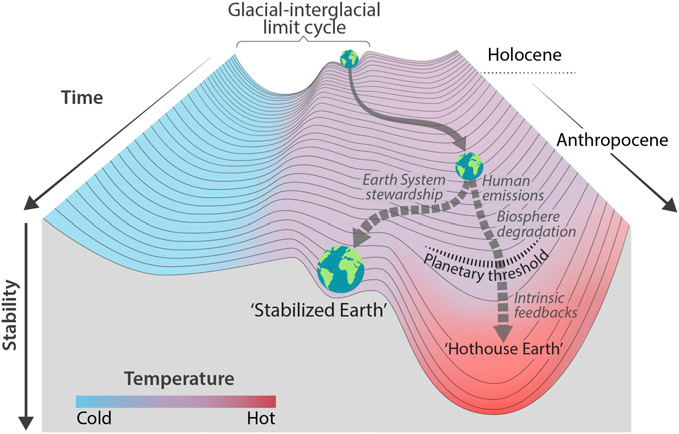}
	\caption{Stability landscape showing the pathway of the Earth System
		out of the Holocene and thus, out of the glacial–interglacial limit cycle and
		to its present position in the hotter Anthropocene. The fork in the
		road in \Cref{f:reclining-eight} is shown here as the two divergent pathways of the
		Earth System in the future (broken arrows). From \textsc{Steffen} and co-authors \cite[Fig. 2]{Steffen:2018}, open access article distributed under Creative Commons Attribution-NonCommercial-NoDerivatives License 4.0 (CC BY-NC-ND).}
	\label{f:steffen-trajectories}
\end{figure}

Admittedly, this is only a fancied simple two--dimensional model to illustrate the emergence of two characteristic time scales. It is a toy model like our scheme in \Cref{f:toy-model}, designed from an educational point of view. Sadly, however, some  (not necessarily representative and perhaps not sufficiently comprehensive) geophysical investigations of climate trajectories yield results that qualitatively remind us of \Cref{f:reclining-eight}, see \Cref{f:steffen-trajectories} from the recent \cite{Steffen:2018}. We quote:
\begin{quote}
	Currently, the Earth System is on a Hothouse Earth pathway driven by human emissions of
	greenhouse gases and biosphere degradation toward a planetary
	threshold at $\sim 2\,^{\circ}{\rm C}$, beyond which
	the system follows an essentially irreversible pathway driven by intrinsic
	biogeophysical feedbacks. The other pathway leads to Stabilized Earth, a
	pathway of Earth System stewardship guided by human-created
	feedbacks to a quasistable, human-maintained \textit{basin} of attraction.
	
	``Stability" (vertical axis) is defined here as the inverse of the potential
	energy of the system. Systems in a highly stable state (deep valley) have
	low potential energy, and considerable energy is required to move them
	out of this stable state. Systems in an unstable state (top of a hill) have
	high potential energy, and they require only a little additional energy to
	push them off the hill and down toward a valley of lower potential energy.
\end{quote}

\subsubsection{Aspects of irreversibility and hysteresis}\label{sss:irreversibility}
\begin{figure}[ht]
	\includegraphics[scale=0.396]{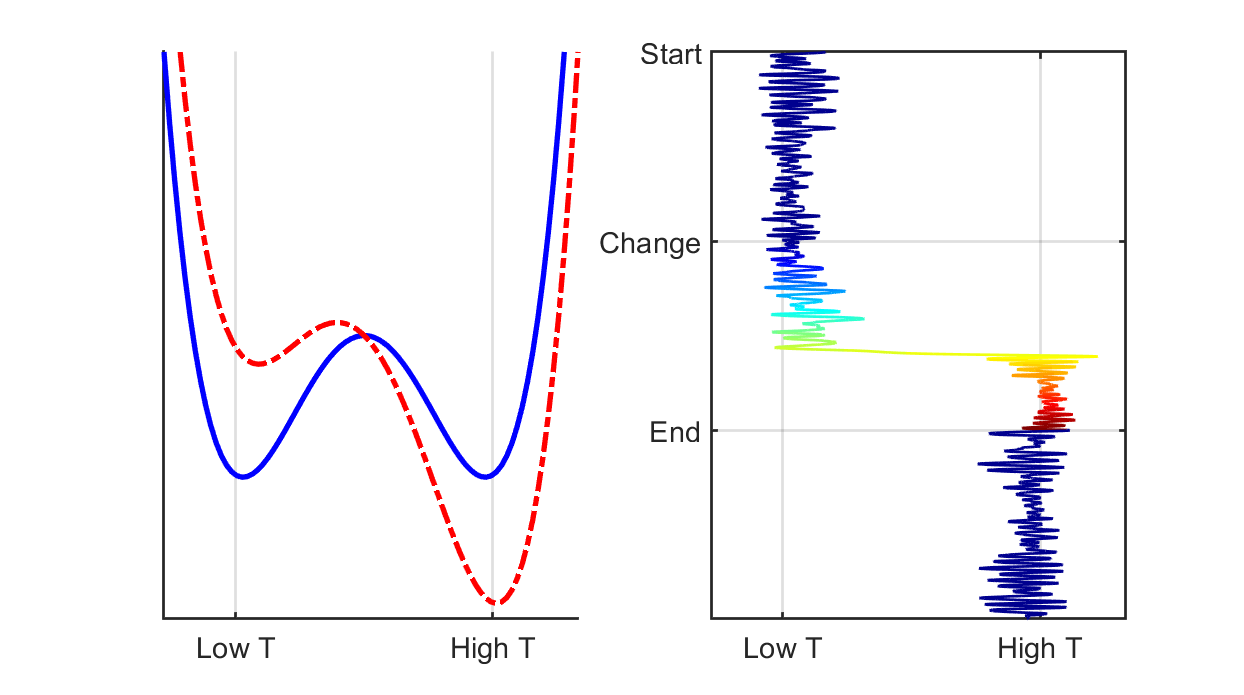}
	\caption{Illustration of stability dependence on temperature. The left-hand panel displays the high stability potential for $\alpha=0$ (blue curve) and a modified potential (red dotted curve) where $T_{\operatorname{Low}}$ lost its stability (see text). The right-hand panel displays a typical Langevin simulation with $\alpha$ being zero initially. At a given time (noted \emph{change}) the parameter $\alpha$ starts increasing with time. Note that in this model, after falling in the {\em basin} at $T_{\operatorname{High}}$, even reversing the potential from red to blue would not permit returning to $T_{\operatorname{Low}}$ from $T_{\operatorname{High}}$.}\label{f:fourth-order-poly}
\end{figure}

So, our simple mathematical toy-model is similar to the system described by \cite{Steffen:2018}. As seen above, it illustrates the time-scale issues at hand. In that one-dimensional system of temperature, $T$, the parameter $\alpha$ depends on the level of anthropogenic emission of greenhouse gasses. For $\alpha = 0$, this system has two minima of equal stability, but different temperature, $T_1$ and $T_2$ with $T_1 < T_2$. This is illustrated in the left panel of \Cref{f:fourth-order-poly} as a full blue line. For larger $\alpha$, the stability in the two minima starts changing such that the high-temperature minimum grows more stable (increasing barrier height for returning to the lower temperature equilibrium), while the low-temperature minimum becomes unstable  (decreasing barrier height for crossing over to the higher temperature equilibrium) and eventually smooths out and disappears, shown as dash-dotted red line in the figure.

In this toy-model, a simulation with stochasticity can illustrate a possible development of the Earth-temperature, and the corresponding stability. The temperature of the Earth is considered to move in accordance with Langevin dynamics (see \Cref{ss:emergence}), with small stochastic jumps in the time-derivative of the temperature dependent on the current gradient of the stability, see, e.g., \cite[Ex. 8.26]{Chandler:1987}. With such a system, the temperature of the Earth will  be likely to stay close to a local minima, although small changes may occur on short time-scales. In the right-hand panel of \Cref{f:fourth-order-poly}, an example is shown where the initial temperature is close to the low-temperature stability minimum. 
At a point in time, denoted \emph{Change} in the figure, the $\alpha$ parameter starts increasing linearly with time (In the figure, the trajectory color changes with the value of $\alpha$). This change signifies a change in the temperature-stability landscape caused by large-scale human intervention in the Earth system. In the simulation, increasing $\alpha$ beyond a tipping point eventually leads the Earth to move from the low-temperature \textit{basin} into the high-temperature \textit{basin}.

The significance of the trajectory of the Earth temperature in \Cref{f:fourth-order-poly} is in the (admittedly only fancied, neither proven nor predicted by the present state of art of climatology) \textit{sudden} change in temperature. While near the low-temperature minimum, oscillations on a short time-scale are somewhat insignificant, and even at the onset of $\alpha$ increasing with time, there seems to be no particular cause for alarm. However, suddenly in this textbook model, the \textit{bump} in the middle of the stability-potential is crossed, and the temperature continues to rise uncontrollably, until the high-temperature \textit{basin} is reached.

Additionally, reversing the human-caused issues, e.g., setting $\alpha = 0$, does not allow for a steady return to the original situation. Indeed, to escape the high-temperature \textit{basin}, $\alpha$ would have become negative, to cross the \textit{bump} going from right to left. Returning would take not just halting the human-caused increase of $\alpha$, but an active effort to produce the opposite effect resulting in a negative $\alpha$. 
The phenomenon is analogous to the hysteresis of elastic deformation and well known from environmental studies, where a deformation/degradation of a system happens almost instantly when the pressure passes a given threshold, while returning to the previous state under relaxation is happening in a much larger time scale, if at all.

%
%
%
%
%

\begin{table}[h]
	\caption{Characteristic time scales of climate change: Anthropogenic emissions occur on top of an active natural carbon cycle that circulates carbon among the reservoirs of the atmosphere, ocean, and terrestrial biosphere on timescales from sub-daily to millennial, while reversing the \textit{clathrate gun} (see below) and exchanges with geologic reservoirs occur at longer timescales, \cite{Archer:2009}.} 
	\label{tab:time-scales-climate-change}
	\rotatebox{90} {\scalebox{0.835}{ {\renewcommand{\arraystretch}{1.9}
				\begin{tabular}
					{l c p{2.6cm} p{5.2cm} c p{4.9cm}}
					No. & $t$ & Name & Description & Size [years] & Effect\\%
					\toprule%
					1.& $t_{\operatorname{civ}}$ & Civilisation time & Holocene, phase of relatively stable postglacial climate & $10^4$ & Development of civilisation\\
					2.& $t_{\operatorname{rad}}$ & Human forced radiation pattern changes  & Impact on the radiative energy balance of
					the Earth due to Anthropocene greenhouse gas emission & instant & Slow, but continuing rise of Earth temperature with yearly variation\\
					3.& $t_{\operatorname{adapt}}$ & Adaptation time & Trans\-for\-ming for sustainability, prevention, mitigation,  & $10^1-10^2$ & Establishing con\-sensus between people, government, stake\-holders; changing mindsets and habits\\
					4.& $t_{\operatorname{sec}}$ & Secondary effects & Release of methane from melting permafrost regions and oceanic methane clathrates & $10^1-10^2$ & Presumably rapid rise of Earth temperature beyond (not yet precisely known) thres\-hold\\
					5.& $t_{\operatorname{e-folding}}$ & E-folding time scale of CO$_2$ & Time for an atmospheric CO$_2$ concentration to decrease to a proportion of $e^{-1},\, \sim 37\%$, of it's original & $5\times 10^1-10^2$ & Misleading expectation that fossil fuel CO$_2$ in the atmosphere was to diminish according to \textit{linear kinetics}\\
					6.& $t_{\operatorname{mean}}$ & Mean lifetime of CO$_2$ & 
					Time of the elevated CO$_2$ concentration of the atmosphere according to carbon cycle models & $10^4-10^5$ & Leftover CO$_2$ in the atmosphere after ocean invasion interacts with the
					land biosphere\\
					7.& $t_{\operatorname{reverse}}$ & Reverse time & Returning to present climate equilibrium orbit & $10^6$ & Swinging back by renewed organic and oceanic uptake
				\end{tabular}
	}}}
\end{table}


\begin{figure}[ht]
	\includegraphics[scale=1.2]{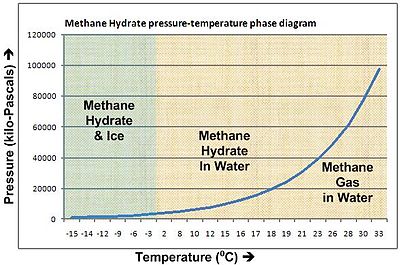}
	\caption{Methane clathrate is released as gas into the surrounding water column or soils when ambient temperature increases. Diagram based on data from \cite{Osegovic-Tatro-Holman:2006}. This work is released into the public domain by Wikipedia contributors, \url{https://commons.wikimedia.org/wiki/File:Methane_Hydrate_phase_diagram.jpg}, licensed under https://creativecommons.org/licenses/by-sa/3.0/.
	}
	\label{f:clathrate-gun-hypothesis}
\end{figure}

\begin{figure}[ht]
	\includegraphics[scale=0.64]{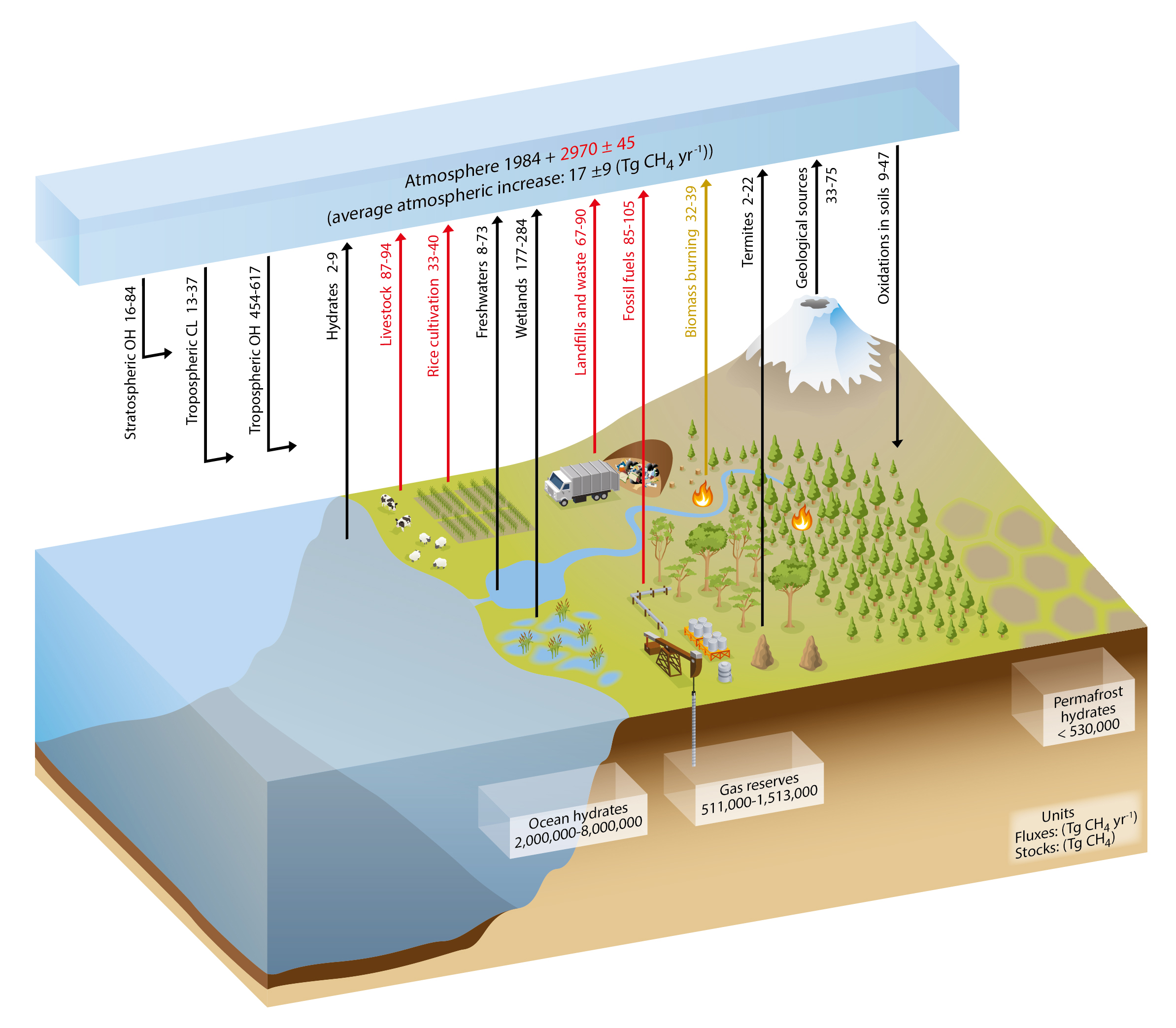}
	\caption{Global methane stocks and fluxes estimated for the 2000--2009 time period. Black arrows denote fluxes that are not directly caused by human activities since 1750, red arrows anthropogenic fluxes, and the light brown arrow a combined natural plus anthropogenic flux. From the Intergovernmental Panel on Climate Change \cite[Fig. 6.2, p. 474]{ClimateChange:2013}. PLSclear Permission Request 26 Sept., 2019}
	\label{f:methane-stocks-and-fluxes}
\end{figure}

\subsection{Multiple time scales in climate change modelling}\label{s:climate-change}

\subsubsection{Historical insight in the stability landscape of Earth}
The Earth was formed 4.5 billion years ago. Many details of Earth history are lost due to plate tectonics. However, there exists geological evidence that the climate has been significantly different from how it has been in recent geological periods where the Earth is predominantly in an  \textit{ice age} \textit{basin} (blue basin on \Cref{f:steffen-trajectories}) in the stability landscape. 
The climate fluctuated into shorter \textit{warm periods} where humans developed civilization eventually (red basin on \Cref{f:steffen-trajectories}). With the dramatic influence that humans have exerted on the climate, it is instructive to ask: how different can the climate of Earth be? By briefly discussing two examples from geology, we learn that the Earth can have widely different climates reflecting several basins in the stability landscape.

\paragraph{Example A. Snowball Earth} Imagine an Earth where ice glaciers stretch all the way from the poles to the equator. The hypothesis is known as \textit{Snowball Earth}, a term coined in a short and highly informative paper  \cite{Kirschvink:1992} by \textit{J. Kirschvink}. With reservations required by the incomplete and imperfect rock record, geologists suggest that the Earth was in a cycle where ice covered Earth several times around 700 million years ago. 
The albedo effect (reflection of light on the ice sheet, see below \Cref{ss:climate-change-albedo}) of ice is a feedback mechanism that can stabilize the Earth in such a cold state. 

However, carbon dioxide has possibly also played a crucial role, and interestingly the period ends with the Cambrian explosion where multi-cellular life appears in abundance. This explosion of life indicates an increase of carbon dioxide production, taking the Earth away from the cold state --- just in time before a further decrease of the temperature would have crystallized eventually produced and released CO$_2$ and kept the Earth in a permanent frozen stage. 

Geologists and paleontologist have evidence suggesting that the planet Earth had been subjected to oscillations between accelerating greenhouse and albedo feedback periods, extinguishing most life on Earth and so creating windows for the development of new types of life, but each until now with a turning point where the opposite feedback mechanism could take over.  Geophysicists and astrophysicists can imagine various control mechanisms like extreme volcanic activity or extreme deviations of Earth's obliquity  that were effective in releasing the feedback mechanisms and had or would possibly guard again against a runaway of one of the two feedback mechanisms. For details see \cite{Hoffman:2011} and the references given there. 

\paragraph{Example B. Warm Earth} Fifty five million years ago, the \textit{Paleocene-Eocene Thermal Maximum} occurred. At this time we would experience the opposite climate on Earth, where palm trees and crocodiles could be found in the Arctic. In this period the global temperature was 8 degrees warmer than today. For reference, the \textit{Medieval Warm Period} (ca. year 1100) and the \textit{Little Ice Age} (ca. year 1600) only represent temperature anomalies of about 0.2 degrees. Thus, these variations in historical times are small compared to the discussed scenarios.
The hypothetical \textit{Hothouse Earth} proposed by \textsc{Steffen} and co-authors \cite{Steffen:2018} is a scenario where the collapse of the Earth's ecosystem caused by human emission of carbon dioxide results in a warm Earth.
One feedback mechanism for this Hothouse Earth basin in the stability landscape is the release of methane clathrates to be discussed in the following section.

\subsubsection{Structural instability --- the example of the clathrate gun hypothesis}\label{ss:climate-change}
In our textbook model above in \Cref{ss:emergence+public}, we mentioned the \textit{clathrate gun hypothesis}, i.e., the threat of a 
giant methane release from the floor of the oceans and permafrost regions when the output variable $T$ of our toy model exceeds a critical and presently still unknown value and the level of greenhouse gases in the atmosphere can no longer be controlled by human activity but will be governed by autonomous processes of heat forced release of greenhouse gases from natural sources. That was purely hypothetical.

However, in parts of the scientific literature it is claimed that we are not in a hypothetical but in a real danger of structural instability of the climate, perhaps of passing thresholds beyond which human efforts to prevent catastrophic climate change will become futile because of feedback loops. 
To give examples for possibly threatening feedback mechanisms, one often points to the clathrate gun hypothesis, to be discussed in this subsection,  and to the diminishing albedo effect, to be discussed in the following subsection. Both feedback mechanisms are based on the emergence of multiple time scales, where the interaction of processes with different characteristic times may lead to structural instability, i.e., accelerating and perhaps uncontrollable runaway effects with unknown tipping points, limit points or limit cycles. 

For the physical science basis and the \textit{Clathrate Gun Hypothesis}, we refer to the comprehensive 2007-- and 2013--reports \cite{ClimateChange:2007, ClimateChange:2013} of the \textit{Intergovernmental Panel on Climate Change} (IPCC). For the key argument we quote the chemist \textsc{J.S. Avery}:
\begin{quote}
	\quad 
	If we look at the distant future, by far the most dangerous feedback loop
	involves methane hydrates or methane clathrates (a partly frozen slushy mix of methane gas
	and ice, usually found in sediments, i.e., crystalline water--based solids physically resembling ice,  in which the host molecule is water and the guest molecule is methane; their detailed formation and decomposition mechanisms are not fully understood, see \cite{Gao-House-Chapman:2005, Reay-et-al:2018}, \textit{added by the authors}).	When organic matter is
	carried into the oceans by rivers, it decays to form methane. The methane
	then combines with water to form hydrate crystals, which are stable at the
	temperatures and pressures which currently exist on ocean floors. However, if
	the temperature rises, the crystals become unstable, and methane gas bubbles
	up to the surface. Methane is a greenhouse gas which is 70 times as potent as
	CO$_2$.
	
	The worrying thing about the methane hydrate deposits on ocean floors
	is the enormous amount of carbon involved: roughly 10,000 gigatons. To put
	this huge amount into perspective, we can remember that the total amount of
	carbon in world CO$_2$ emissions since 1751 has only been 337 gigatons.
	A runaway, exponentially increasing, feedback loop involving methane hydrates
	could lead to one of the great geological extinction events that have
	periodically wiped out most of the animals and plants then living.\newline From \cite[Section 4.6]{Avery:2018}, reprinted by permission of \copyright The Danish Peace Academy.
\end{quote}
See also \Cref{f:clathrate-gun-hypothesis} of the methane dissolution depending on pressure and temperature. 

When talking about the possible release of methane clathrates, we have to take into regard the short half--life time of seven years of methane in the atmosphere and to distinguish between clathrate deposits at different sea levels. As a matter of fact, methane clathrates are common constituents of the shallow marine geosphere, they occur in deep sedimentary structures and form outcrops on the ocean floor. For estimates of the global methane stocks and annual fluxes see \Cref{f:methane-stocks-and-fluxes}.

Let us consider a clathrate deposit at a seafloor at 1000 m depth. The pressure at that depth is  100 atmospheres [at] $\sim$ 10,000 kilopascal [kps]. 
Apparently, even a substantial increase of the Global Mean Sea Level (GMSL), say by 10--15 m (most recent expectations in \cite[Ch. 4, Section 4.2]{ClimateChange:2019} are in the range of 0.5--2.3 m for the year 2100), will only add a negligible 1 at to that pressure, while a local temperature increase say by 8\degree C (most recent expectations in \cite[Sections 3.2.1.2.1 and 5.2]{ClimateChange:2019}) from presently 5\degree C average in tropical waters, and so surpassing the phase threshold of 13\degree C at the pressure of 100 at, will release the clathrates. 

Clearly, a release in shallow coastal waters happens already and will accelerate with any increase in atmospheric and ocean temperature. 

The situation is completely different at deep layers of the oceans. At the average 4000 m depth of the oceans, a temperature increase of 23\degree C would be required for the release, and at levels down to the deep rifts and reefs at 12,000 m depth, a release due to middle temperature increase is just unthinkable.

Then, how should one model the climate change potential of methane (CH$_4$)? As always in physics, we may proceed by isolating the object of interest, here the CH$_4$ stocks, flows and radiative effects. Hence, we begin with the following primitive compartment model for the impact of CH$_4$ on climate change. As a first approximation, the CH$_4$ impact is perceived as additive in relation to the CO$_2$ impact.

\begin{figure}[htbp]\centering
	
	\begin{tikzpicture}[scale=0.75, every node/.style={scale=0.75}]
	\node[rectangle,draw,fill=boxColor,minimum size = 1cm] (c) at (0,1) {$c$};
	\node[rectangle,draw,fill=boxColor,minimum size = 1cm] (m) at (0,-1) {$m$};
	\node[rectangle,draw,fill=boxColor,minimum size = 1cm] (heatC) at (3,1) {Heat($c$)};
	\node[rectangle,draw,fill=boxColor,minimum size = 1cm] (heatM) at (3,-1) {Heat($m$)};
	\node[circle,draw,fill=boxColor] (add) at (5,0) {\textbf{+}};
	\node[rectangle,draw,fill=boxColor,minimum size = 1cm,align = center] (totTemp) at (8,0) {$H(c,m)$\\$=H(c)+H(m)$};
	\draw[-Latex] (c.east) -- (heatC.west);
	\draw[-Latex] (m.east) -- (heatM.west);
	\draw[-Latex] (heatC.east) -- (add.north west);
	\draw[-Latex] (heatM.east) -- (add.south west);
	\draw[-Latex] (add.east) -- (totTemp.west);
	\node at ($(c.east)!0.5!(heatC.west)$) [above] {$\kappa$};
	\node at ($(m.east)!0.5!(heatM.west)$) [above] {$78 \kappa$};
	\end{tikzpicture}
	
	\caption{Superposition of the radiative climate forcing of atmospheric CO$_2$ (upper branch) and CH$_4$ (lower branch) considered as two independent processes. Here $c$ denotes an excess concentration in relation to the pre-industrial level of CO$_2$ at time $t_0$\/. That $c$ is kept fixed, say over a period of 20 years, and induces by radiative forcing with rate $\kappa$ an additional heat uptake $H(c)$ of absorbed solar energy on the Earth. Correspondingly, $m$ denotes an excess concentration of CH$_4$ degrading exponentially with a half--life of 7 years and inducing an additional heat uptake $H(c)$ of absorbed solar energy with an almost two decades more potent rate 84$\kappa$, resulting in a total heat increase of $H(c,m)=H(c)+H(m)$.}
	\label{fig:TikzDrawingSimple}
\end{figure}
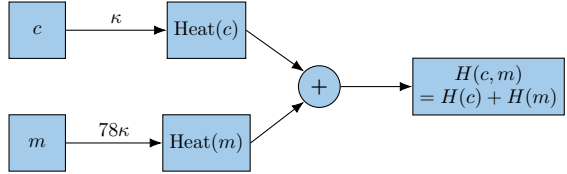

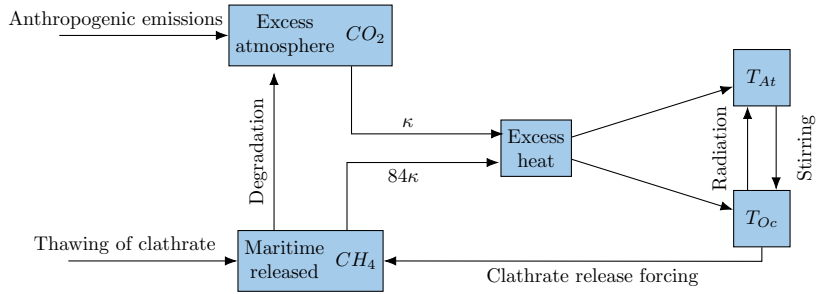
\begin{figure}[htbp]\centering
	
	\begin{tikzpicture}[scale=0.75, every node/.style={scale=0.75}]
	\node[rectangle,draw,fill=boxColor,minimum size = 1cm] (c) at (-4,2) {$\begin{matrix} \text{Excess} \\ \text{atmosphere} \end{matrix} \; \; CO_2$};
	\node[rectangle,draw,fill=boxColor,minimum size = 1cm] (m) at (-4,-2) {$\begin{matrix} \text{Maritime} \\ \text{released} \end{matrix} \; \; CH_4$};
	\node[rectangle,draw,fill=boxColor,minimum size = 1cm,align = center] (heat) at (0,0) {Excess \\ heat};
	\node[rectangle,draw,fill=boxColor,minimum size = 1cm] (tempA) at (4,1.25) {$T_{At}$}; 
	\node[rectangle,draw,fill=boxColor,minimum size = 1cm] (tempO) at (4,-1.25) {$T_{Oc}$};
	\draw[-Latex] ($(m.north)!0.5!(m.north west)$) -- ($(m.north)!0.5!(m.north west)+(0,2.8)$);
	\draw[-Latex] ($(m.north)!0.5!(m.north east)$) -- ($(m.north)!0.5!(m.north east)+(0,1.2)$) -> ($(m.north)!0.5!(m.north east)+(2.7,1.2)$);
	\draw[-Latex] ($(c.south)!0.5!(c.south east)$) -- ($(c.south)!0.5!(c.south east)+(0,-1.2)$) -> ($(c.south)!0.5!(c.south east)+(2.7,-1.2)$);
	\draw[-Latex] (heat) -- (tempA);
	\draw[-Latex] (heat) -- (tempO);
	\draw[-Latex] ($(tempO.north)!0.5!(tempO.north west)$) -- ($(tempA.south)!0.5!(tempA.south west)$);
	\draw[-Latex] ($(tempA.south)!0.5!(tempA.south east)$) -- ($(tempO.north)!0.5!(tempO.north east)$);
	\draw[-Latex] (tempO.south) -- ($(tempO.south)+(0,-0.25)$) -- (m.east);
	\draw[-Latex] ($(m.west) + (-3,0)$) -- (m.west);
	\draw[-Latex] ($(c.west) + (-3,0)$) -- (c.west);
	\node[above] at ($(m.west) + (-2,0)$) {Thawing of clathrate};
	\node[above] at ($(c.west) + (-2,0)$) {Anthropogenic emissions};
	\node[above,rotate = 90] at ($(m.north)!0.5!(m.north west)+(0,1.4)$) {Degradation};
	\node[above] at ($(c.south)!0.5!(c.south east)+(1,-1.2)$) {$\kappa$};
	\node[below] at ($(m.north)!0.5!(m.north east)+(1,1.2)$) {$84 \kappa$};
	\node[above,rotate = 90] at ($(tempA.south west)!0.5!(tempO.north west)$) {Radiation};
	\node[below,rotate = 90] at ($(tempA.south east)!0.5!(tempO.north east)$) {Stirring};
	\node[below] at ($(tempO.south)+(-3,-0.25)$) {Clathrate release forcing};
	
	\end{tikzpicture}
	
	\caption{Interaction/Coupling of two processes with different characteristic times. The essential new feature is dividing the temperature in atmospheric and oceanic (and soil) temperatures and introducing a feedback by making the clathrate release dependent of the temperature of oceans and                soils.}\label{fig:TikzDrawingAdvanced}
\end{figure}

Clearly, assuming mutual independence of the two processes leaves the superposition model of \Cref{fig:TikzDrawingSimple} qualitatively misleading, contrary to our theoretically more solidly founded interaction model sketched below in \Cref{fig:TikzDrawingAdvanced}. It seems, however, that the primitive superposition model fits nicely with our historical and Earth--historical data.

Here are the key data that are established: As mentioned above, we have vague data of giant CH$_4$ deposits in the Earth crust: in permafrost surface regions, in and beneath the continental slopes and deeper at the floors of the oceans. Drilling for commercial exploitation begins to provide a clearer picture at least for regions with depths of 1000-2000 m, see \cite{Milkov:2004, Li-et-al:2018, Zhang-et-al:2018}.

Much more precise knowledge is available about atmospheric CH$_4$. Some of the mass extinction of early Earth history (though not the last one that gave way for mammals) are explained by the release of CH$_4$ from the Earth crust due to catastrophic astrophysical collisions, geological eruptions, oceanic heat waves or tectonic faults. For the last 400 ky ice core records indicate oscillations of atmospheric CH$_4$ only between $\sim$ 350 and 750 ppb, most probably coinciding with glacial and interglacial periods. Since the beginning of the industrial revolution around 1750 the concentration increased to 1850 ppb, with a rate of $\sim$ 10 ppb/10 years in recent years (and a stagnation around 2000, probably due to a high energy price that enforced a more economic handling of natural gas production and distribution).

Atmospheric CH$_4$ mostly degrades by oxidation to CO$_2$ and H$_2$O and has, as mentioned before, a half--life of 7 years. It has absorption peaks at wavelength 7.7 and 3.3 $\mu$m, i.e., in the lower regions of the IR radiation due to the heat emission from the Earth's surface and exactly where carbon dioxide and water vapor have only low or no absorption. There are different methods to estimate the Global Warming Potential (GWP) of excesses in CH$_4$ relative to the GPW of excesses of CO$_2$ (\textit{excesses} relative to the radiative heat balance of the centuries before industrialization) out of direct radiative forcing and indirect forcing due to degradation. The most recent estimates for CH$_4$ (discussed in \cite[Table 1]{Reay-et-al:2018}) yield a GWP of $\sim$ 28 over 100 years and 84 over 20 years. Comparing with the corresponding CO$_2$ data and with a little calculation, one finds that about one quarter or one-third of the Global Mean Temperature increase since pre-industrial time can be explained by CH$_4$ increase. 

These data fit nicely with the primitive model of \Cref{fig:TikzDrawingSimple}: they suggest that, historically, the release of methane has mostly been independent of global mean temperature, be it by natural disasters or by agricultural and industrial activity. 

However, it is almost a textbook example of the necessity of multiscale modelling and simulation that we should not trust the primitive model, no matter how nicely it fits with the given data, when we actually are aware of possible interactions of the two radiative processes. Alarming is perhaps not so much the clathrate-gun hypothesis alone, but what we know from \textit{physical chemistry} about the role of the temperature in the phase diagram of methane, what becomes evident in the recent \textit{exploratory drilling} for natural gas in coastal waters (see above), and what types of runaway effects we can derive \textit{mathematically} from the interaction of linear processes with different characteristic times. 

Therefore we close this section by a schematic drawing of an integrated compartment model for the radiative forcing of climate change by interaction of the methane and the carbon dioxide processes. It is speculative since nobody has data about the position, depth and volume of the clathrate deposits in the permafrost regions and beneath the floor of the oceans. But it is realistic, since we know the simple mechanisms.

One word more about the structural instability typically associated with the interaction of processes with a multiplicity of characteristic times. The mathematical possibility of runaway effects, however, is only one side of the described feedback. Equally well, we can, as always in control theory, point to the potentially stabilizing effects of feedback mechanisms for else structurally unstable processes. The observed oscillations between higher and lower atmospheric CH$_4$ levels in glacial and interglacial periods may have had such stabilizing effects for an else highly volatile climate development.

%
%
%

\subsubsection{The feedback mechanisms of thawing ice sheets, rising sea level and diminishing albedo}\label{ss:climate-change-albedo}
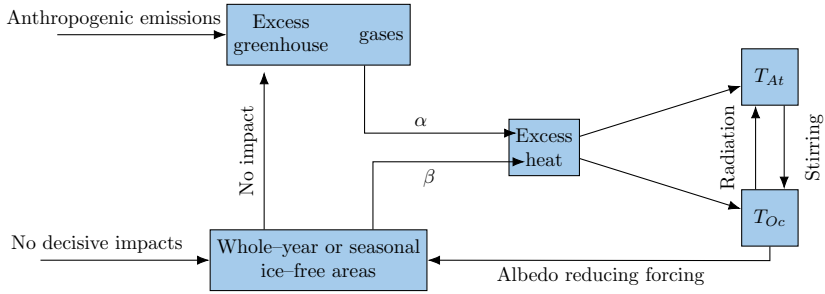
\begin{figure}[htbp]\centering
	
	\begin{tikzpicture}[scale=0.75, every node/.style={scale=0.75}]
	\node[rectangle,draw,fill=boxColor,minimum size = 1cm] (c) at (-4,2) {$\begin{matrix} \text{Excess} \\ \text{greenhouse} \end{matrix}$ \; \; gases};
	\node[rectangle,draw,fill=boxColor,minimum size = 1cm] (m) at (-4,-2) {$\begin{matrix} \text{Whole--year or seasonal} \\ \text{ice--free areas} \end{matrix}$};
	\node[rectangle,draw,fill=boxColor,minimum size = 1cm,align = center] (heat) at (0,0) {Excess \\ heat};
	\node[rectangle,draw,fill=boxColor,minimum size = 1cm] (tempA) at (4,1.25) {$T_{At}$}; 
	\node[rectangle,draw,fill=boxColor,minimum size = 1cm] (tempO) at (4,-1.25) {$T_{Oc}$};
	\draw[-Latex] ($(m.north)!0.5!(m.north west)$) -- ($(m.north)!0.5!(m.north west)+(0,2.8)$);
	\draw[-Latex] ($(m.north)!0.5!(m.north east)$) -- ($(m.north)!0.5!(m.north east)+(0,1.2)$) -> ($(m.north)!0.5!(m.north east)+(2.7,1.2)$);
	\draw[-Latex] ($(c.south)!0.5!(c.south east)$) -- ($(c.south)!0.5!(c.south east)+(0,-1.2)$) -> ($(c.south)!0.5!(c.south east)+(2.7,-1.2)$);
	\draw[-Latex] (heat) -- (tempA);
	\draw[-Latex] (heat) -- (tempO);
	\draw[-Latex] ($(tempO.north)!0.5!(tempO.north west)$) -- ($(tempA.south)!0.5!(tempA.south west)$);
	\draw[-Latex] ($(tempA.south)!0.5!(tempA.south east)$) -- ($(tempO.north)!0.5!(tempO.north east)$);
	\draw[-Latex] (tempO.south) -- ($(tempO.south)+(0,-0.25)$) -- (m.east);
	\draw[-Latex] ($(m.west) + (-3,0)$) -- (m.west);
	\draw[-Latex] ($(c.west) + (-3,0)$) -- (c.west);
	\node[above] at ($(m.west) + (-2,0)$) {No decisive impacts};
	\node[above] at ($(c.west) + (-2,0)$) {Anthropogenic emissions};
	\node[above,rotate = 90] at ($(m.north)!0.5!(m.north west)+(0,1.4)$) {No impact};
	\node[above] at ($(c.south)!0.5!(c.south east)+(1,-1.2)$) {$\alpha$};
	\node[below] at ($(m.north)!0.5!(m.north east)+(1,1.2)$) {$\beta$};
	\node[above,rotate = 90] at ($(tempA.south west)!0.5!(tempO.north west)$) {Radiation};
	\node[below,rotate = 90] at ($(tempA.south east)!0.5!(tempO.north east)$) {Stirring};
	\node[below] at ($(tempO.south)+(-3,-0.25)$) {Albedo reducing forcing};
	
	\end{tikzpicture}
	
	\caption{Interaction/Coupling of the radiative heating of the atmosphere and the oceans of a supposed rate $\alpha$  with the induced reduction of the Earth's albedo, inducing a decrease of the cooling effects by a rate $\-beta$. The decisive difference to \Cref{fig:TikzDrawingAdvanced} is that in this model the change of the albedo is driven only by climate change and, in difference to the methane model, external drivers can be neglected.}\label{fig:TikzDrawingAlbedo}
\end{figure}

The observable increase of global mean temperature with its special features in subpolar regions drives the thawing of the ice sheets and the rise of the sea level. Both the decrease of the ice covered areas and the decrease of land areas reduce the albedo of the Earth and accelerate climate change --- similarly with the increase of methane emission. There is an important difference though: Contrary to the preceding discussion of methane, it seems now that a superposition model, i.e., treating the processes as independent, is meaningless.  Applying an interaction model similar to \Cref{fig:TikzDrawingAdvanced} is mandatory, see \Cref{fig:TikzDrawingAlbedo} with a compartment model that is simpler than that of \Cref{fig:TikzDrawingAdvanced}.

\subsection{Challenges, failures, and misconceptions in climate change modelling}
To sum up, we emphasize a few common misconceptions in climate modelling when disregarding the emergence of a multiplicity of time scales. Such disorientation can happen also in environmental administration and in climate change mitigation and adaptation grass-roots  movements, when stakeholders occasionally solely follow feelings and political trends and  focus only on short-range or only on long-range effects. For the underlying general mathematical problems of multiscale sampling, modelling and simulation we refer to \textsc{J.D. Logan}'s textbook \cite{Logan:2013}, passages of the encyclopedia \cite{Engquist:2015}, learned journals like SIAM's \textit{Multiscale Modeling and Simulation} (MMS) and monographs like \cite{E:2011, Engquist-et-al:2005, Engquist-et-al:2009, Horstemeyer:2012, Majda:2016}. 

\subsubsection{Sampling problems}\label{sss:sampling}
Properties of systems can become time dependent in the sense that what you measure depends on the time scale of the measurement.  We shall touch upon that below in \Cref{s:matter,s:life,s:society}.

\subsubsection{Truncation errors in multiscale numerical simulation}
Leaving the multiscale problems in climate \textit{modelling} aside, we shall give just one example of the intricacies of multiscale \textit{computational methods} in climate modelling from every-day simulation experience in atmospheric science,  following the geophysicist \textsc{R. Klein} in \cite[p. 1002 and p. 1004]{Klein:2015}. Roughly speaking, we distinguish between diabatic and adiabatic temperature changes. Diabatic changes are very slow and at small rate --- but irreversible and so decisive for temperature changes in the long run, while adiabatic changes, being more frequent and at larger rates, are reversible and so negligible in the long run. Hence, there are two different characteristic times, $t_d$ for diabatic changes and $t_a$ for adiabatic changes with $t_d\gg t_a$\/. Truncation errors from the discretization of the adiabatic processes become of an irreversible character and can dominate in the long run over the essential diabatic changes. Worst of all, we can not do without simulating the adiabatic processes: they are needed to calculate the diabatic changes. 
\subsubsection{The bias of multi--model based projections} To solve multiscale problems, a natural first--order approximation is the decomposition of the problem in a multitude of submodels, each with its own characteristic scale, and then patching the results by a averaging process. 
The Danish meteorologists \textsc{Madsen, Langen, Boberg and Christensen} point to the systematic failure of that way of dealing with multiscale problems in \textit{multi-model based projections}: In \cite{Madsen-et-al:2017}, they show an {inflated} uncertainty in {multimodel--based regional climate projections}. Roughly speaking, the complexity of atmospheric physics does not permit precise global and longterm climate simulations. Therefore, regional longterm projections are typically based on patching multiple models together to obtain the geographical distribution of the multimodel mean results. Trivially, that procedure runs into the probabilistic intricacies of taking means of non-comparable magnitudes. \textit{Consequently}, as the Danes write, \textit{the risk of unwanted impacts may be overestimated} (or underestimated, {our insertion}) \textit{at larger scales as climate change impacts will never be realized as the worst} (or best, our insertion) \textit{case everywhere}. 

\subsubsection{Unfounded linearizations} A related common failure of dealing with multiscale problems is approximating the underlying equations by \textit{linearization}. 
That is, e.g., the case in thoughtless use of the concept of Global Warming Potential (GWP). In \cite[Section 2.10]{ClimateChange:2007}, the comprehensive 2007-IPCC report on the physical science basis of climate change compares the anticipated climate change impact of a compound $i$ (say methane or an aerosol) with the anticipated climate change impact of the reference substance $r:=CO_2$ by setting
\begin{equation}\label{e:gwp}
GWP_i \ :=\ \frac{\int_0^{TH} RF_i(t)\, dt}{\int_0^{TH} RF_r(t)\, dt}\ =\
\frac{\int_0^{TH} a_i\cdot [C_i(t)]\, dt}{\int_0^{TH} a_r\cdot [C_r(t)]\, dt},
\end{equation} 
where $TH$ denotes the choice of a time horizon, important for evaluating differences in the degradation/ocean- and land-depositing processes; $RF_i$ and $RF_r$ the global mean radiative forcing  of components $i,r$ with $RF_i< 0$ for $i$ aerosol; $a_i$ and $a_r$ the radiative forcing per unit mass increase in atmospheric abundance  of components $i,r$ (radiative efficiency), and $[C_i(t)]$ and $[C_r(t)]$ the time-dependent abundance of the components. Note that the radiative efficiency is considered as being scale independent, i.e., the pattern of absorption and scattering is considered as fixed and so the radiative forcing as linear in the concentrations. 

Equation \eqref{e:gwp} may be useful to tune multi-component abatement strategies by providing numerical values for
the trade-off between emissions of different forcing agents, in particular after the minor corrections made in the more recent comprehensive 2013-IPCC report on the physical science basis of climate change \cite[Section TS.3.8 and Section 8.7]{ClimateChange:2013}.
However, the nominator and denominator of \eqref{e:gwp} itself are anticipated; they are fancied and do not yield appropriate impact functions but would be gravely misleading:  
Clearly, the greenhouse effect of a thin layer of CO$_2$ molecules can be both calculated and measured in a laboratory, contrary to the greenhouse effect of the 700 km thick troposphere, stratosphere, mesosphere, and thermosphere layers. Any assumption of approximatively proportionality will lead astray due to the non--linear radiative interaction. 

One has to worry that there is no longer a characteristic time or a characteristic temperature difference to observe. We may have already fabricated the irrevocable preconditions for a hot--bed path towards large-scale climate changes following a set pattern and being yet beyond a tipping point. 

\subsubsection{Ill--posed problems and the \textit{butterfly} effect} 

As discussed in \cite[Paragraph 11.10.1.2]{ClimateChange:2007}, another source of uncertainty originates from the \textit{ill--posedness} of the initial conditions, boundary conditions and coefficients of atmospheric equations. In climate change modelling, however, it can be misleading to emphasize the so-called \textit{butterfly} effect, i.e., a supposed extremely high dependence of the outcome of a dynamical process on a small variation of the initial values: In mathematics, evolutionary processes over manifolds \textit{with boundary} that are subjected to strong boundary conditions (like the external radiative forcing of the climate on Earth) will typically be governed by the boundary conditions in the long run and not by the butterfly effect that plays a role in dynamical processes only on short time scales (besides its nasty consequences also in the long run in iterative numerical schemes).

\subsubsection{The \textit{atomism} of modern science}\label{sss:atomism}
Half a century ago, environmental scientists, experts in municipal waste management like \textsc{Barry Commoner} (1917--2012), were the first pointing to recurrent failures in science-based approaches to ecological problems in agriculture and environment. Perhaps the most shocking example was the well-intended science-based introduction of DDT to check insect pests, which in the long run \textit{caused} insect pests by eliminating also the insect predators --- and many similar cases of 
well-intended science-based failures, e.g., in sewage disposal or the design of power systems, see \cite{Commoner:1992} or the condensed review \cite{Commoner:1971}:
\begin{quote}
	If modern technology has failed, there must be something wrong as well with our science, which generates technology. Modern science operates well as long as the system of interest is not complex. We can understand the physical relationship between two particles, but add a third particle and the problem becomes extraordinary difficult. Modern science has only poor methods for dealing with systems that are characterized by complex interactions... The tendency to \textit{atomize reality} (our emphasis) is a fundamental fault of modern science. (l.c., p. 177-178)  
\end{quote}

\subsubsection{Discarding secondary effects and slipping across thresholds} In our context, an implementation of that forlorn atomism  is focusing on processes only in one of the multiple time scales and neglecting processes dominated by concurring and immanent other relevant characteristic time scales like \textit{secondary effects}, as discussed in our \Cref{ss:climate-change}, or \textit{slipping involuntarily and inadvertently} into another orbit, as sketched in \Cref{ss:emergence+public} and exemplified further below in \Cref{s:matter} in materials science and in \Cref{s:life} in biomedicine.

\subsubsection{Overparametrisation}\label{sss:overparametrisation}
In  classical mathematical physics, we oppose over\-parametrisation, whether it appears in the introduction of parameters without physical meaning or in the use of excessively many parameters. In environmental science, climate science, materials science, bioscience and mathematical economics, it seems to us, that overparametrisation may be, next to the opposite traditional atomistic reductionism, one of the worst --- and most common --- failures, see also \Cref{sss:modelling-living-tissue} below.

\section{Multiple time scales of matter --- Viscosity of soft materials}\label{s:matter}

\subsection{Classes of multiscale problems}
\begin{figure}[ht]
	\includegraphics[scale=0.84]{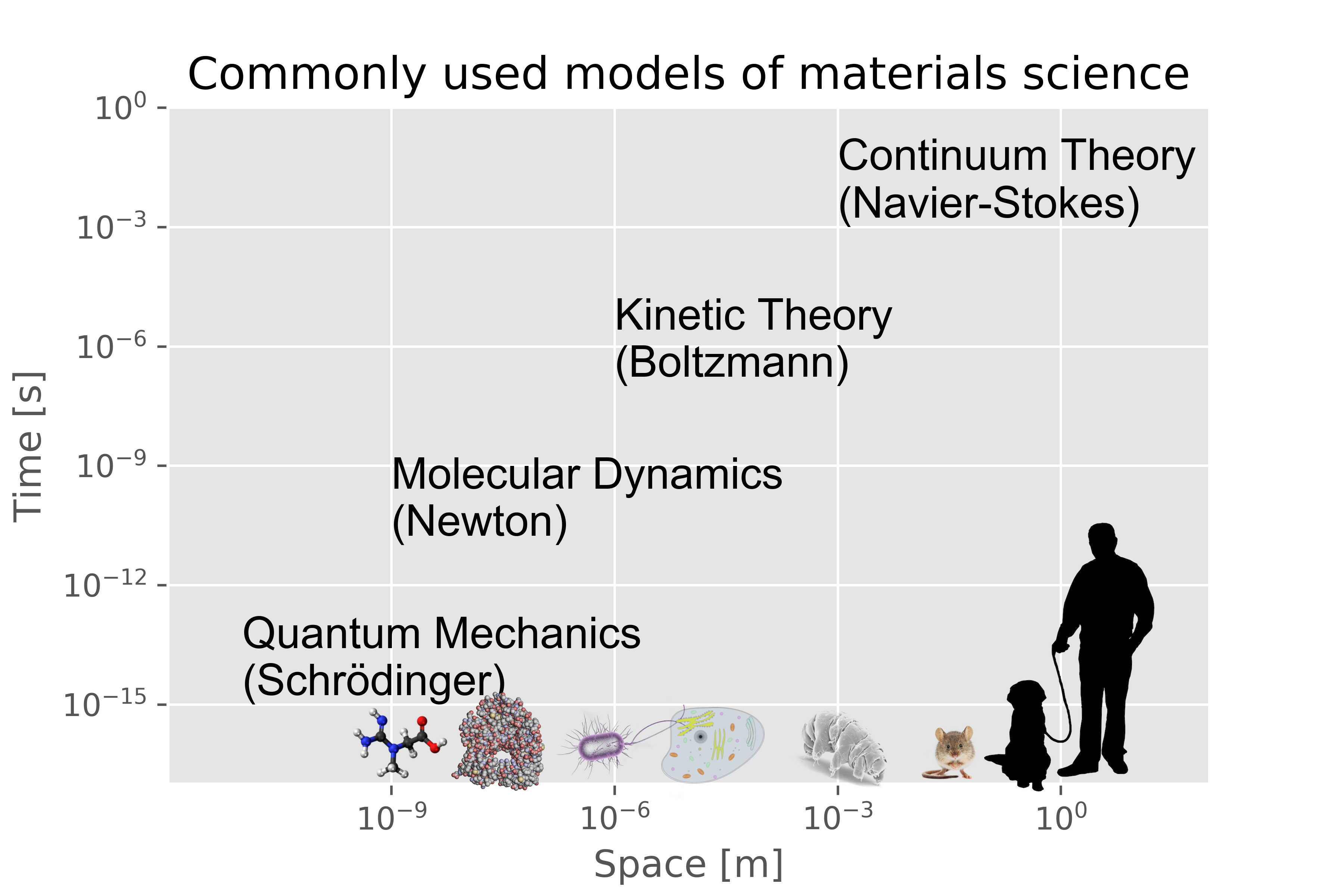}
	\caption{Commonly used models of physics on different scales. Confinement to one selected space and time scale and one type of model (one first principle) may explain the success of mathematical modelling and simulation in large parts of physics and engineering. Traditionally, that has made these fields ``simple" fields from a multiscale point of view, contrary to life sciences and social sciences that have to do without a sieve of first principles and so have no naturally given separation of scales. New challenges like climate change and new observation and simulation facilities, however, have widened the gate to multiscale modelling and simulation also in physics and engineering. Inspired by  \textsc{E} and \textsc{Engquist} \cite[Fig. 1]{E-Engquist:2003}.}
	\label{f:laws-and-scales}
\end{figure}

Most problems in materials science have multiple time scales. A chemical reaction, for example, may begin slowly and the concentration changes little over a long time; then, the reaction may suddenly go to completion with a large change in concentration
over a short time. There are two time scales involved in such a process. 
Another example occurs in fluid flow, where the processes of heat diffusion, advection, and possible chemical reaction all have different scales. The processes at different time scales are governed by physical laws of different character, see \Cref{f:laws-and-scales}.

The mathematicians \textsc{W. E} and \textsc{B. Engquist} distinguish between two classes of multiscale problems, see \cite[Section 1.4.3]{E:2011} and \cite[p. 1068f]{E-Engquist:2003}. \textit{Type A problems} are problems with localized defects around which microscopic models have to be used; elsewhere one can use some macroscopic models. As example they mention the crack propagation in solids. \textit{Type B problems} are those for which the microscale model is needed everywhere either as a
supplement to or as the replacement of the macroscale model. This occurs, for example,
when the microscale model is needed to supply the missing constitutive relation in the
macroscale model. Below in \Cref{s:life} we shall present a type B model that describes the insulin secretion of a glucose stimulated beta cell by a macroscale model in the time range of minutes and the space range of $\mu$m, but depends on a microscale model of the electrical input of Ca$^{++}$ oscillations in microdomains in the time range of seconds and the space range of nm.

In this section we present a kind of \textit{type C multiscale model}, where there is no multiplicity of models and scales to begin with and the multiple time scales emerge in the course of the model application.

\subsection{The emergence of multiplicity of time-scales in liquid dynamics}
\label{ss:emergence} 

\begin{figure}
	\includegraphics[width=13.cm]{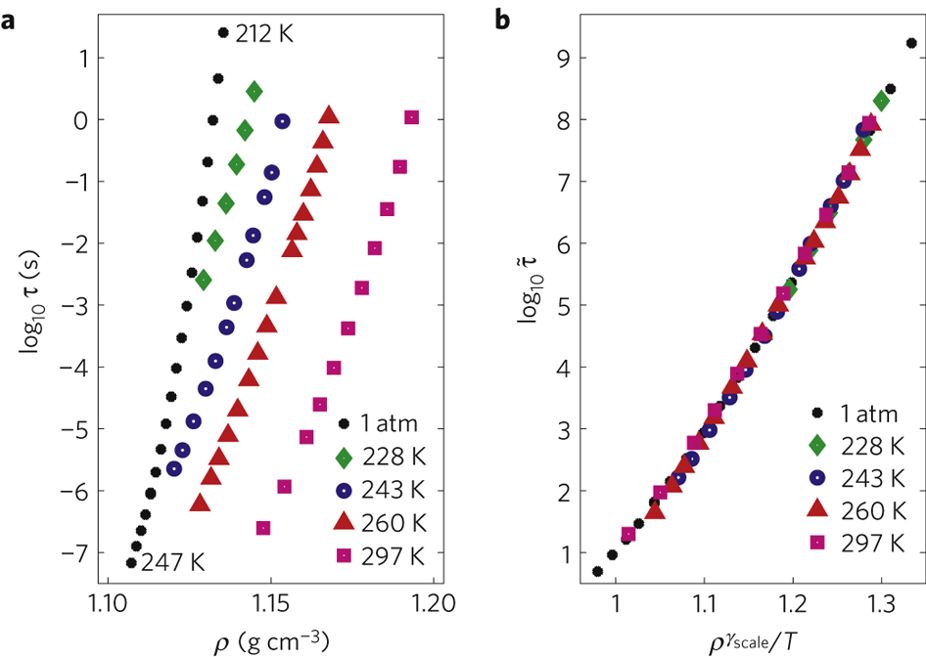}
	\caption{(a) Measurements of the \textit{dielectric relaxation time} $\tau$, 
		as a function of mass density $\rho$ (while in the text we use the number density: the molar mass gives the proportionality between the two densities), of the silicone oil DC704. The measured time $\tau$ is related to the rotation time of molecules. Measurements are done along isotherms of four different choices of the temperature $T$ and along the atmospheric pressure isobar (1 atm). The relaxation time at ambient pressure (black dots) changes more than eight orders over a narrow temperature span of only 35 degrees. (b) The \textit{reduced} (i.e., dimensionless) relaxation times $\tilde\tau:=\tau \rho^{1/3}\sqrt{m/T}$, where $m$ denotes the mass of the molecule. Interestingly, the different $\tilde{\tau}$ collapse into a single master curve when plotted against $\Gamma=\rho^{\gamma_\textrm{scale}}/T$, where $\gamma_\textrm{scale}=6$ is a scaling exponent. This indicates a hidden scale invariance that collapses the phase diagram from two variables ($\rho$ and $T$) into a single ($\Gamma$).  In the text, we investigate a model where particles interact via inverse power-laws: $r^{-n}$ where $r$ denotes the distance between two pairs. This model also possesses this scale invariance with $\gamma_\text{scale}=n/d$ where $d$ is the dimension of space. Thus, we only have to investigate a single state variable. We choose temperature. Reproduced from \cite[Fig. 1]{Gundermann:2011} with the publisher's permission.}
	\label{f:Gundermann}
\end{figure}

When liquids are cooled, dynamics may suddenly become dramatically slow. As an example, \Cref{f:Gundermann} shows results of measurements on a silicone oil (chosen since it is chemically stable resulting in reproducible measurements). 

\subsubsection{Relaxation time as function of mass density}
Specifically, the measured quantity is the \textit{dielectric relaxation time}. In layman terms, it describes how fast the molecules rotate. Ignoring interactions between molecules, a \textit{back of the envelope} calculation suggests that the rotational time should be in the order of 1 picosecond [ps]. However, the measured times are significantly slower due to collective dynamics.
The sluggish dynamics reflect the emergence of a slow time-scale in the system \cite{Donth:1982,Sastry:1998,Garrahan:2002,Dyre:2006,Biroli:2013}.
Below we will look into answering the question ``Why are the dynamics of cold liquids surprisingly slow?'' by studying a model liquid.

From a reductionist viewpoint one should apply Quantum Mechanics to understand the dynamics of liquids (lower left corner in \Cref{f:laws-and-scales}).
In that theoretical framework, the state of the liquid is described by a multiparticle wave-function $|\Psi \rangle$. This mathematical object contains information about all relevant subatomic particles --- electrons, neutrons and protons. For a given Hamiltonian $\mathcal{\widehat H}$, the time propagation of the wave-function  can in principle be computed by solving the equations of motions suggested by \textsc{E. Schr{\"o}dinger} in 1925: $i|\dot\Psi \rangle=\mathcal{\widehat H}|\Psi \rangle$ (here the dot refers to a time derivative). Unfortunately, only a few textbook examples like the harmonic oscillator or a particle in an infinite square well can be solved analytically. 
Today, computers are routinely used to solve the equations of motions of atomic systems on the picoseconds time scale using clever approximations. 
This timescale is long enough to understand liquid dynamics at high temperatures. However, longer time-scales are needed to understand the dynamics of cold liquids.
Thus, we cannot hope to solve the Schr{\"o}dinger equation explicitly. 
Instead we will address the question using a classical potential that approximates the true Quantum Mechanical energy surface and dynamics.

We investigate a classical Hamiltonian using one of the numerical integration methods generally referred to as \textit{Molecular Dynamics} \cite{Frenkel:2002}:
Consider $N$ particles on a \textit{d}-dimensional torus (i.e. a  periodic \textit{d}-dimensional box) of volume $V=L^d$, where $L$ is the side length. For simplicity we study a $d=2$ dimensional liquid. Let the $d\times N$ dimensional collective coordinate be ${\bf R} := \{{\bf r}_1,{\bf r}_1,\ldots,{\bf r}_N\}$, so the potential energy function is $U({\bf R})$ (defined in the paragraph below). The (classical) Hamiltonian $\mathcal{H}$ is the sum of the potential and the kinetic energy: $\mathcal{H}({\bf R},\dot {\bf R}) = U({\bf R}) + K(\dot {\bf R})$, where $K(\dot {\bf R}) :=\frac{1}{2}\sum_i^N m_i|{\bf v}_i|^2$, $m_i$ denotes the mass and ${\bf v}_i:=\dot{\bf r}_i$ denotes the velocity of particle $i$.

The dynamics of the system is computed numerically by solving Newtons classical equations of motion using a leap-frog algorithm: If $\Delta t$ is a time step and ${\bf F}_i=-\nabla_iU$ is the force on particle $i$, then the next velocity and position in an adjacent timestep is found to be
\[
{\bf v}_{t+\Delta t/2}^{(i)}\ =\ {\bf v}_{t-\Delta t/2}^{(i)}+{\bf F}_i\Delta t/m_i \text{ and }  {\bf r}_{t+\Delta t}^{(i)}\ =\ {\bf r}_t^{(i)}+{\bf v}_{t+\Delta t/2}^{(i)}\Delta t,
\] 
correspondingly. This integration scheme is symplectic and the same trajectory is generated if time is reversed. Thus, there is no systematic drift of the total energy (except from numerical truncation of floating points), contrary to the popular fourth order Runge-Kutta (RK4) integration scheme.

The \textit{kinetic temperature} of a system is
\[
T\ :=\ \frac 1{k_B}\langle\sum_i^N(m_i |{\bf v}_i|^2)/N_f\rangle, 
\]
where $\langle \cdots \rangle$ is a time average, $k_B$ denotes the Boltzmann constant and $N_f:=dN-d$ is the number of degrees of freedom in the system (the removal of $d$ degrees of freedom accounts for the fixed total momentum). The temperature is determined by the initial positions and velocities of the particles. Alternatively we can control the temperature by coupling our system to a heat bath with some temperature as done with a Langevin thermostat: Imagine a thin gas (the heat bath) interacting weakly with particles in the system. The particles in the gas will apply a drag force, and random kicks to particles of the system. In the above mentioned algorithm we can model this by computing the force on particle $i$ as 
\[
{\bf F}_i=-\nabla_iU+\gamma \bar v_i +\sqrt{2\gamma k_B T} \mathcal{R}(t),
\]
where $\gamma$ determines the coupling with the heat bath, $\bar v_i$ denotes the velocity of the particle, and $\mathcal{R}(t)$ is a delta-correlated Gaussian process with zero-mean: $\langle\mathcal{R}(t)\mathcal{R}(t')\rangle=\delta(t-t')$ with the distribution $P(\mathcal{R}):=\exp(-\mathcal{R}^2/2)/\sqrt{2\pi}$. 

We need to define a potential energy function, $U({\bf R})$, of a model liquid not prone to crystallization (since we are interested in the liquid state). Inspired by the Kob-Andersen binary inverse power-law (KABIP) model presented in \cite{Pedersen:2010} we use a model where the potential energy function is a sum of inverse power laws in the pair distances: $U({\bf R})=\sum_{i>j}^N \varepsilon u(|{\bf r}_j-{\bf r}_i|/\sigma\sigma_{ij})$, where the pair energy function for a given dimensionless pair diameter $\sigma_{ij}$ is $u(r) = r^{-18}-1.5^{-18}$ for $r<1.5$ and zero otherwise ($\varepsilon$ and $\sigma$ are discussed below). The truncation of the potential at 1.5 makes computations faster, since forces only have to be computed between neighbors. Fortunately, since $1.5^{-18}\ll1$, the truncation does not much influence results at the investigated temperatures. The interaction parameters between different types of pairs are $\sigma_{AA}=1.1$ and $\sigma_{AB}=\sigma_{BB}=0.9$. The parameters $\varepsilon$ and $\sigma$ set an energy- and a length scale, respectively.  All particles are given the same mass $m_i=m$. Results are presented in units derived from $\varepsilon$, $\sigma$, $m$, and the Boltzmann constant $k_B$.

We shall investigate a system of $N=1600$ particles at number density $\rho=N/V=\sigma^{-d}$ consisting to 70\% of the larger A particles and to 30\% of the smaller B particles (i.e. the system size is $L=40\sigma$). As before, $d$ denotes the dimension of the space, set to $d:=2$ for numerical simplicity.
The model is implemented into the RUMD software package \cite{rumd} that utilizes graphics cards for fast computations. The code also uses a \textit{neighbor list} and a \textit{cell list} resulting in an algorithm where the computational time only scales as the number of particles $N$ (not $N^2$ as it would be expected if the force on particle $i$ depends on the positions of all other particles). We refer readers to a standard textbook on Molecular Dynamics for more details:  see \cite{Frenkel:2002}. 

\subsubsection{Low temperature dynamics}
Below we will investigate the dynamics as a function of the temperature with a focus on low-temperature dynamics. However, let us first do a {\it back of the envelope} calculation where we use a mean-field approximation, i.e., our calculation will only evolve averages such as the mean density of particles ($\rho$) and the temperature.
A characteristic time is given as the average time it takes two particles to encounter. Let us assume that this is when a particle has traveled 10\% of an inter-particle distance $l_0=0.1\sqrt[d]{1/\rho}$. The average velocity is $v_0=\sqrt{dk_BT/m}$ (assuming the system is large, $N_f=N$).
Thus, the inter-particle collision time is expected to be in the order of 
\[
t_0\ =\ l_0/v_0\ =\ 0.1\sqrt[d]{1/\rho}\sqrt{m/dk_BT}\ \fequal{d=2}\ 0.1\sqrt{m/2\rho k_BT} 
\]
in our case.
For short times, $t\ll t_0$, the particles are expected to move ballistically: ${\bf r}_i(t)={\bf r}_i(0)+{\bf v}_i(0)t$. For long times,  $t\gg t_0$, particles will have many encounters and the movement becomes diffusive: $\langle|{\bf r}_i(t)-{\bf r}_i(0)|^2\rangle = 2dDt$, where $R^2\equiv\langle|{\bf r}_i(t)-{\bf r}_i(0)|^2\rangle$ is the mean squared displacement and $D$ the diffusion constant.

\begin{figure}[ht]
	\includegraphics[width=6.4cm]{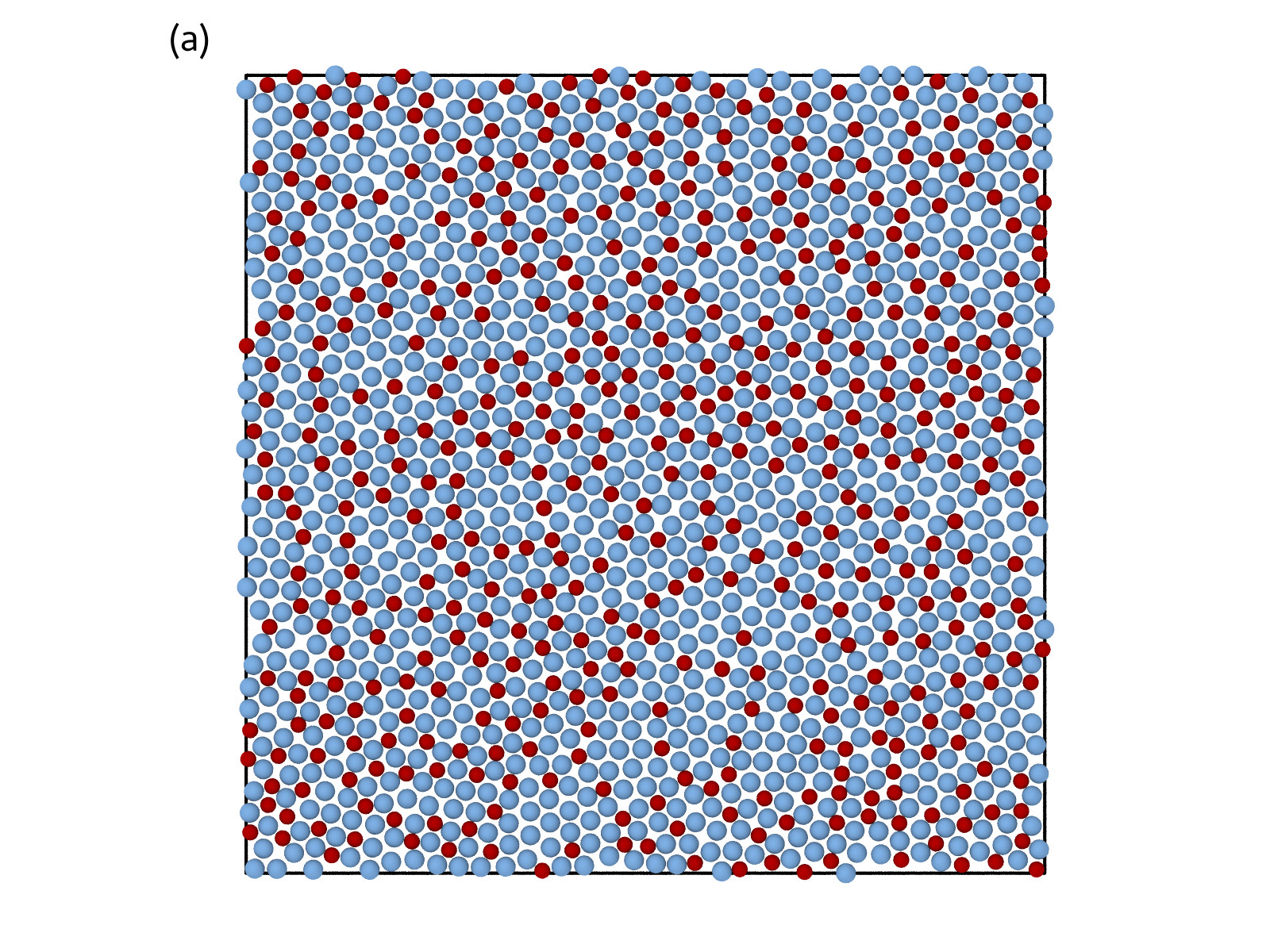}
	\includegraphics[width=6.5cm]{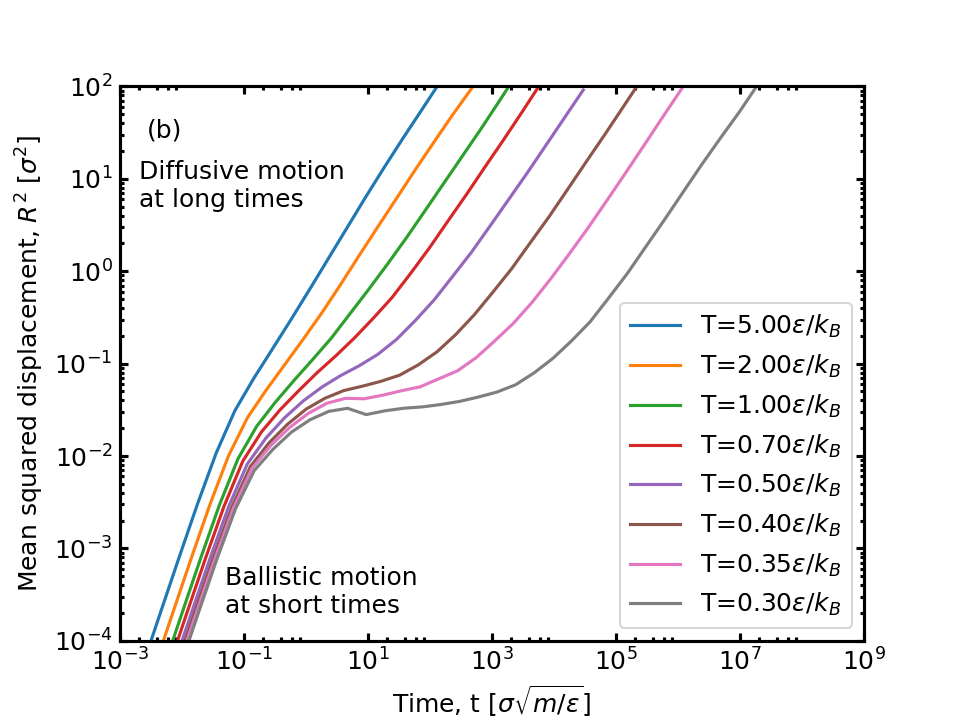}
	\includegraphics[width=6.4cm]{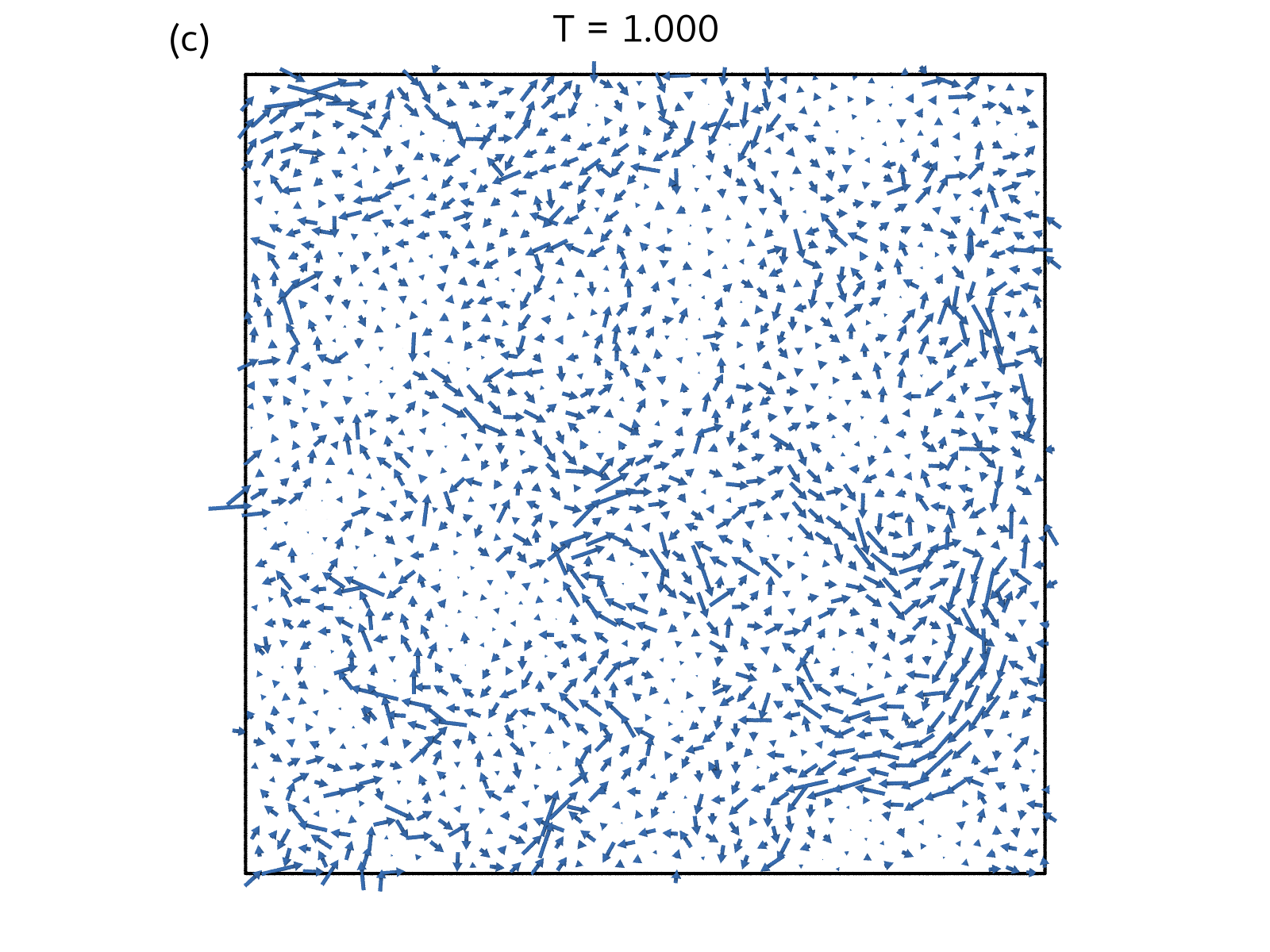}
	\includegraphics[width=6.5cm]{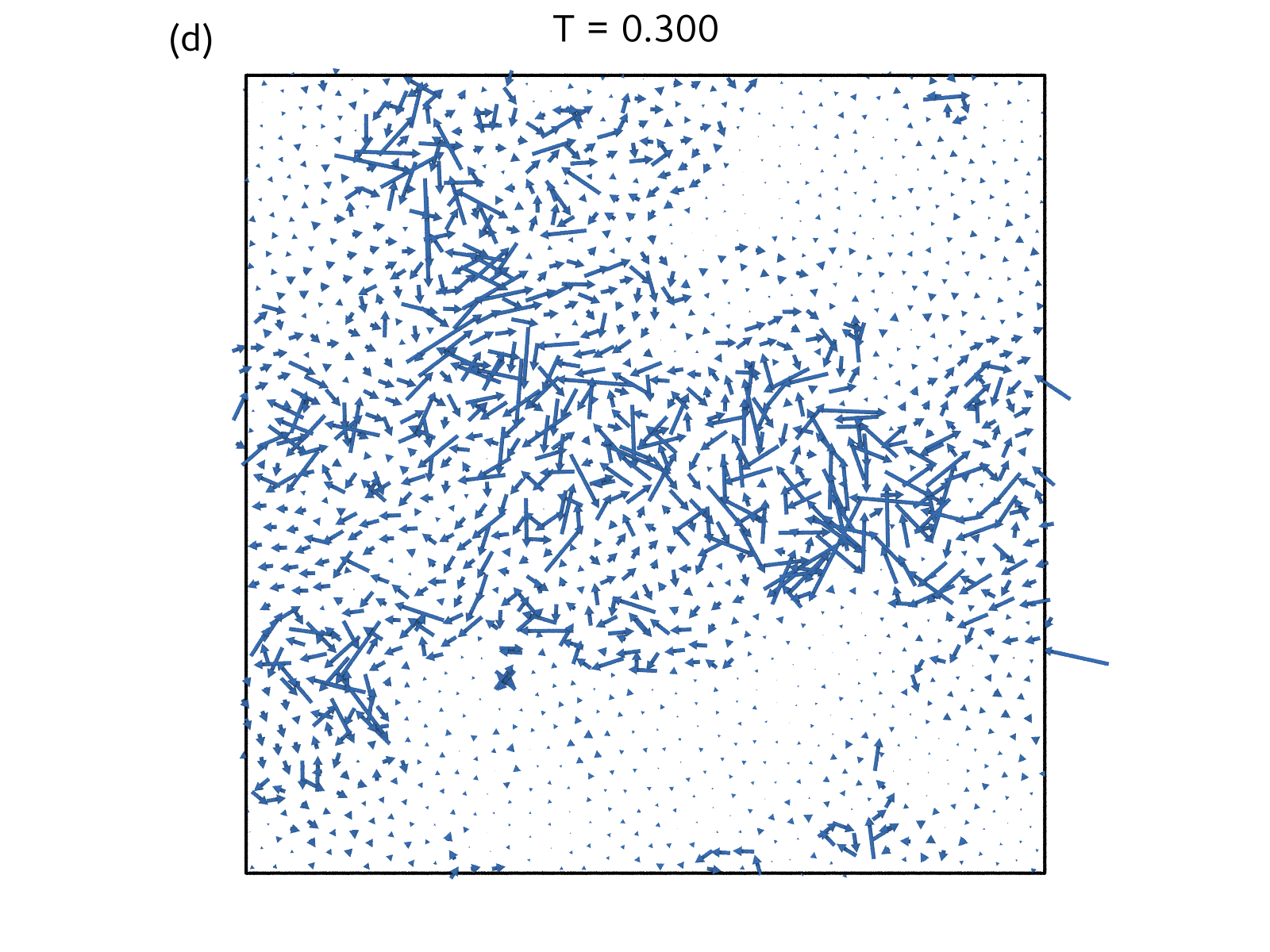}
	\caption{Two dimensional model of a liquid consisting of large A particles (blue) and smaller B particles (red). (a) Representative configuration at $T=0.3\varepsilon/k_B$. (b) The mean squared displacement, $R^2$ at a range of temperatures. A slow time-scale emerges at low temperatures signaled by the appearance of a caging plateau in the mean squared displacement. (c) Displacement vectors at temperature $T=\varepsilon/k_B$ and (d) $T=0.3\varepsilon/k_B$, where the mean squared displacement is $R^2=\sigma^2$ for the A particles.}
	\label{f:ipl}
\end{figure}

For our model at temperature $T=\varepsilon/k_B$\/, the characteristic time is
\[
t_0\ =\ 0.1\sigma\sqrt{m/2\varepsilon}\ \simeq\ 0.07\sigma\sqrt{m /\varepsilon}.
\]
Values for a molecular liquid like the silicone oil DC704 (\Cref{f:Gundermann}), are in the order of $\varepsilon \simeq 1 $ kcal/mol, $\sigma \simeq 1$ nm, $m \simeq 100$ u resulting in the time-scale $t_0\simeq$ 0.3 ps. Chemical details of a particular molecule change $\varepsilon$, $\sigma$ and $m$ resulting in changes of $t_0$ within approximately one order of magnitude. Thus, the fact that the actual relaxation time measured (as exemplified in \Cref{f:Gundermann}) is many orders of magnitude slower, signals the emergence of a slower timescale.

\begin{figure}[h]
	\includegraphics[width=6.cm]{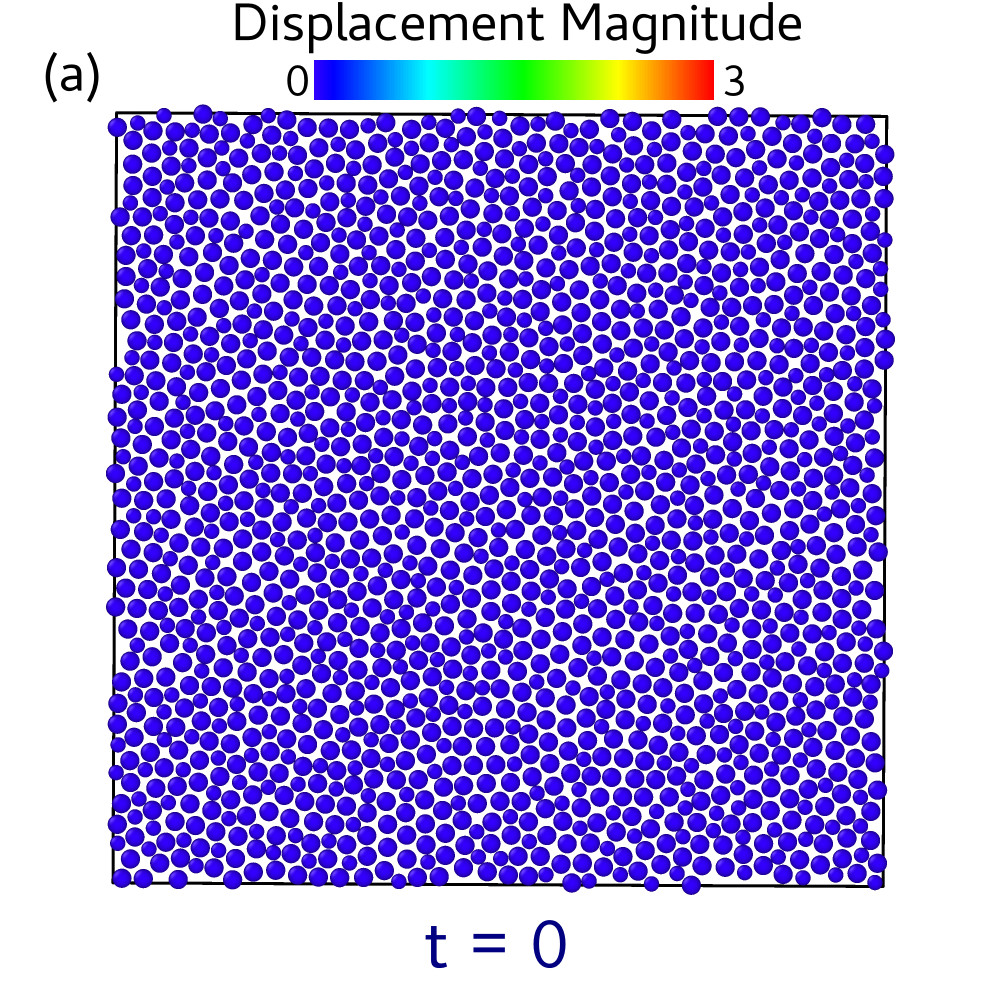}
	\includegraphics[width=6.cm]{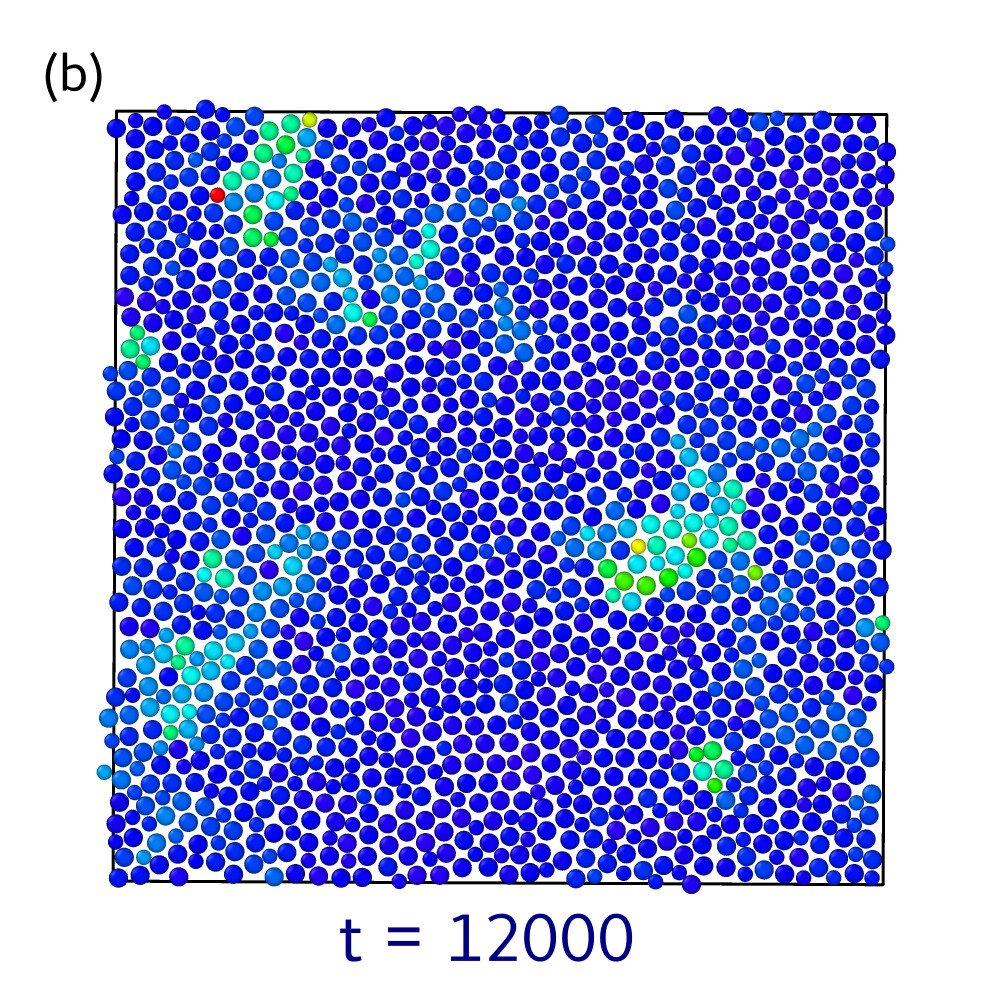}
	\includegraphics[width=6.4cm]{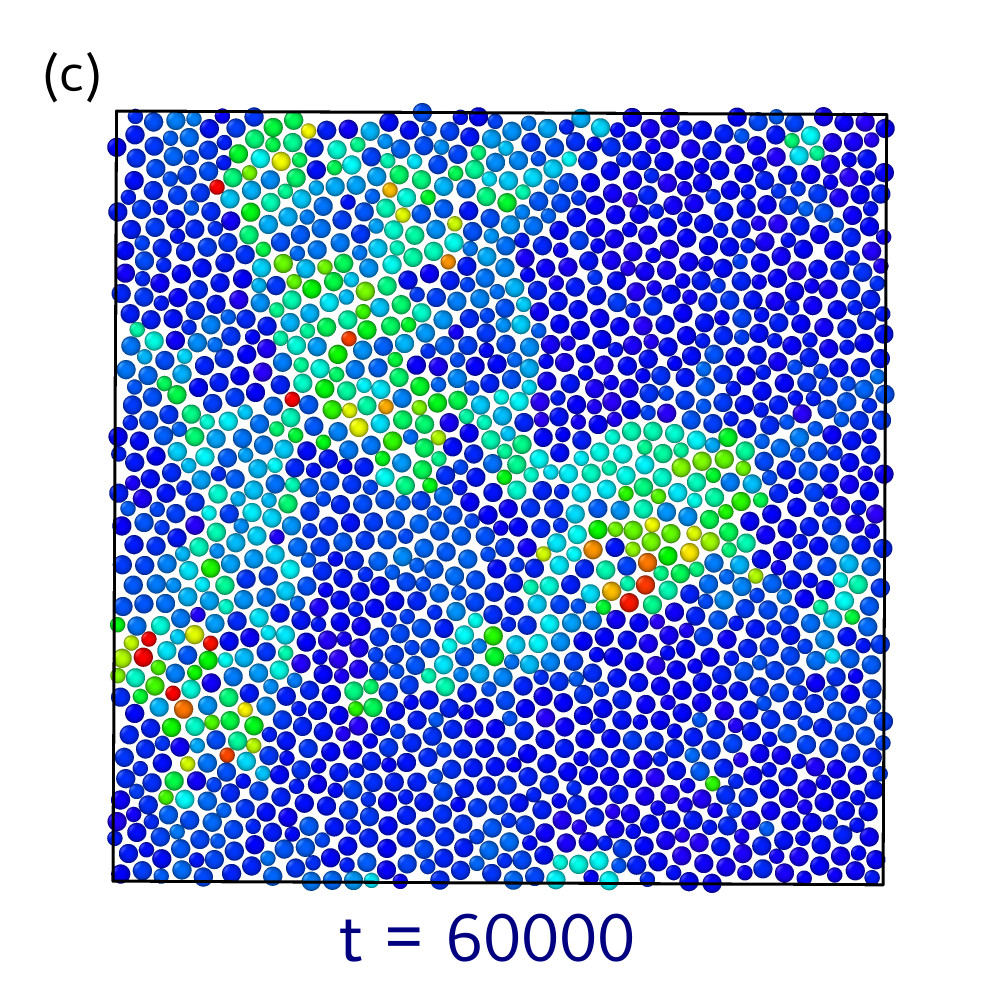}
	\includegraphics[width=6.4cm]{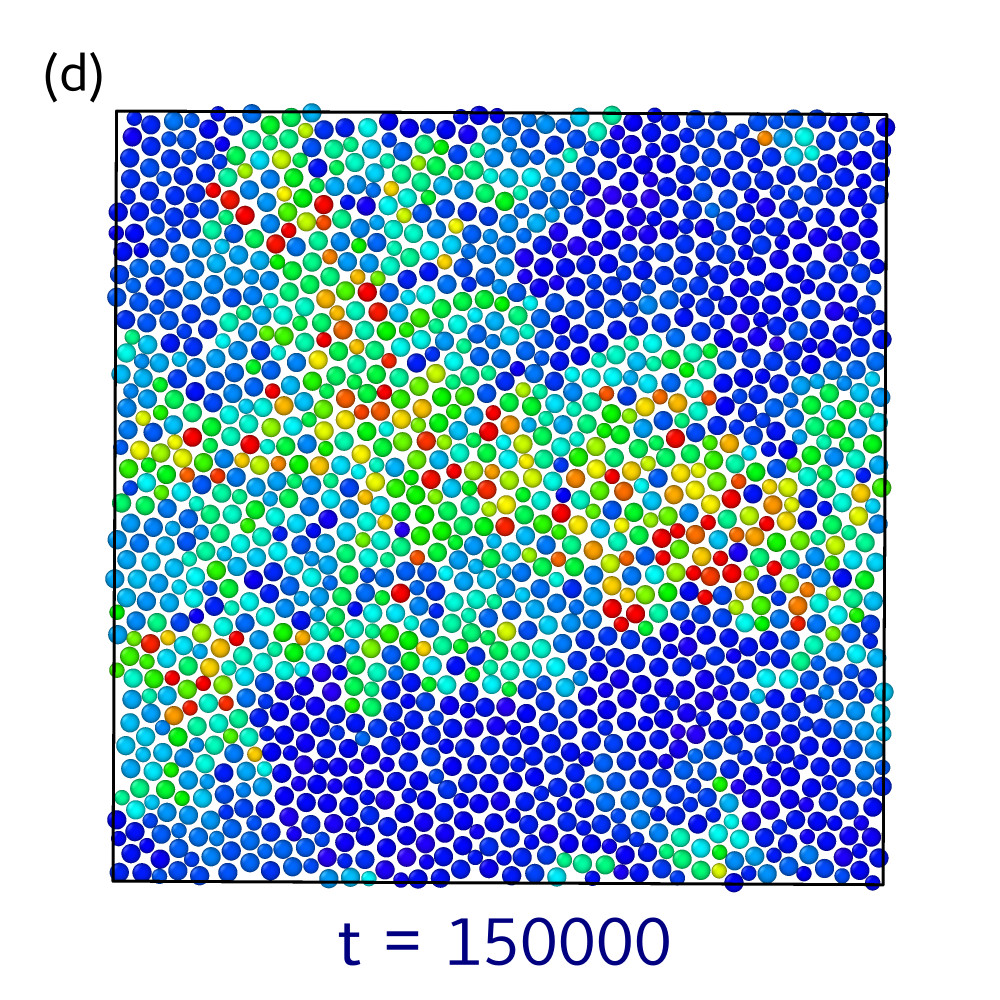}
	\caption{The magnitude of particles' displacement after a time-interval of $t=0$ (a), $t=12000\sigma\sqrt{m /\varepsilon}$ (b), $t=60000\sigma\sqrt{m /\varepsilon}$ (c) and $t=150000\sigma\sqrt{m /\varepsilon}$ (c) at temperature $T=0.3\varepsilon/k_B$. The dynamics are spatially heterogeneous, where a few particles move far. Clusters of particles move together in an avalanche that in turn triggers avalanches in the vicinity.}
	\label{f:displacement}
\end{figure}

\Cref{f:ipl}a shows a representative configuration from the molecular dynamics of the simple model of a liquid, and \Cref{f:ipl}b shows the mean squared displacement, $R^2$, at a range of temperatures. At high temperatures ($T>\varepsilon/k_B$), we have a simple behaviour from ballistic to diffusive motion.
At low temperatures ($T<\varepsilon/k_B$), a slow time-scale emerges. 
Molecules stay caged for a long time, seen as a plateau in $R^2$ in \Cref{f:ipl}b. This emergence of a slow time-scale is related to dynamical heterogeneity. \Cref{f:ipl}c shows the displacement vectors of particles when $R^2=\sigma^2$ at $T=\varepsilon/k_B$. This can be compared with the displacement vectors at low temperatures shown on \Cref{f:ipl}d. 
At low temperatures, about half of the particles have moved significantly, while the remaining have not moved. Moreover, at low temperatures flow events occur in sudden avalanches where many particles move collectively. \Cref{f:displacement} shows the magnitude of the displacements at different times. After a time-interval of $t=12000\sigma\sqrt{m /\varepsilon}$ (\Cref{f:displacement}b), a few regions of particles have moved in a localized cluster. The regions of mobility facilitate dynamics in nearby regions (\Cref{f:displacement}c), explaining the dynamical heterogeneity (Figs. \ref{f:ipl}d and \ref{f:displacement}d).

\subsubsection{Measuring multiple time scales in materials}\label{sss:sampling-in-materials}

Most materials are neither purely viscous (Newtonian liquid) nor purely elastic (elastic solid), but something in between. This means that the properties of these materials become time dependent, i.e., what you measure depends on the time scale of the measurement. Examples of such materials include plastics, rubbers, bitumen, and glasses. The dynamical properties of these materials can cover more than 16 orders of magnitude in time, which makes measuring their properties a tremendous challenge for experimenters - no single experiment can cover this large range. We refer to \cite{Hecksher-et-al:2019} on how we measure broadband mechanical spectra of supercooled liquids and on the possible connection between short and long time properties of visco-elastic materials.

\subsubsection{\textit{Natura non facit saltus} --- refuted and confirmed}

Nature doesn't make jumps (saltus acc. pl., $u$--declination). That claim goes back to Greek Antiquity's nature philosophers, renewed by the German mathematician and nature philosopher \textsc{Gottfried Wilhelm Leibniz} (1646--1716), later reformulated and disseminated by the Swedish biologist and father of the modern taxonomy \textsc{Carl Linnaeus} (1707--1778), underlying the evolutionary theory of \textsc{Charles Darwin} (1809--1882), and (blatantly erroneously) generalized to economics by the British mathematician and economist \textsc{Alfred Marshall} (1842--1924). The jumps of quantum mechanics and gene mutations have refuted this old claim for molecular and smaller distances but could not shutter a common belief in the continuity of changes on larger and more common scales. 50 years ago, however, the biochemist \textsc{Manfred Eigen} (1927--2019) explained precisely why the common belief must be refuted also on the macroscopic level: 
\begin{quote}
Chance has its origin in the vagueness of these elementary events... Under special conditions, however, there can also be an escalating rocking of the elementary processes and thus a macroscopic representation of the vagueness of the microscopic dice game. \cite[p. 35]{Eigen:1975}, our translation	
\end{quote} 

In \Cref{sss:landau-Langevin,sss:climate-trajectories,sss:irreversibility} we gave a simple mathematical model of  \textsc{Eigen}'s \textit{escalating rocking} that can be realised by Landau-Langevin diffusion in laboratory. It shows the necessity of dramatic jumps in some systems. Due to the advances in geometric analysis of the 1960s, in particular the catastrophe theory \cite{Thom:1989} of \textsc{Ren{\'e} Thom}, the why and how of the appearance of jumps in state space under continuous change of control variables is mathematically well understood, e.g., by the seven elementary catastrophes for systems governed by a potential like in biological and environmental morphology. See also recent discrete simulations of sudden morphological changes by biophysicist \textsc{Kim Sneppen} and collaborators, e.g., in \cite{Schneppen:2017}.

As shown in this section, we see \textsc{Eigen}'s \textit{escalating rocking} also  in the liquid dynamics at low temperatures. Interestingly, these dynamics  have some parallels to the climate of Earth. Think of the one coordinate in the $\bf R$ vector as the temperature somewhere, 
and the remainder as other parameters that are important for the Earth system. In both cases we will see that the temperature is fluctuating around some local fixed point. However, at some time in the development an \textit{avalanche} will occur, and the observed parameter will change a lot in a short time. In the dynamics both of soft materials and of the climate of Earth, the origin of structural instability is a strong feedback coupling to the remaining part of the parameter space. That refutes \textsc{Leibniz}'s dictum which otherwise is confirmed in high-temperature dynamics, where remaining particles/parameters can be treated as a mean-field.

\section{Multiple time scales of life}\label{s:life} 
In environmental and life sciences, multi-scale processes are the norm.
Spatial scales vary over as much as 15 decades of magnitude as we progress
from processes involving genes, proteins, cells, organs, organisms, communities,
and ecosystems; time scales vary from times that it takes for a protein to fold to
times for evolution to occur. Several scales can occur in the same problem. 

\subsection{Multiplicity of time scales in cell physiology and public health}

Basically, the challenges for sampling, modelling and simulation are similar to the problems addressed in the preceding sections. To begin with, we shortly emphasise the need to work with a broad range of different time scales in sampling to avoid self-deception (as discussed before in \Cref{sss:sampling,sss:sampling-in-materials} for climate modelling and materials science), and explain the need to reduce the number of different time scales in the modelling process to the most meaningful for a given problem to avoid leviathan non-transparent models. 

\subsubsection{Emergence of different processes in measurements of living tissue under different time scales}\label{sss:sampling-living-tissue}
When studying biological processes at the cellular and sub-cellular scale, temporal information has traditionally not been available, since the tissue had to be fixated in order to study the cells under a microscope. Recent technological developments have unlocked the possibility to see biological processes unfold in real time, leading us to new discoveries and revision of textbook knowledge. Time-lapse imaging however comes with some caveats and pitfalls: Biological processes vary greatly in both temporal and spatial scales, making the time and spatial scale choice instrumental for what is observed. That is quite similar to what we noticed above in \Cref{sss:sampling-in-materials} for time scale dependent measurements of visco-elastic properties of materials. In materials science, the range of the required time scales may be much larger than in the study of living tissue on the cell and subcellular level. On the contrary, in living tissue there are many more concurrent, but qualitatively different processes to follow than in materials science.   Furthermore, for living tissue high temporal frequency observation comes at the cost of perturbing the phenomenon we are studying, e.g., by heating and other misleading signals.  As a case in point, we refer to a classical study \cite{Puri&Hebrock:2007} and its recent revision \cite{Nyeng-et-al:2019} with observations of the formation of the pancreatic ductal system, which is a system of tubes that transport enzymes out of the pancreas and into the intestine.

Interestingly, also in epidemiological investigations, sampling in \textit{different} time scales may be mandatory. As an example we mention the need to evaluate childhood vaccine programs at various time scales: Vaccines are powerful tools against infectious diseases as evidenced by global eradication of smallpox 40 years ago.  Infants are now vaccinated against common childhood diseases (like measles, pertussis).   Measles, with a basic reproduction number $R_0=16$, requires $>95$\% vaccine coverage to break transmission, and many countries with effective programs were declared measles-free.  However, measles has since returned, fuelled by a growing population of unvaccinated adults without natural immunity; a dynamic predicted mathematically long ago.   
Similarly, vaccine program control of common childhood killers such as pneumococcal pneumonia took place in a setting of rapidly increasing human development, and substantial vaccine benefits ensued.  However, in a medium perspective, problematic strain replacement led to an upgrade of Pneumococcal Conjugate Vaccine (PCV) only a decade into the program. Finally, in a longer time perspective, a dramatic reduction in pneumococcal disease mortality was achieved in the pre-vaccine era, due to benefits of improving economy.
For vaccine program effects in various time perspectives and the use of mathematical modelling to evaluate long-term effects of childhood vaccine programs we refer to \cite{Simonsen-et-al:2019-2, Simonsen-et-al:2019}.

\subsubsection{Mathematical imperative: choose the \textit{essential} time scales in public health administration}\label{sss:modelling-living-tissue}
In infectious (i.e., communicable) diseases many different characteristic times exist. Contrary to common belief of public health administrators (widespread also among environmental administrators), who expect more reliable practical advise from more complex models, credible mathematical modelling and applicable numerical simulation have to avoid \textit{over-parametrisation} and leviathan complexity: Reliability requires transparency which again requires the selection of a few dominant time scales and to discard marginal ones. That choice can be different from disease to disease, ranging from behavioural time lengths to incubation periods, from characteristic times of mutations in biological agents and in human populations, see \cite{ViggoAndreasen-et-al:2015}. 

Interestingly, in modelling diseases one always \textit{tries} to separate the time scales, supposing a \textit{superposition} of the different underlying processes. That is legitimate for some diseases, e.g.,  for modelling the spread of gonorrhoea and permits highly reliable numerical simulations and effective public health administration, comparable to the well-established modelling and simulation of fish ecology for international fishing regulation. For other diseases like the measles, malaria and the flu, the highlight is the \textit{interaction} (coupling) of the different processes.

\subsubsection{Time scale problems in two worked examples}

In the following two subsection we shall address the emergence of multiple time scales in two fields of human cell physiology, the production of blood in the human body with new light on the rise of malignancies in slow-fast processes, and the insulin secretion of pancreatic beta--cells, where the recognition of two different characteristic times plays a role in diabetes diagnosis and therapy. Like in many other physiological problems, in cancer and diabetes research multiple spatial and time scales are intertwined and the theoretical and application challenges \textit{push} the research  towards nano geometry, see \cite{Boo:2012, Sneyd-et-al:2016}. Moreover, advances on the technological side \textit{pull} towards multiscale analysis: even for a single cell a wide range of observational means have become available, with length scales from \AA\ in electron microscopy to $\mu$m in confocal fluorescence microscopy and  multifocal multiphoton imaging, and a corresponding wide range of time scales.

\subsection{Multiscale models of the production of blood in the human body}

An illustrative example is the production of blood in the human body and how slow processes in an otherwise fast system can lead to haematopoitic malignacies such as leukemia or myeloproliferative neoplasms (MPNs). While the majority of the cells that constitute the blood have a lifespan in the order of days, and reconstitution following loss of blood is of a similar magnitude, MPNs develop on a much greater timescale, estimated as about a decade \cite{Andersen-et-al:2017}. Understanding exactly how these malignancies arise and develop on the slow timescale is key to an efficient treatment.

Some blood cancers such as the MPN malignancies are believed to emerge from mutations in the haematopoeitic stem cells, located in the bone marrow. A random mutation in one such stem cell causes it to be dysfunctional, e.g. to produce an excess of a particular sub-type of cell or to be non-reactive to signals limiting its reproduction. As the stem cell self-renews, a fitness advantage from the mutation can lead to the mutation-type becoming the dominant type of stem cell.

\begin{figure}[ht]\centering
	\includegraphics[width =  \linewidth]{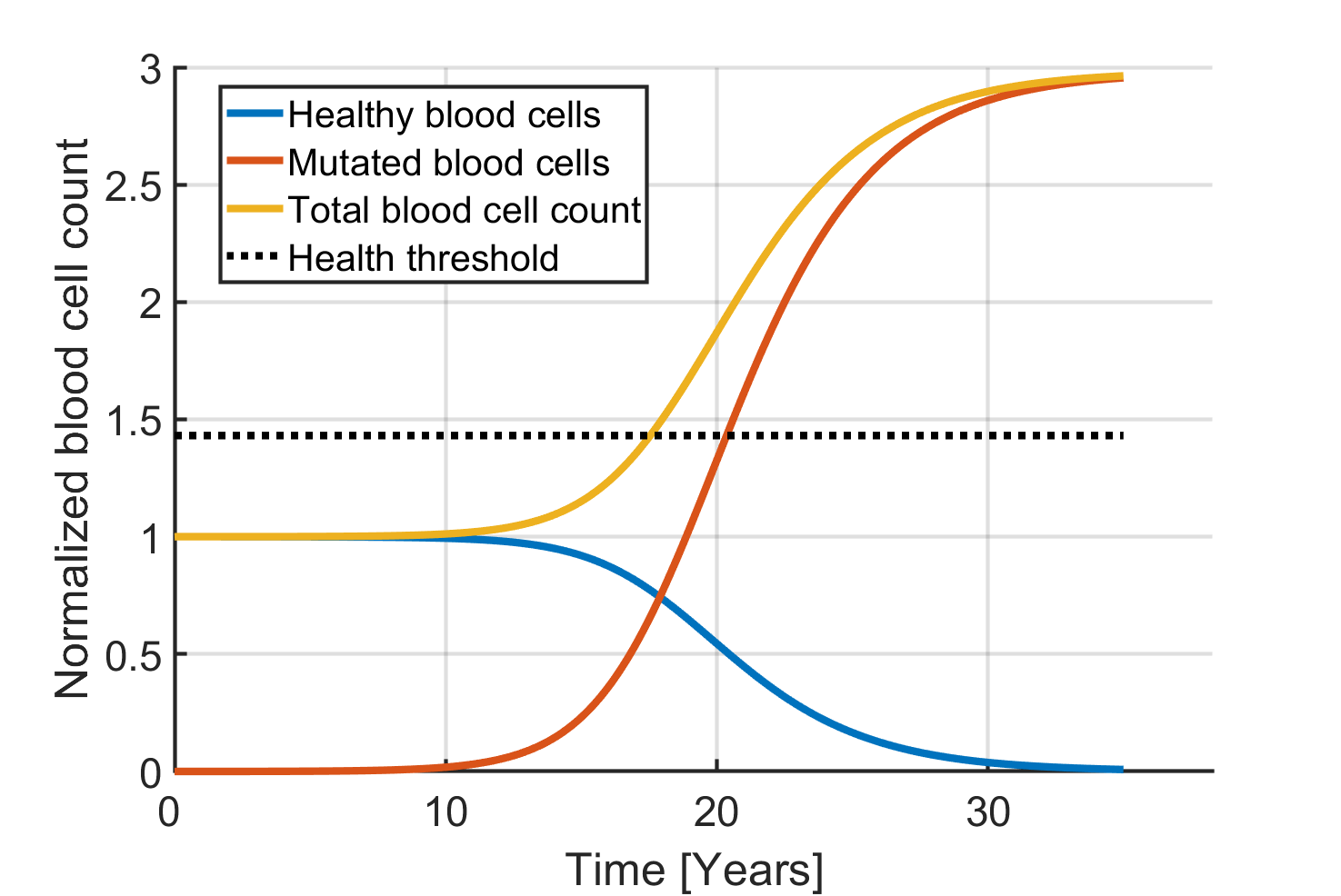}
	\caption{Development of blood cell count in the model of \cite{Andersen-et-al:2017} following mutation of a single stem cell. The dotted boundaries are estimates of the interval within which an individual would be considered healthy and where the risk of complications is low.}\label{fig:cancitisBloodCount}
\end{figure}
\cite{Andersen-et-al:2017} describes a system of six coupled ordinary differential equations, in an effort to model the blood producing system of the human body, and the development of MPNs. The model combines the behavior of the stem cells with a feedback from the blood through an abstract measure of inflammation of the immune system. Since MPNs are rare, the random mutation of stem cells are expected to occur on a timescale greater than the average human life-time. As such, the rate of mutation can be considered effectively zero, and instead a single mutated cell is added initially. 

\Cref{fig:cancitisBloodCount} displays how the blood cell count develops in the model. For the count of various types of blood cells, there is typically a threshold, which, when exceeded, is grounds for further investigations of the patient. While above the threshold the risks of complications such as thrombosis is also expected to be greatly increased. The figure displays an estimate of this threshold as a black dotted line at around 143\% \cite{Beck:2008,De:2014}.
Interestingly, the model predicts that this threshold is not exceeded until about 17.5 years after the initial mutation of a stem cell. In addition, this is the same point at which the mutated blood cells constitute about half of the total blood cell count. 
%
Even though the blood production is capable of fast reconstitution after blood loss and although the main constituents of the blood have a fast turnover-rate, the time-scale of the development of MPNs is much slower. 

Once a stage is reached were the disease is immediately noticeable, the mutated cells make up such a large part of the blood that treatment must lead to drastic changes to negate the harm done within the two-decade long progression of the disease. 

%

\begin{figure}[h]
	\includegraphics[scale=0.89]{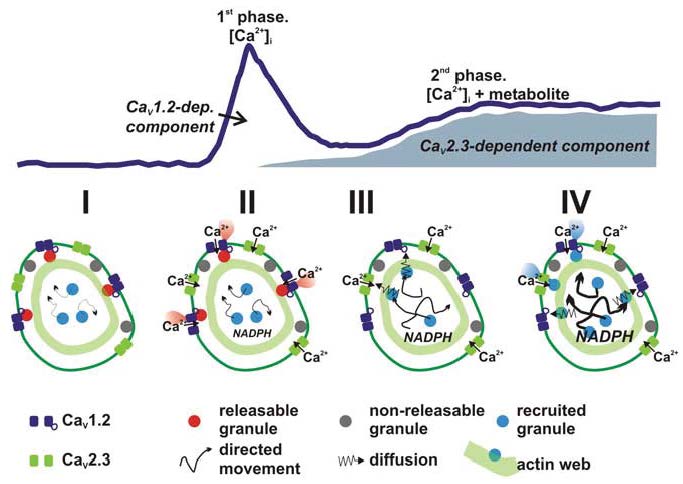}	\caption{Voltage-gated Ca$^{2+}$ channel (Ca$_V$) activation and insulin granule movement during phasic
		insulin secretion.Insulin secretion is elicited by an increase
		of the extracellular glucose concentration. It is triggered with a delay of $\sim$ 1 min (the time needed
		for glucose to be metabolized). It develops in a multiphasic
		time course over 30--60 minutes with a peak after about 5 minutes, followed by an intermediate decline and the main burst afterwards.
		From \cite[Fig. 2.5, p. 40]{Renstrom2011}, reprinted by permission from Springer Nature.}
	\label{f:insulin-secretion}
\end{figure}

\subsection{Regulated exocytosis in pancreatic beta--cells}
\label{s:biphasic-insulin}
Another illustrative example of the emergence of multiple time scales in cell physiology is the biphasic insulin secretion of pancreatic $\beta$--cells. 

\subsubsection{Discovery of biphasic insulin  secretion}\label{ss:discovery-biphasic-secretion}

50 years ago, the biochemist \textsc{G.M Grodsky} and coworkers demonstrated in \cite{Grodsky-et-al:1968}, that glucose-induced insulin secretion in response to a step increase in blood glucose concentrations follows a biphasic time course consisting of a rapid and transient first phase followed by a slowly developing and sustained second phase, see \Cref{f:insulin-secretion}.

In some aspects, it reminds of the ominous feature of two characteristic times in climate change: by constant forcing, 
\begin{enumerate}
	\item \textit{to begin with} nothing extraordinary is observable; 
	\item \textit{then for a short while} consequences become
	\begin{itemize}
		\item \textit{first} ever more observable and 
		\item \textit{further-on} go less-and-less noted in spite of continuing forcing and continuing and now substantial aggregation of consequences
	\end{itemize}
	\item \textit{until} (in the climate change situation) a threshold is reached and feedback mechanisms take over.
\end{enumerate}
It is known that the loss of the first phase is correlated with diabetes. Hence, as emphasized in \textsc{P. Rorsman}'s and \textsc{E. Renstr{\"o}m}'s review \cite{Rorsman-Renstrom:2003} and the monograph \cite{Beta:2011}, understanding the reason for the biphasic feature of (normal) insulin secretion is wanted for better diagnosis and treatment of dominant variants of diabetes mellitus type 1 and type 2. 

\subsubsection{Mathematical models of the exocytosis cascade}\label{ss:exocytosis-cascade}
Mathematically, it is easy to reproduce the biphasic feature in a black box model, alone with two compartments, i.e., just two coupled differential equations with suitably tuned coefficients. See \textsc{Grodsky} et al. in \cite{Grodsky-et-al:1970} and the follow-up literature. Unfortunately, all these two-compartments models have rates that cannot be interpreted in physiological, biomedical, biochemical or bio-electrical terms, nor measured independently. Hence these models have the same methodological status as the toy model we suggested in \Cref{s:toy-models}. 

The mathematician \textsc{A. Sherman} and collaborators provided in  \cite{Sherman-et-al:2008} an alternative mathematical model of the regulated exocytosis, now with coefficients that, in principle, can be assigned biomedical empirical evidence, see \Cref{tab:parameters}. It is well known that the $\beta$--cells are electrically active, and use electrical activity to transduce an increase in glucose metabolism to calcium influx, which triggers insulin release.

\textit{Sherman}'s model is based particularly on the electrostatic differences between two compartments: the microdomains, i.e., channel regions on the cell surface (the plasma membrane) where most of the secretion is supposed to happen, and the interior cell liquid (cytosol): A glucose stimulus, e.g., modeled by a step or a train of five alternating square-pulses, generates a concentration $C_{\md}$ in the microdomains of Ca$^{2+}$ ions that can reach $\mu$M level, and, in the model, a two decades lower concentration $C_{\operatorname{i}}$ in the cytosol.

\begin{table}[ht]
	\caption{Kinetic parameters of the EC model at the resting state, from \cite{Sherman-et-al:2008}}
	\label{tab:parameters}
	%
	\centering
	\begin{tabular}
		{r l | r l | r l}
		Parameter & Value & Parameter &Value & Parameter &Value\\%
		\toprule%
		$k_1$& 20 $\mu$M$\ii$ s$\ii$ & $k_{-1}$ & 100 s$\ii$ & $r_1$ & 0.6 s$\ii$\\
		$r_{-1}$& 1.0 s$\ii$ & $r_2^0$ & 0.006 s$\ii$ & $r_{-2}$ & 0.001 s$\ii$\\
		$r_3^0$& 1.205  s$\ii$ & $r_{-3}$ & 0.0001 s$\ii$ & $u_1$ & 2000 s$\ii$\\
		$u_2$& 3.0  s$\ii$ & $u_{3}$ & 0.02 s$\ii$ & $K_p$ & 2.3 $\mu$M\\
	\end{tabular}
\end{table}

Next, the authors model the movement of  the 10,000-15,000 insulin vesicles (granules) in the interior of a $\beta$--cell to the plasma membrane and preparing, docking, making a fusion pore and releasing the insulin  content by the different and changing number of insulin vesicles in eight compartments, representing eight different places and preparation stages of the vesicles, driven by the dynamics of the concentration of Ca$^{2+}$ ions.

The authors introduce the following system of 8 coupled ordinary differential equations to model the exocyosis cascade (EC). The pools $N_j, j=1,\dots 4$ aggregate vesicles of different maturity and vicinity to the cell membrane; the pools $N_5$ and $N_6$ describe the re-supply; and the two variables $N_F,N_R$ count the vesicles that have fusioned with the plasma membrane, respectively, the number that have widened the fusion pore and released their insulin content:
\begin{align}
\begin{split}
\dot N_1 \ &=\ -[3k_1C_{\md}(t) + r_{-1}] N_1 + k_1N_2 + r_1N_5\/,\\
\dot N_2 \ &=\ 3k_1C_{\md}(t)N_1 - [2k_1C_{\md}(t) + k_{-1}]N_2 + 2k_{-1}N_3\/,\\
\dot N_3 \ &=\ 2k_1C_{\md}(t)N_2 - [k_1C_{\md}(t) + 2k_{-1}]N_3 + 3k_1N_4\/,\\
\dot N_4  \ &=\ k_1C_{\md}(t)N_3 - [3k_{-1} + u_1]N_4\/,\\
\dot N_5 \ &=\ r_1N_1 - [r_1 + r_{-2}]N_5 + r_2N_6\/,\\
\dot N_6  \ &=\  r_3 + r_{-2}N_5 - [r_{-3} + r_2]N_6\/,\\
\dot N_F \ &=\ u_1N_4- u_2N_F\/,\\
\dot N_R \ &=\ u_2N_F- u_3N_R\/.
\end{split}
\end{align}
Both the resupply and the priming steps are assumed to depend
on $C_{\operatorname{i}}$ using
\[
r_2\ :=\ r_2^0C_{\operatorname{i}}(t) / [C_{\operatorname{i}}(t) + K_p]\ \tand\
r_3\ :=\ r_3^0C_{\operatorname{i}}(t) /[C_{\operatorname{i}}(t) + K_p].
\]
The parameters chosen in \cite{Sherman-et-al:2008} are given in \Cref{tab:parameters}. With those rates, the numerical simulations reproduce the emergence of two time scales, 5 minutes for the first phase and hours for the second, and explain it, not surprisingly, by the saying "First things first", i.e., the model and the data confirm the preconception that two different mechanisms operate during the two phases, with transient first-phase secretion due to exocytosis of granules that were docked at the plasma membrane, and the second sustained phase due largely to resupply of granules. 

Since then, a stream of new experimental data has demanded new explanations of specific aspects of the biphasic secretion and led to reconsidering the preceding electrostatic model and to various modifications, e.g., see \cite{Pedersen:2009, Pedersen-Sherman:2009, Pedersen-et-al:2014}.

A different approach is given in \cite{ApApBoKo:2011}. The authors model the morphology and dynamics of the
making of the fusion pore or porosome as a cup-shaped lipoprotein structure (a dimple
or pit) on the cytosol side of the plasma membrane and describe the formation
of the dimple by a free boundary problem. They discuss the various forces acting
and analyse the magnetic character of the wandering electromagnetic field wave
produced by intracellular spatially distributed pulsating (and well-observed) release
and binding of Ca$^{++}$ ions anteceding the bilayer membrane vesicle fusion of exocytosis. This approach explains the energy efficiency of the dimple formation prior
to hemifusion and fusion pore and the observed flickering in secretion. It provides
a frame to relate characteristic time lengths of exocytosis to the frequency, amplitude and direction of propagation of the underlying electromagnetic field wave.


\subsubsection{Strength and limitations of simplistic models --- a first conclusion}
Like in climate modelling, the electrodynamic model of the regulated exocytosis is a simplistic model with all its strength and limitations. 
We must distinguish between the aspects of the model, which are based on first principles and/or undeniable empirical evidence and/or theory--based numerical simulations --- and the assumptions and approximations which are required where conclusive evidence still is missing. 
Then one can only hope that our mathematical and physical assumptions and conclusions are reasonable as long as we neither have  a comprehensive theoretical basis, nor reliable numerical simulations nor experimental programs to verify – or falsify the assumptions made. Clearly one would wish to have models, both for climate change and for cell physiology, that are sufficiently detailed to guide action in prevention, attenuation, treatment and cure. Whether such a model is ever obtainable, be it for climate change or cell physiology, is not clear with view upon the immense complexity of the agents involved, see also the caveat above in \Cref{sss:modelling-living-tissue}.

Therefore, it seems so important to us in all otherwise well--founded and well--intended approximations and refinements not to lose sight of the possible emergence of multiple time scales.

\section{Multiple time scales of economy --- short and long waves}\label{s:society}
Multiscale modelling and simulation is most elaborated in \textit{materials science}, in integrated computational engineering of solids, and in the simulation of soft materials, as shown in \Cref{s:matter}.
It is also well--developed in modern \textit{mathematical physiology and epidemiology} with the simultaneous treatment of multiple spatial and temporal scales, as shown in \Cref{s:life}. 
In present main-stream \textit{economics}, multiscale modelling and simulation have perhaps not achieved a similar attention, albeit   
in economics, multiscale modelling and simulation date back about 150-200 years.

\subsection{Theory models and empirical data}

Economic activities are spread in space and time. Space scales are ranging from the small distances between workplace and home, to the possibly large distances between birthplace and workplace, and between places of production and use or consumption of parts and services. Time scales range from the seconds in financial decisions over working hour, day, week or month, to sowing and harvesting periods; process time for a particular manufactured good; lead time from ordering to delivery; terms for employing workforce, amortization of fixed assets, borrowing, investing, and accounting; and time frames for input-output tables and planning horizons in enterprises or on national level. In principle, all these different time scales can be inserted in a variety of economic multiscale models, serving different goals --- better or less successful. 

\subsubsection{The sampling problem}\label{sss:sampling-in-economics}

Like in climate change modelling in \Cref{sss:sampling}, in materials science in \Cref{sss:sampling-in-materials},  and in biomedicine and public health in \Cref{sss:sampling-living-tissue}, also in economics is is well known that different choices of registration intervals can suggest different views of ongoing processes, see \cite[Section 2.3]{Juselius:2006}: If, e.g., you analyse interest rates, you can choose quarterly, monthly, weekly or daily observations and do a statistical analysis the results of which, while not independent, will give you a different perspective on the problem at hand. For example, daily data will contain information about co-movements in the data over the very short horizon (adjustment is probably not much longer than a day if even that). The stochastic trends in such data are likely to be associated with deviations from long-run equilibria measured for example with monthly or quarterly information over longer periods where the stochastic driving trends are associated with macroeconomic imbalances. (Communicated by K. Juselius).

\subsubsection{Bridging economic thinking and data gathering: filtering different characteristic time lengths}
Most of these discrete  variables can adequately be approximated by continuous variables. Even so, as in the preceding sections, our question is when and how \textit{different characteristic time scales} emerge, no matter if the models are in discrete or continuous time. On the way to answering that question, we indicate how multiscale \textit{thinking} continues to reveal new economic phenomena and to explore new concepts and innovative computational paradigms. We emphasize \textit{thinking} and \textit{data gathering} contrary to \textit{modelling} and \textit{simulation}: clearly, in the world of real economics one hardly finds convincing non--trivial predictions based on mathematical modelling and numerical simulation as witnessed, e.g., by the disasters of 1928/2008 and at other turning points. As in climate change modelling and simulation, understanding the basic lows, familiarity with the empirical evidence and adapting the mindset to the multiscale character of the very problems may be more important than proliferation and trust in seemingly precise predictions. Therefore, in this section
our emphasis is on multiscale data gathering and analysis.

In \Cref{ss:ups-and-downs}
we introduce to the search for characteristic times in the development of society, culture and economics, first in \Cref{para:general-cycles} by reviewing general concepts of periods in history and then in \Cref{para:economic-cycles-giants} by recalling the rise of interest for multiscale problems among outstanding economic thinkers of the last 100 years. 

In \Cref{ss:two-kind-of-cycles} we address the emergence of multiple time scales in economic cycles and show how neglecting these multiscale aspects must lead astray. In  \Cref{para:keynes} we explain the suggestive power of short term fluctuations that are easy to detect, to model and to simulate. In \Cref{para:schumpeter} we show that the more decisive long-term cyclic behaviour is not easily approachable. Alone the detection requires advanced numerical, geometrical and statistical tools like penalized splines or wavelets, and credible modelling and simulation seem impossible due to the unknown magnitude and influence of background variables.

In \Cref{ss:spatio-temporal} we shall shortly mention basic multiscale spatio--temporal problems. 

\subsection{Time scales in society}\label{ss:ups-and-downs}
The efforts of integrating microeconomics into the macroeconomics and macrodynamics of capitalism yield many mathematical challenges. For our survey we have chosen the emergence of two different characteristic times in economic cycles. As 
\begin{itemize}
	\item in \textit{climate modelling}, where the small characteristic time of direct effects of radiative forcing is immediate while the giant characteristic time of reversing secondary effects only is accessible by theoretical means, 
	
	\item in data of \textit{economic growth} only the small-scale fluctuations of business cycles are immediately sensible at a time scale of 6-8 years and mathematically easy to model, estimate and simulate, while the in many aspects more decisive long-term cycles on a time scale of 40-60 years require theoretical means to become visible.
\end{itemize}

\subsubsection{Cyclic changes in society, economics, and culture}\label{para:general-cycles}
The American physicist--philosopher and semiotics--logician \textsc{C.S. Peirce} argued in \cite[Evolutionary Love, pp. 361--374]{Peirce:2012} for a characteristic length of ca. 800 years for distinguishable phases in the slow evolution of rationality through teaching, disseminating, also geographically, canonizing, disputing, testing, challenging, falsifications, running out of steam and giving way to new mindsets and new cultures. He identified dramatic breaks in Greece around 400 BCE, in Byzantium around 400 CE, the predominance of scholasticism since 1200 CE (and radically new thoughts based on his semiotics and ``pragmaticism" emerging around our present time). 

The German  historian and philosopher of history \textsc{O. Spengler} whose interests included mathematics, speculated independently in \cite{Spengler:1918} in the same direction with approximately the same times scales  as \textsc{Peirce}, pointing to shared features in rise, culmination and decline of different epochs in history.

The US-American artists and historians \textsc{W. Strauss} and \textsc{N. Howe} reformulated \textsc{Spengler}'s approach in biologistical terms in their \textit{Generational Theory} \cite{Strauss-Howe:1997}. 
Roughly speaking, they take an average life to be 80 years, and consisting of four periods of $\sim$ 20 years: Childhood, Young adult, Midlife, Elderhood. A generation is an aggregate of people born every $\sim$ 20 years. 
They describe a four--stage eternally repeating cycle of social or mood eras of approximately 20 years' length which they call \textit{turnings}: The High, The Awakening, The Unraveling and The Crisis. Hence, in that setting each generation experiences four turnings. Etc.

While the preceding approaches to the emergence of multiple time scales in society and culture, inspiring as they may be, are hard to verify in a rigorous manner, there are many more hard data and easier quantifiable multiple time scales in economic theory: more precisely, in the description and analyzes of macro-economic cycles in capitalist economies we can follow the emergence of multiple time scales.

\subsubsection{Economic cycles} \label{para:economic-cycles-giants} 
The ups and downs in economy have attracted the interest of many thinkers. It seems that these oscillations are a characteristic feature of \textit{capitalism}, with a mass of people free to sell their work and to buy consumer goods, and classes of property or capital owners free to determine the direction of their land or capital use for making profits. Both freedoms can though be restricted by state regulation and unions and other alliances within parts of the working masses and groups of entrepreneurs. 

In history, other economic systems have also had their crises due to weather, demographic factors (diseases, migration, birth surplus) and war. For the misery of crises in capitalism, it was common to blame demographic factors as well, the unrestrained reproduction of the working masses, as the English cleric \textsc{T.R. Malthus} (1766--1834) put it --- until the investigations by \textsc{K.~Marx} (1818--1883) gave an explanation for the cyclic occurrence of crises in capitalism solely due to the internal logic of its functioning, see, in particular, \cite[Chapters 22--25]{Marx:1887}.

Since then, a handful of mostly Austrian--British--American giants of economic thinking, clever mathematical modelers and profound empirical analysts, have investigated the macrodynamics of capitalism more closely, finally arriving at a clear view upon the emergence of \textit{two characteristic times} in the continuing evolution of capitalism.
\begin{itemize}
	\item \textsc{J.M. Keynes} (1883--1946) confirmed and explained in \cite{Keynes:1936} the phenomenon of \textit{business cycles} as a \textit{macroeconomic} generalization of the trivial pork and cattle cycles that describe the phenomenon of cyclical fluctuations of supply and prices in livestock markets, see \Cref{para:keynes}.
	
	\item \textsc{J.A. Schumpeter} (1883--1950) described in the monumental \cite{Schumpeter:1939} the superposition of waves of different length in the evolution of capitalism. He showed how technological innovations had driven the longterm waves, baptized by him as \textit{Kondratieff waves} in honor of the Soviet economist \textsc{N. Kondratiev} who earlier had observed waves of different length in global capitalist economy. As emphasized in \cite[p. 126]{Freeman:2009}, "the central point of [\textsc{Schumpeter}'s] whole life work [is]: that capitalism can only be understood as an evolutionary process of continuous innovation and `creative destruction'", see also \cite{Freeman:1996}. Not surprising for aficionados of multiscale modelling and multiscale data collection, the view upon the process of creative destruction would be blurred by neglecting the multiscale aspects of real economies and freezing all but a few variables. Worse, illusions would be generated about the possibility to draw policy conclusions from such naive data and simplistic models.
	
	\item \textsc{A.F. Burns} (1904--1987) was a researcher, a consultant for various US-governments, and the Chairman of the Federal Reserve for eight years. He tried to extract concrete advice for economic regulations from \textsc{Schumpeter}'s work on economic growth and technological innovation in the aftermath of The Great Depression (1928-1940), and to eliminate \textsc{Keynes}' and \textsc{Schumpeter}'s accompanying criticism of capitalism, see \cite{Burns:1969} and the seminal joint work \cite{Burns-Mitchell:1946} with his mentor (open to socialist ideas) \textsc{W.C. Mitchell} (1874--1948).
	
	\item \textsc{R.M. Goodwin} (1913--1996), as a mathematician, selected the Great Ratios (see also \Cref{tab:economic-standard-notations}):
	\begin{itemize}
		\item The \textit{employment rate}  on the external labor market is defined as $e:= L/L^s$\/, where $L$ denotes the employment and $L^s$ the labor supply. Note that $e$ is positively correlated or for simplicity identified with the rate $u$ of capacity utilization of firms.
		\item
		The \textit{share} of wages in national income is defined as $v:= \tfrac{WL}{pY}$\/,
		where $W, L, p, Y$ denote the nominal wages, the employment, the price level, and the aggregate income.  
	\end{itemize}
	These secularly trendless magnitudes turned out to be essential for describing,  modelling and simulating the economic cycles in a rigorous mathematical and socio-politically meaningful way by combining basic thoughts of \textsc{Marx, Keynes} and \textsc{Schumpeter}. \textsc{Goodwin} announced his approach in \cite{Goodwin:1951} and elaborated it in a series of contributions, partly reprinted in \cite{Goodwin:1989}. 
	
	\item \textsc{M. Friedman} (1912--2006) put monetary politics into the frame of long waves, thereby addressing old questions of the labor movement regarding the interrelations between employment, wages and inflation, see \cite{Friedman:1968}. It seems, though, as explained in \cite{Flaschel:2009} ``that a Marxian reinterpretation of the baseline Monetarist model of inflation, stagflation, and disinflation may be more to the point from a factual viewpoint than Friedman's initial and later attempts to explain these phenomena against the background of a Walrasian (i.e., neoclassical) understanding of the working of the economy".
\end{itemize}

\begin{table}[t]
	\caption{Standard notations for modelling Keynesian trade waves in discrete time, from \cite[pp. xiii-xiv]{Flaschel:2009}.} 
	\label{tab:economic-standard-notations}
	\rotatebox{00} {\scalebox{0.81}{ {\renewcommand{\arraystretch}{1.9}
				\begin{tabular}
					{l l p{12.6cm}}
					Symbol & Description \\%
					\toprule%
					$C$& Aggregate planned consumption\\
					$I$& Aggregate planned investment\\
					$K$ & Capital stock\\
					$L, L^{\operatorname{sppl}}$ & Employment, labor supply\\
					$W$ & Nominal wages\\
					$Y$ & Aggregate income (= supply, i.g., with time lag)\\
					$c$ & Marginal propensity to consume\\
					$e:=L/L^{\operatorname{sppl}}$ & Rate of employment\\
					$p$ & Price level\\
					$s:=1-c$ & Savings rate\\
					$u$ & Rate of capacity utilization of firms\\
					$v:= \frac{WL}{pY}$ & Share of wages\\
					$\nu:=K/Y$ & Capital coefficient\\
					$C_t, I_t, Y_t$ etc. & Discrete time values of $C, I, Y$\\
					$Y_0, p_0$ etc. & Steady state values (or short--run equilibrium values)
				\end{tabular}
	}}}
\end{table}


\subsection{The emergence of two different characteristic times in economic cycles}\label{ss:two-kind-of-cycles}

\subsubsection{Keynes' business cycles}\label{para:keynes}
\textsc{Keynes}' thoughts about short term business cycles carry an irresistable elegance. His intimate relations with the empirical and political side of economics are impressive: ``Our criticism of the accepted classical theory of economics has consisted not so much in finding logical flaws in its analysis as in pointing out that its tacit assumptions are seldom or never satisfied, with the result that it cannot solve the economic problems of the actual world." \cite[Chapter 24]{Keynes:1936}. The supporting mathematical expressions and arguments are characterized by outsatnding operationality and transparency. All that may have blocked partially for an early awareness of the economic longterm cycles. 

Before addressing the longterm cycles (here called \textit{Schumpeterian}), we present a few basic facts about the short-term cycles (here called \textit{Keynesian}). Following the main stream of economic literature, we give a \textit{discrete} time Keynesian (or Hicksian) Trade Cycle Model, as discussed and simplified in \cite[Section 3.7]{Flaschel:2009}. For a mathematically more elegant model in continuous time in the tradition of \textsc{Lotka--Volterra} environmental models we refer to \cite{Flaschel-Kauermann-Teuber:2005}.

\paragraph{A. The ground model of \textit{C-I-Y}--interaction} As always in modelling and simulation, \textit{the first task} is the choice of a few basic variables and of simple approximative relations between them. Using the notations of \Cref{tab:economic-standard-notations}, the multiplier--accelerator interaction between consumption, investment and income can be 
modeled by the following three equations
\begin{align}
C_t\ &=\ c\, Y_{t-1}\,,\label{e:keynes-consumption}\\ 
I_t\ &=\ \nu\, (Y_{t-1} - Y_{t-2})\,\tand\label{e:keynes-investment}\\ 
Y_t\ &=\ C_t + I_t + A \,.\label{e:keynes-income}
\end{align}
Hence, we have a consumption (and so savings) function which is lagged by one period. Furthermore, investors are here purely looking backward by using
the last observed change in sales instead of the currently expected one for their
investment decision.  \Cref{e:keynes-income} finally describes goods--market equilibrium with an additional term $A$, which is assumed to be positive and
stands for \textit{autonomous demand}.

Plugging \Cref{e:keynes-consumption} and \Cref{e:keynes-investment} in \Cref{e:keynes-income} yields a non-homogeneous difference equation of order two with constant coefficients
\begin{equation}\label{e:basic-differences}
Y_t - (c+\nu)Y_{t-1}+\nu Y_{t-2}-A\ =\ 0.
\end{equation}
In the standardized variable $Z :=Y - \frac 1s\, A$, i.e., by subtracting the steady state solution $Y_0=\frac 1s A$ with $s:=1-c$ from the variable $Y$, the preceding equation takes the homogeneous form
\begin{equation}\label{e:basic-differences-homo}
P(Z)\ \stackrel{!}{\equiv}\ 0 \quad\text{ for }\quad P(Z)(t)\ :=\  Z_t - (c+\nu)Z_{t-1}+\nu Z_{t-2}\/.
\end{equation}
The textbook solution of \Cref{e:basic-differences-homo} (see \textsc{N{\"o}rlund}'s classic \cite[X.5, Paragraph 156]{Norlund:1924} of 1924 for linear difference equations of arbitrary positive order with constant coefficients) is obtained by first noting that the coefficient of the highest order term $Z_t$ is 1 and of the lowest order term $Z_{t-2}$ is not vanishing, hence any solution of  \Cref{e:basic-differences-homo} must be of the form $Z(t)=\la^t\, w(t)$ for a suitable $\la\in\CC$ and a suitable function $w$. Then $P(Z)\equiv 0$ implies $w\equiv 0$ and $\la$ a root of the characteristic equation $\la^2 - (c+\nu)\la+\nu  = 0$. 
%
We focus on the case $c< 2\sqrt{\nu}-\nu$ yielding a pair of complex conjugate $\la_{1,2} = |\la|(\cos \theta + i\/\sin\theta)$ and, finally, a real cyclic solution to \Cref{e:basic-differences-homo}
\begin{equation}\label{e:real-cyclic}
Z(t) \ =\ |\la|^t \left(\delta\, \cos(\theta\/ t -\e)\right) ,
\end{equation}
where $\e,\delta$ are given by initial conditions (values that describe the displacement
from the equilibrium $Z = 0$ if such a displacement occurs). Let us further assume that $|\la|>1$ to avoid the implosive behavior of the economy, but so accepting explosive behavior, i.e., cyclic fluctuations with explosively increasing amplitudes and constant period of $\tau= 2\pi/\theta$. To sum up, that will occur when  $c< 2\sqrt{\nu}-\nu$ and $\nu > 1$ hold true simultaneously, that is for accelerator
coefficients that are larger than one and marginal propensities to consume which are
sufficiently low.

Note that this solution $Z$ of \Cref{e:real-cyclic} and the corresponding solution $Y$ of \Cref{e:basic-differences} 
only provide so far an explosive cycle around a stationary level of national
income like a forced oscillator with one characteristic time solely, the period $\tau$, totally neglecting the multiscale character of an economy. 

\paragraph{B. Introducing exogenous growth --- the unrestricted model} For the emergence of a second characteristic time, \textit{the next task} is to allow steady growth in such an approach, e.g., by adding a trend
component 
\begin{equation}\label{e:increasing-autonomous}
A_t\ =\ (1+g)^t\/A,
\end{equation}
that is to assume that autonomous expenditures grow with a constant rate $g$ over time. In \cite[pp. 83f]{Flaschel:2009}, \textsc{Flaschel} gives an elegant explanation of how to derive a general cyclic solution of the system given by equations \eqref{e:keynes-consumption}, \eqref{e:keynes-investment}, \eqref{e:keynes-income} and \eqref{e:increasing-autonomous} from the solution \eqref{e:real-cyclic},  just exploiting  the linearity of the relations:
\begin{equation}\label{e:real-cyclic-general}
Y(t) \ =\ (1+g)^t \frac{(1+g)^2}{(1+g)(s+g)-\nu g}\/A\ +\ |\la_1|^t \left(\delta \cos(\theta t - \e )\right),
\end{equation}
where $\theta$ is determined by $\la_1/|\la_1| = \cos\theta +i \sin\theta$, (i.e., $\cos\theta = \frac{c+\nu}{2\sqrt{\nu}}$) and
where $\delta$ and $\e$ are the two initial conditions needed to supply a unique solution for our inhomogeneous difference equation of order 2. 

\begin{figure}[h]
	\includegraphics[scale=0.32]{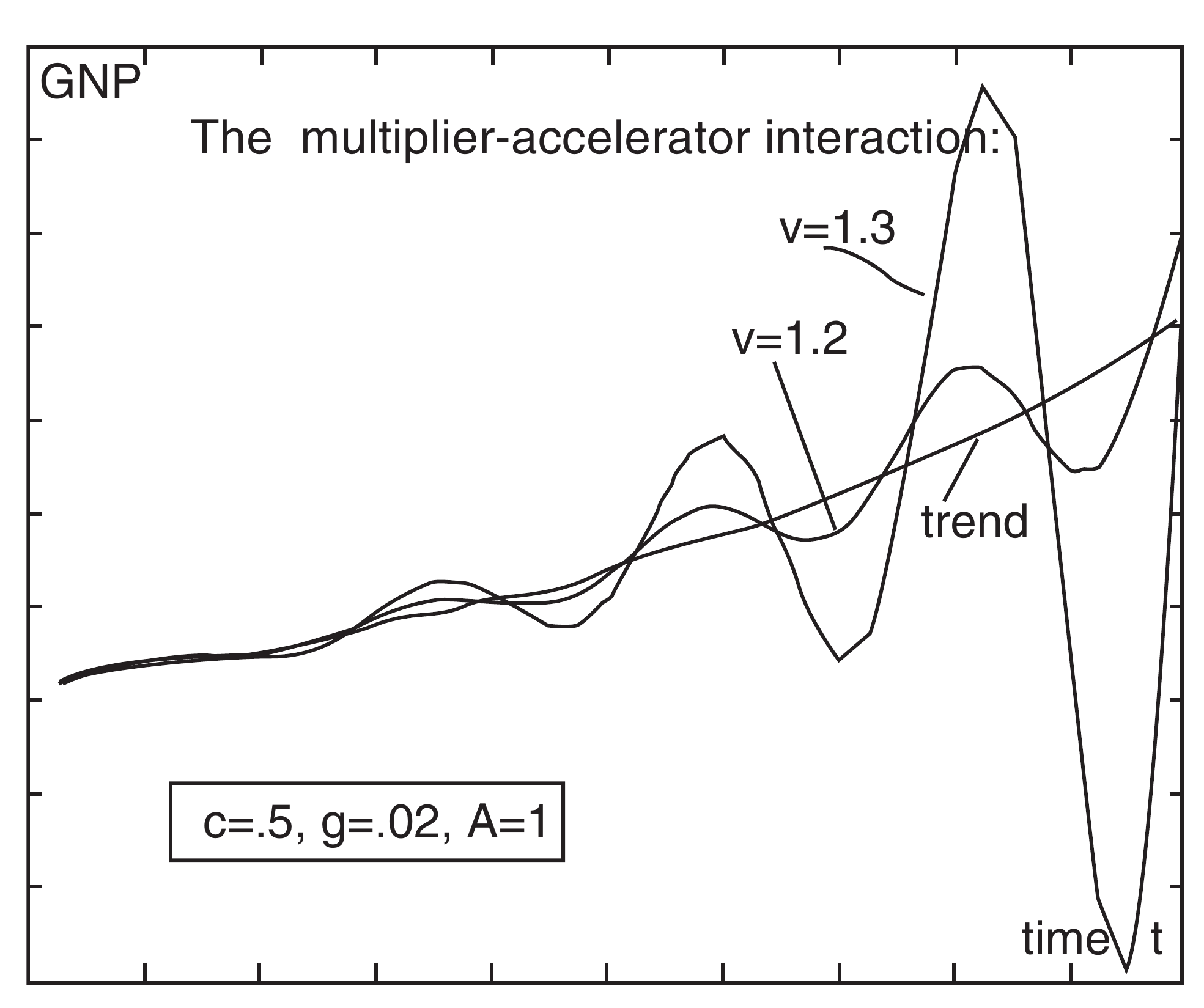}
	\includegraphics[scale=0.32]{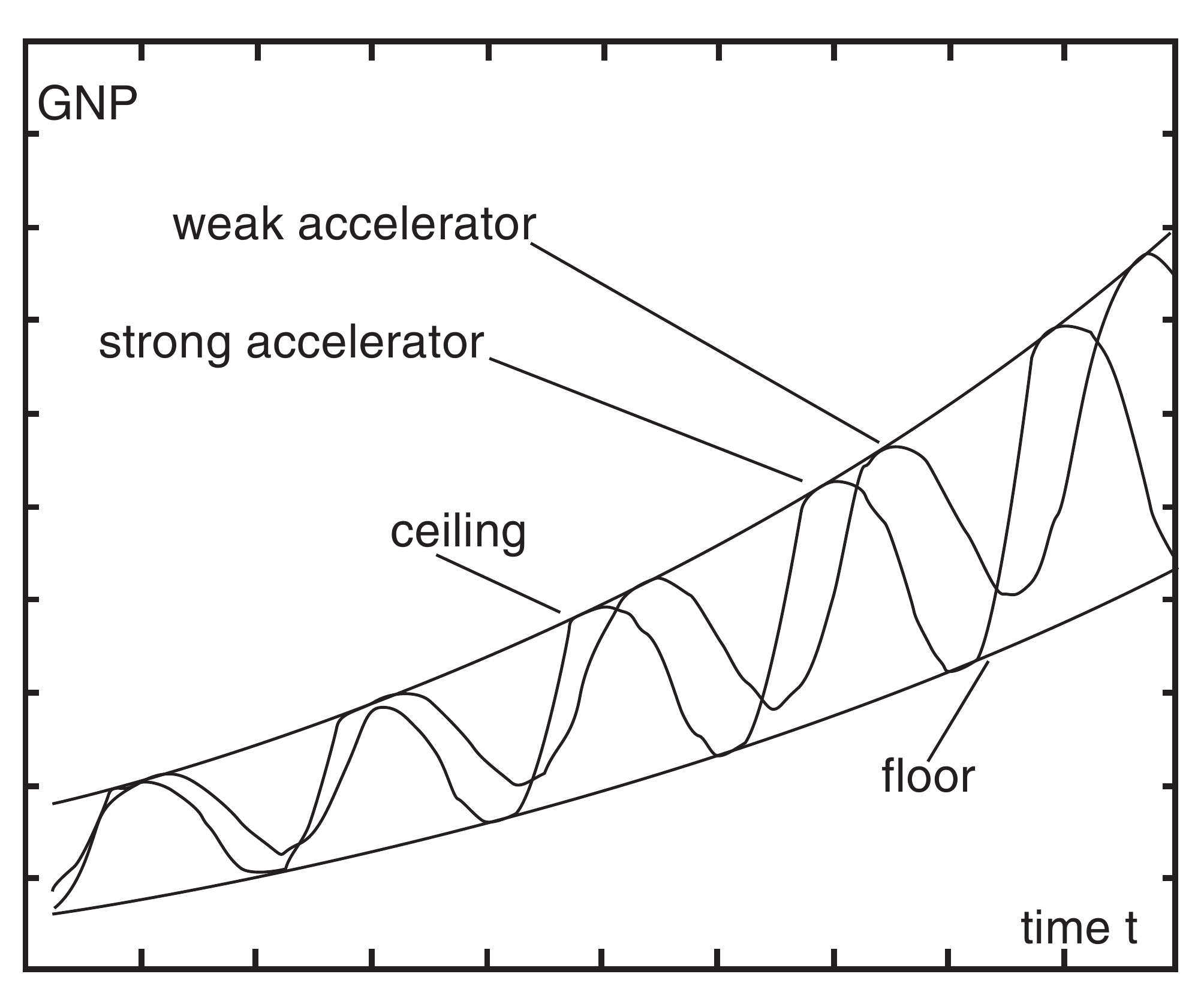}\newline
	\includegraphics[scale=0.33]{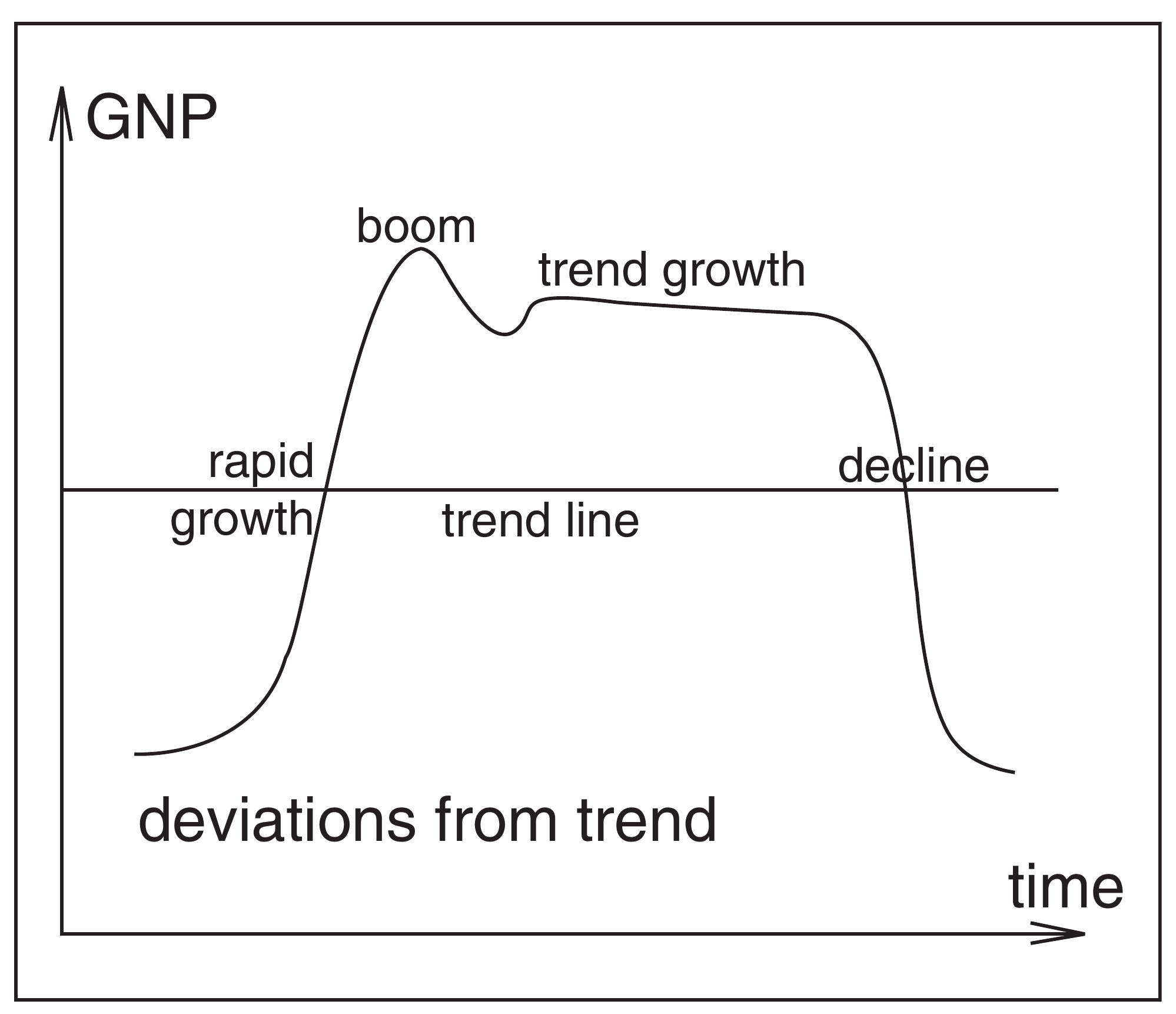}
	\caption{a) The dynamics of the solution \eqref{e:real-cyclic-general} of the unrestricted multiplier--accelerator model for typical parameters. b) The dynamics of the damped solutions of the restricted multiplier--accelerator model. c) The stylized business cycle of the 1950s and 1960s. All figures from \cite[pp. 85f]{Flaschel:2009}.}
	\label{f:hicks-cycles}
\end{figure}

To sum up, \Cref{e:real-cyclic-general} gives the circular flow of income of the
unlimited multiplier–accelerator model of \textit{C-I-Y}--interaction with exogenous growth.
A possible result of this goods–market interaction is provided in \Cref{f:hicks-cycles}a. This
figure shows the case of a cyclically explosive multiplier–-accelerator interaction,
which comes about for accelerator coefficients $\nu > 1$ and marginal propensities to
consume $c$ that are sufficiently low.

\paragraph{C. Damping the oscillations in the restricted model --- still futile} As \Cref{f:hicks-cycles}a immediately shows, there is a \textit{third task}, namely adding extra forces to this model
to keep its dynamics within reasonable bounds. The simplest way of doing this is to
add ceilings and floors like in modelling the damped elastic spring. This will give rise
to a damped type of behavior, where the ceiling, the floor, or both
delimiters can be operative (\Cref{f:hicks-cycles}b).

\textsc{Flaschel} has the following disenchanted comments in \cite[p. 85]{Flaschel:2009}: ``These simulations in particular show ... that ceilings (and floors) are only very briefly operative during
each cycle... The cycle ... therefore bears no close
resemblance to the form of the business cycle that was (and to some extend still is)
believed to be typical (at least) for the fifties and the sixties of (the 20th) century," referring to our \Cref{f:hicks-cycles}c. 

\textsc{Flaschel} points to several other shortcomings of such simple modelling and advocates the incorporation of prices and inflation to gain realism. Later on, in the same monograph (p. 275) he criticizes the choice of certain assumptions in mainstream economic modelling as, axiomatically seen, to be wrong ``so that complicated additional constructions (epicycles) become
necessary to reconcile this approach with the facts". He cites consentingly a question once raised by \textsc{J. Fuhrer}, the Executive Vice President \& Senior Policy Advisor of the  Federal Reserve Bank of Boston regarding monetary models:
	``Are we adding `epicycles' to a dead model?"

We shall not judge, but rather immediately proceed to the \textit{empirical} side. We will see clearly that short-term and long-term fluctuations appear simultaneously in the \textit{data}, no matter how ambivalent or even unreliable and implausible the \textit{mathematical modelling} and the \textit{numerical simulations} may appear.
\bigskip

\bigskip

\begin{figure}[h]
	\includegraphics[scale=1.025]{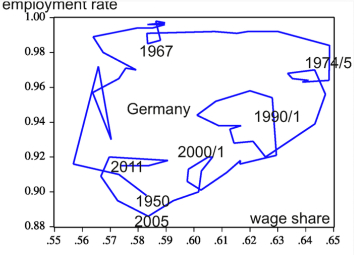}
	\caption{Exploring the postwar economy 1950-2011 in Germany with bivariate loops using penalized spline regression. The main loop is the ca. 60-years long  \textsc{Schumpeter}--wave with six ca. eight-years \textsc{Keynes}--business cycles embedded. Reprinted by courtesy of \textsc{P. Flaschel}, Bielefeld, based on data provided by \textsc{C. Groth}, Copenhagen}.
	\label{f:kondratiev+keynes-cycles_DE}
\end{figure}

\subsubsection{Schumpeter--cycles}\label{para:schumpeter}
As emphasized before, long waves are not always visible in economic data: 
\begin{description}
	\item[Short--phase waves] Production data like GDP will mostly disclose only a longterm growth trend with \textit{short-term oscillations} of the type of \textsc{Keynes}' business cycles. 
	\item[Long--phase waves] Happily, the coordinate change to \textsc{Goodwin}'s great ratios $e:=L/L^{\operatorname{sppl}}, v:=\frac{WL}{pV}$ (see \Cref{tab:economic-standard-notations}) shows the evolution of the {employment rate} of workers in the sphere of production $e(t)$ and its consequences for the {income distribution} between capital and labor $v(t)$ over \textit{longer periods}. 
\end{description}
To explore the emergence of two characteristic time scales in these two variables, we refer to \cite{Kauermann-Teuber-Flaschel:2012}: they assume that $y(t) := (e(t) , v(t))$ follows a \textit{long term trend} $c(t) =(c_1(t),c_2(t))$ and a \textit{business cycle fluctuation} $g(t) = (g_1(t), g_2(t))$, and model $y(t) = c(t) + g(t) + \e(t)$ with an error term $\e(t)$. In that way, the data  can point to two essential aspects of economic crises in capitalism that otherwise are nowadays disguised behind GDP data, namely 
\begin{enumerate}
	\item the periodic \textit{creative destruction} of the means and processes of agricultural and industrial production and of rendering services, and
	\item the damping potential of democratic regulation.
\end{enumerate}
It seems that these two aspects (or \textit{mechanisms}) are operative both in short--phase and long--phase waves.  

\paragraph{D. The unregulated economy at \textsc{Marx}' time} Roughly speaking, the unregulated capitalism at \textsc{Marx}' time led to deep economic crises every 6--10 years. There was no theoretical need to distinguish long waves and short waves: the short waves were sufficiently brutal not only for the working population but also for the land and capital owners to provide some creative destruction in an apparent manner.

\paragraph{E. The continuously regulated economy under strong democratic influences} That changed with the foundation and growing power of trade unions of working people that gave 
\begin{description}
	\item[Long--phase waves] much larger time span for the essential ups and downs, pictured in the long cycles, 
	\item[Short--phase waves]  while it supported some crisis management to get over the short cycles. 
\end{description}
The paramount example is provided by the postwar economy 1950-2011 in Germany (first Western Germany and West Berlin, then, since 1990, the unified Germany). After the break-up of the biggest German capital concentrations by the victory powers of World War II, some substantial rights were granted to the working people and there was an agreement between the major political parties to respect these rights and to hinder excessive influence of the capital side. Moreover, in collective negotiations between unions and management, there were often an invisible third social partner present, namely the social rights granted in the Eastern part of Germany.     

Our \Cref{f:kondratiev+keynes-cycles_DE} was found by exploring the German data with \textit{bivariate loops} using \textit{penalized spline regression}, as explained in \cite{Kauermann-Teuber-Flaschel:2012}. It is based on data provided by \textsc{C. Groth} and \textsc{J.B. Madsen}, Copenhagen, see also \cite{Groth-Madsen:2016}. Of course, the raw data for such a long period give a cloud which can be difficult to interpret in the presence of multiple time scales, as emphasized above in our Introduction. Following the mentioned stochastic geometry approach, one arrives at a \textit{wage share} $\times$ \textit{employment rate} diagram showing a main loop of the ca. 60-years long--wave of the \textsc{Kondratiev--Schumpeter--Goodwin--Friedman} (you name it) clockwise oriented cycle with six ca. eight-years \textsc{Keynes--Hicks}ian also clockwise oriented business cycles embedded.  

\begin{figure}[h]
	\includegraphics[scale=0.62]{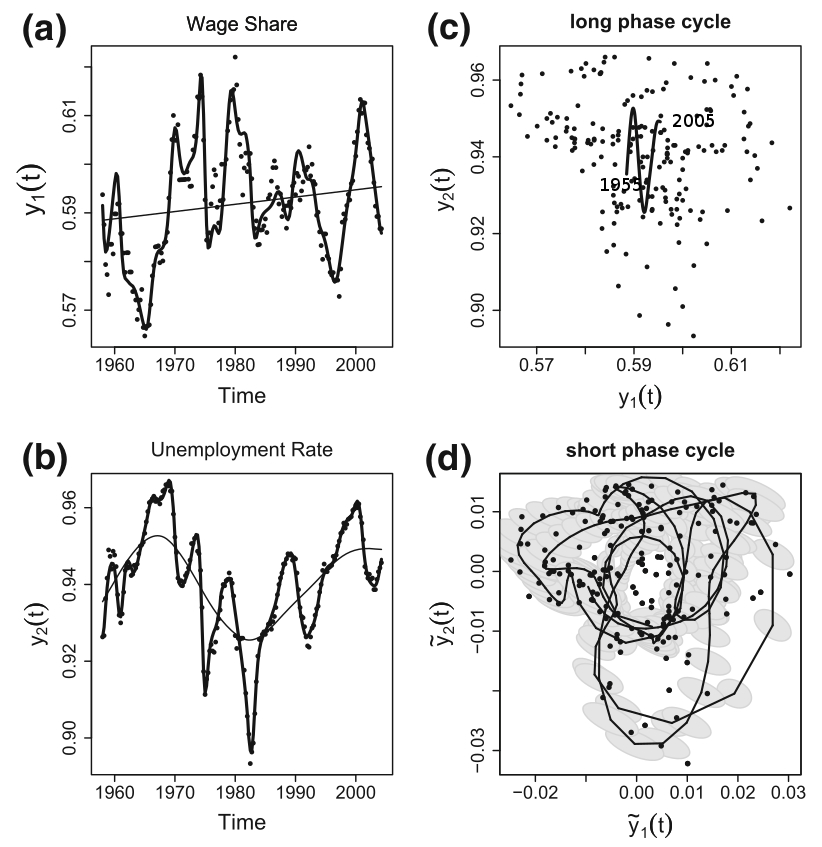}
	\caption{Exploring the US economy 1955-2005 with bivariate loops using penalized spline regression. Plot \textbf{a} and \textbf{b} show the time series wage share $y_1(t)$ and employment rate $y_2(t)$, with long--phase estimate (solid line) and final estimation (bold solid line). Plot \textbf{c} shows the observation cloud with the estimated zigzag line instead of an expected long--phase cycle and \textbf{d}
		shows the de-trended time series with estimated short--phase business cycle (solid line) and its confidence region (grey shaded area). Reprinted from \cite[Fig. 4]{Kauermann-Teuber-Flaschel:2012}}.
	\label{f:kondratiev+keynes-cycles_US}
\end{figure}

\begin{figure}[h]
	\includegraphics[scale=1.6]{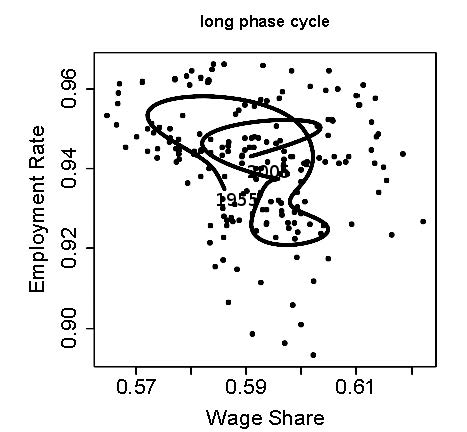}
	\caption{The plot shows the observations with
		a numerically forced long phase cycle (solid line). Courtesy by \textsc{P. Flaschel}}.
	\label{f:kondratiev+keynes-cycles_US_numerically-forced}
\end{figure}

\paragraph{F. An economy with rapid shifts between regulation and de--regulation} 
In \cite{Kauermann-Teuber-Flaschel:2012}, \textsc{Kauermann}, \textsc{Teuber} and \textsc{Flaschel} found that the corresponding time series of the US economy for the period 1955--2005 yield a different \Cref{f:kondratiev+keynes-cycles_US}: The long-phase wave in (c) seems to dipsy-doodle and lacks any remembrance of a cycle, presumably due to the state interventions of price and wage-stops of the \textsc{Nixon} administration and the abolition of union influence by the \textsc{Reagan} administration. Probably for the same reason, the \textsc{Keynes} business cycles in (d) show strong erratic fluctuations. Of course, one could force the data numerically to yield a curve resembling a cycle, see \Cref{f:kondratiev+keynes-cycles_US_numerically-forced}. Globally, the cycle form is misleading for the US economy when compared with the statistically more reliable \Cref{f:kondratiev+keynes-cycles_US}.c. However, we can read the swivels of the US economy easier in the numerically forced plot.

We must conclude, that the fluctuations of a real economy reveal the emergence of at least two different and each important time scales, as seen above in material sciences and cell physiology and claimed to be decisive in understanding climate change. Like always in multiscale modelling, there can be strong arguments for focusing on short term oscillations and what to do about them, as well as on a secular trend and what to do about it. Focus on claims from one side are stifling discussion and can lead astray. We borrow the wording from \textsc{Andrew Smithers}, the London based founder of Smithers \& Co., which provides economics-based asset allocation advice to over 100 fund management companies. In a letter to Financial Times \cite{Smithers:2019}, he commented on raising concerns about an evolving secular stagnation (e.g., in \cite{Summers:2015, Rachel+Summers:2019, Wolf:2019}) and warned: ``claims of secular stagnation stifle serious discussion and thus inhibit the chances of improved policies". 
Note that the concerns regarding climate change expressed above in \Cref{s:emergence-in-math}  are mostly about neglecting the secondary and, to be feared, secular effects of greenhouse gas emissions in the scientific literature and public perception due to -- understandable -- focus on the Paris accord and other short range estimates of immediate consequences of radiative forcing for the next few decades. \textsc{Smithers}' message is that regarding the global economy the challenges of dealing with emerging multiple time scales may be reversed.

\subsection{Multiscale spatio--temporal problems in economics} \label{ss:spatio-temporal}
At the end of this section we turn to the intricacies of multiple time scales in location modelling with its own multiscale challenges.

Multiscale spatial economy is the subject of mathematical location theory. It had its golden years in the middle of the 19th century during the railroad revolution of transport, and was revived by the analysis of trade patterns and location of economic activity during the ongoing globalization of the economy: 
\begin{quote}
	``Patterns of trade and location have always been key issues in the economic debate. What are the effects of free trade and globalization? What are the driving forces behind worldwide urbanization?" (From the citation of awarding \textit{The Sveriges Riksbank Prize in 	Economic Sciences in Memory of Alfred Nobel} 2008 to \textsc{P. Krugman})
\end{quote}
A multi--disciplinary example of high numerical complexity is to uncover the spatio--temporal patterns and dynamics of urban waste generation and management in a metropolitan area like Shanghai by a high spatial resolution material stock and flow analysis, see \cite{Liu-et-al:2018}. 

In the context of this survey, the essential point of the multiscale problems in location theory can be formulated via the emergence of multiple time scales: In the epoch of the building of railroads, e.g., the characteristic length for economic activity was approximately $s=50$ km and the typical velocity $v=50$ km/h, yielding a characteristic time $t_{\operatorname{trade}}=s/v$ of one hour, while the characteristic time $t_{\operatorname{build}}$ for building a new connection typically was larger by a factor $1.7 \times 10^5$, namely around two years. In nucleus, that is the way multiscale problems arouse in location theory.

For the history and the technical details of multiscale modelling of the spatial allocation of resources we refer to \cite{Puu:2003}, see also \cite{Krugman:1993, Fujita:2001}.

%
%
%
%
%
\section{Discussion and conclusions}\label{s:conclusions}

The paper was initiated by public and scientific controversies around the multiscale problems of environmental and climate change modelling. We have shown strongly related multiscale problems all around. In this multidisciplinary survey we confirmed the universality of multiple time scales in mathematical modelling by taking a closer look at striking examples from our own research. In particular, we explained how multiple time scales can emerge from seemingly simple models, i.e., mathematical models where neither time lags nor characteristic differences between the length of oscillations are put into the form and coefficients of the equations beforehand. We demonstrated also that a relatively small variation of the coefficients can suppress --- or support --- possible time scale differences.  

\subsection{Multiple time scales and communication skills}
It is well-known that multiscale problems present challenges not only to our modelling and simulation tools but also to our communication skills. 
In \cite[Preface]{Horstemeyer:2012}, \textsc{M. Horstemeyer} tells the following illuminating story with regard to Integrated Computational
Materials Engineering (ICME) and the use of multiscale modelling in engineering design:
\begin{quote}
	While working at Sandia National Laboratories 
	(laboratories with the mission to maintain the reliability of US American nuclear weapon systems, \textit{added by the authors}) in the mid-1990s, there was
	a meeting of an engineering mechanic, physicist, and materials scientist, and
	they were talking about stress. At the end of the meeting, they had all agreed
	that they understood each other's position. After the meeting, I interviewed
	each person separate from the others and asked what he or she thought about
	when the stress discussion came about. The physicist talked about pressure,
	pressure, pressure. The materials scientist talked about strain, strain, strain.
	And the engineering mechanics researcher talked about second-rank tensor,
	second-rank tensor, second-rank tensor. They had thought that they communicated,
	but they really did not because the paradigm of each one's discipline
	skewed his or her semantical communication. This is often the case for interdisciplinary
	researchers, so one has to be careful when discussing multiscale
	modelling or history modelling from process to performance using the ICME
	tools with others who were trained under a different paradigm.
	
	Because of these different paradigms, I decided shortly after those interviews
	to perform simulations at all the different length scales and to try to
	understand the pertinent cause-effect relationships with the hope that I could
	understand the bridging concepts...
\end{quote}

When \textit{experts} with different backgrounds have difficulties catching each other's characteristic spatial or time scales,
it seems appropriate to close our survey with a view on the challenges of communicating multiple time scales to a \textit{wider audience}.

\subsection{Our results: Models and simulations with a multiplicity of time scales can be of different character}

In this review we elaborated universal aspects of models with multiple time scales in different contexts. Shared numerical challenges and common approaches and solutions of problems with multiple characteristic times hide our main finding: 
(see \Cref{tab:multiplicity-diagram}): 
some essential differences regarding the origin, emergence and meaning of the differences between the temporal scales in different mathematical models and numerical simulations. 

We analysed five cases.
Our point of departure was the distinction between the characteristic times of four selected processes underlying \textit{climate change modelling}: (i) the emission of greenhouse gases; (ii) the induced radiative changes of the atmosphere; (iii)  the surface oceanic uptake of greenhouse gases; (iv) and the deep oceanic release of methane clathrates. In principle, the anthropogenic part of (i) follows the time scale of political decisions. The  complexity and non-linearity in (ii) makes it difficult to determine a tipping point when the linear radiative model no longer is a reliable approximation. According to elementary chemistry, the oceanic uptake in (iii) will be accelerated with every increase of the greenhouse gases, while geology tells us that tens of millions of years may be needed for stabilization and a return to present climatic conditions. Regarding (iv), the heat uptake of the oceans can be observed at a few check points, but globally in no meaningful way estimated or predicted. Increase of the pressure, the factor counteracting the temperature driven clathrate release, is easily measurable by the increase of the oceans' levels and calculable by the expected melting of the polar ice caps. Roughly speaking, the multiscale character of climate change modelling is determined by the superposition of four interrelated, but qualitatively different processes with different, and for the most unknown, characteristic times. As mentioned before, that uncertainty is even more outspoken when one tries taking into regard the multiple spatial scales entering regional models of climate change.

\begin{table}[h]
\caption{Multi-temporality and structural instability: shared features and 		different character of multiscale mathematical modelling and simulation of climate, materials, life, and economy
} 
\label{tab:multiplicity-diagram}
	\centering
	\rotatebox{90} {\includegraphics[scale=0.62]{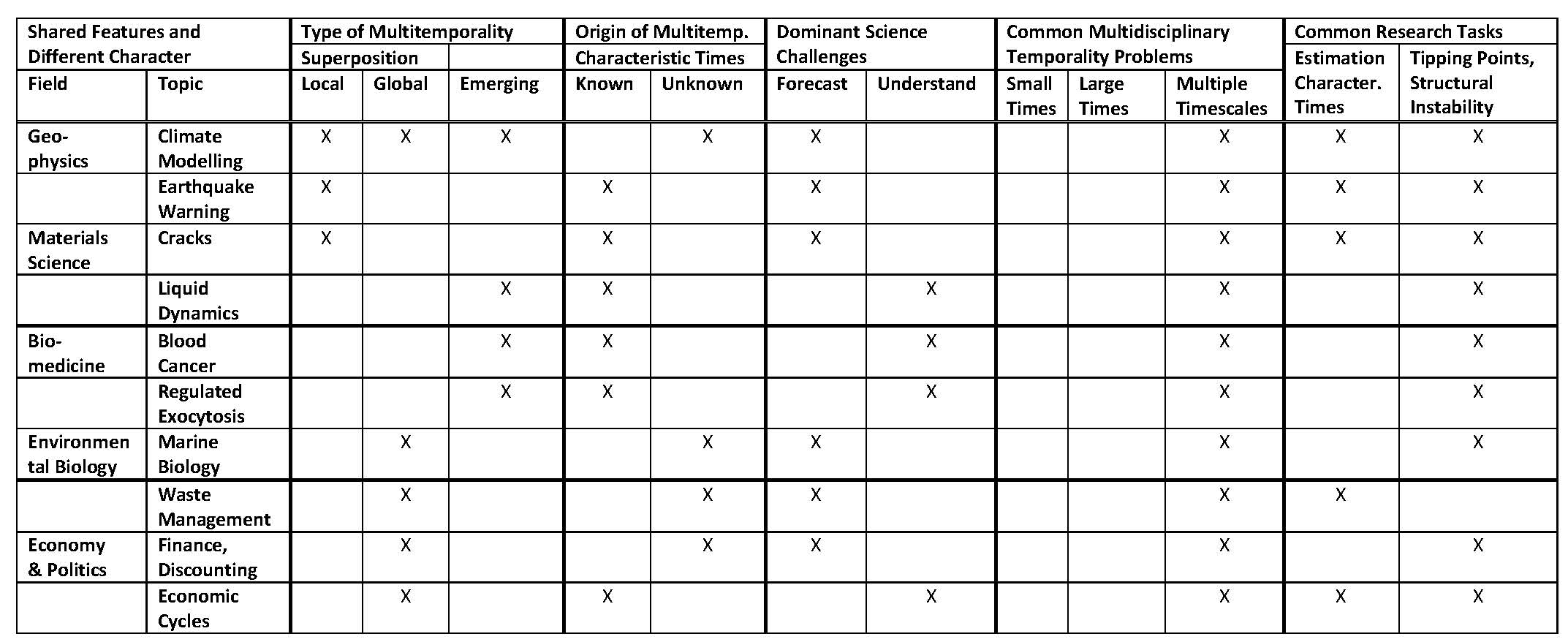}}
\end{table}

Our second case was the glass transition of \textit{disordered materials}. Given a potential for one type of matter, we obtained the surprising emergence of a second characteristic time for the forming of patterns besides the smaller time of instant order changes under changes of pressure and/or temperature. Here the multiplicity of time scales is not the result of a superposition of different processes but emerges from one single model and simulation over evolving time. In principle, there are other processes at work, e.g., of quantum chemistry underlying the perpetual position changes of the electrons of the considered molecules. In materials science, however, these processes need not be incorporated as a separate process with its own characteristic time, but can be approximated sufficiently precisely by a single potential function for each chemical type of molecule - contrary to the situation with climate change where the reduction to only one or two processes would be misleading, no matter how well intended it may be for public dissemination.

Our third case, the development of a certain type of \textit{blood cancer}, was similar to the glass transition: there is one model of the slow mixing of mutated and healthy blood cells without alarming symptoms until the disease is diagnosed late in the disease progression. Such a delay can cause difficulties for an efficient treatment.

Our fourth case, the biphasic \textit{insulin secretion} of pancreatic beta-cells, seemingly exhibited a similar behaviour, namely the emergence of two different characteristic times  by the running of one homogeneous process over time. Closer investigation revealed, however, that there were at least two clearly distinguishable processes with different characteristic times to base the modelling upon.

Our fifth case, the distinction between two characteristic times in macroeconomic waves, looks like a simple superposition of the short Keynesian business cycles, driven by excess or shortcoming in wages, respectively, in employment, and the long Schumpeterian cycles, driven by secular democratic or technological innovations. But not so: the data indicate an interaction of the two formally different looking processes. That interaction may provide for a relative shortness of the Keynes cycles and a structural stability of the Schumpeter wave over decades, quite similar as in our speculation in \Cref{ss:climate-change} regarding a possible stabilizing effect of the interaction of methane and carbon dioxide levels for the climate in Earth history.

Our findings indicate, that political and societal focus on the climate crisis and environmental degradation must be supplemented by a more comprehensive understanding of the intricacies, of the ups and downs of systems with a multiplicity of time scales. Without a wider insight among environmental scientists, grass-roots and political actors, it might be difficult to find and maintain over an extended period the right answer to \textsc{Inger Andersen}, the Executive Director of the United Nations Environment Programme, when she deplores in \cite[p. xiii]{EGR:2019} ``Our collective failure to act strongly and early" and praises ``Political and societal focus on the climate crisis is at an all-time high, with youth movements holding us to account" and giving us ``stark choice: set in motion the radical transformations we need now, or face the consequences  of a planet radically altered by climate change."

\subsection{The morale}

Mitigating environmental and climate change hazards depends decisively on the broad support of an informed public. In our contribution, we point to the multiple time scales in climate change and sustainable development. Greenhouse gases accumulate rapidly in the atmosphere with immediate changes in the radiation pattern, while secondary effects develop slowly like the release of methane from the oceans and permafrost regions and, in the opposite direction, the binding and storage of heat and CO$_2$ in the oceans. Such huge differences between characteristic time lengths provide not only difficulties in mathematical modelling, statistical sampling, and numerical simulation but can become misleading in communicating threats and solutions. The evidence submitted here from multiscale modelling and simulation in other fields (materials science, bio sciences and economy) indicates that disregard of multiple time scales of a problem can either induce overestimation in the short run and underestimation in the end, i.e., the ominous cry wolf effect, and/or underestimation in the short run and overestimation in the end, resulting in fatalistic forfeiting or preposterous activism.  

\appendix
\section{Communicating the emergence of multiple time scales}\label{ss:communicating}

\subsection{The public disregard}
Regarding multi-scale problems we register a wide gap between 
\begin{enumerate}[(A)]
	\item the mathematical proficiency in modelling and simulation of multiscale systems, and 
	\item the public disregard even of the most elementary multiscale aspects, like the emergence of multiple time scales.
\end{enumerate}
Claim (A) is evident from 
\begin{itemize}
	\item the enormous literature, both
	\begin{itemize}
		\item learned journals, 
		\item specialized textbooks as \cite{E:2011, Engquist-et-al:2005, Engquist-et-al:2009, Horstemeyer:2012, Majda:2016, Majda:2018} and 120 other textbooks and scientific monographs with the entry \textit{multiscale} in the title, according to zbMATH \url{https://zbmath.org/?q=ti%3Amultiscale+%26+dt%3Ab&p=1}, and 
			\item a variety of survey articles like \cite{E-Engquist:2003, Konig:2016};
		\end{itemize}
		\item the new paradigm of modelling (often required by the intricacies of the numerical simulation of multiscale problems), namely doing the mathematical modelling, i.e., the choice of the relevant equations on the different scales, in parallel with designing the numerical algorithms as elaborated by E and \textsc{Engquist} in \cite[Outlook, p. 1069]{E-Engquist:2003}: ``\dots in much of computational mathematics,
			we are used to taking for granted that the
			models are given, they are the ultimate truth, and
			our task is to provide methods to analyze and solve
			them. This shields us from the frontiers of science
			where phenomena are analyzed and models are
			formulated.
			
			Multiscale, multiphysics modelling brings in a
			new paradigm. Here the problems are given, and a
			variety of mathematical models at different levels
			of detail can be considered. The right equation is
			selected during the process of computation
			according to the accuracy needs. This brings mathematical
			analysis and computation closer to the
			actual scientific and engineering problems. It may
			no longer be necessary to wait for scientists to
			develop simplified equations before computational
			modelling can be done. This is an exciting new
			opportunity for computational science and for
			applied mathematics. It will bring applied mathematics
			closer to other fields of mathematics, as, for
			example, mathematical physics and probability
			theory. It will also bring these fields closer to
			the frontiers of science."

		\item the multiscale modelling and simulation for design verification and validation purposes of nuclear weapons that permitted the \textit{Treaty on the Limitation of Underground Nuclear Weapon Tests}, also known as the \textit{Threshold Test Ban Treaty} (TTBT) to enter into force in 1990, see what \textsc{Mark Horstemeyer} recalls in \cite[Section 1.3, pp. 4f]{Horstemeyer:2012} about that side of the history and reliability of multiscale modelling and simulation. 
	\end{itemize}
	
	Claim (B) is evident from a large study \cite{Santos-et-al:2016} of 2016. Assessing student perceptions and comprehension of climate change
	in Portuguese higher education institutions and surveying studies from other countries, the authors found 
	\begin{enumerate}
		\item a lack of interrelation between the common attention to Climate Change (CC) in general terms and the personal or political attitudes of the respondents;
		\item a different, but mostly low level of physical understanding;
		\item absence of any feeling for dynamics and characteristic time scales, e.g., when the respondents were mixing the presence, the near future, and further developments.	
	\end{enumerate}
	So much for the academic youth. For quite another cohort, investors on the financial markets, we can derive a similar disregard of the multiple time scales of climate change and transforming for sustainability. In \cite{Allen:2018}, \textsc{Kate Allen}, a capital markets correspondent for the \textit{Financial Times} reports:
	\begin{quote}
		Sales of green bonds are stuttering after several years of rapid growth. In the three months to the end of September 2018, issuers around the world sold \$31.6bn of green-labelled debt, according to research by credit rating agency Moody's. That is 30 per cent lower than the tally for the second quarter, and 18 per cent down on the \$38.5bn sold in the same quarter of 2017.
	\end{quote}
	Moody's had originally forecast that green bond sales in 2018 would hit \$250bn, a considerable
	increase from last year's record \$163bn.
	
	In the annual \textit{BP Statistical Review of World Energy}, \textsc{S. Dale}, British Petrol's chief economist gives a related picture of the growth of world coal consumption in 2017 by 1\% after annual decrease since 2013, a corresponding increase by 1.5\% of CO$_2$ emissions from energy consumption,  and the lack of almost any improvement in the
	power sector fuel mix over the past 20 years. ``The share of coal in the power
	sector in 1998 was 38\% --- exactly the same as in 2017", states \textsc{Dale} 
	in \cite{Dale:2018}; and the price per ton of CO$_2$ is on the order of 1\$, 
	while it should be 25\$ per ton of CO$_2$ \textit{now} to keep the temperature
	increase below $2\12\/^{\circ}{\rm C}$ in 2050 according to \cite[p. 316]{Neuhaus:2013}.
	
	In \cite{Wolf:2018}, the Financial Times' chief economics commentator \textsc{M. Wolf} searched for an explanation of that ``shameful" behaviour of investors. Not surprisingly, he too found a disregard of the multiple time scales at the heart of the problem: 
	\begin{quote}
		In all, we need to shift the world on to a different investment and
		growth path right now. This is more technically possible than we
		used to think. But it is politically highly challenging. Above all,
		climate change involves huge distributional issues — between rich
		countries and poor ones, between countries that caused the
		problem and those that did not, between countries that matter for
		the solution and those that do not and,\textit{ not least, between people
			today, who make the decisions, and people tomorrow, who suffer
			the results} (emphasized by the authors). The natural tendencies are either to do nothing, while
		insisting there is no problem, or to agree there is a problem, while
		merely pretending to act. It is not clear which form of obfuscation is
		worse.
	\end{quote}
This is not the place to discuss whether this is a clash between economic rationality (more precisely, capitalists' profit orientation) and environmental, science based arguments, as proponents of the Climate Justice Movement may argue, see, e.g.,  \cite{Bek-Thomsen-et-al:2017, Bullard-Muller:2012, Jacobsen:2018}  --- or just, as, e.g.,  \textsc{M. Porter}, a Harvard authority on competitive strategy, or \textsc{W. Nordhaus}, the 2017 Nobel laureate in economic theory, may claim with \textsc{Wolf}, a common disregard of shared values in multiple time scales, see \cite{Porter-Kramer:2011} and \cite[Ch. 26, ``Prisoners of the present"]{Neuhaus:2013}. 
	
	\subsection{Can scientists reach the public}
	To communicate the emergence of multiple time scales, we may draw on experts in science communication. The British study \cite{Bowater-Yeoman:2012} distinguishes between three phases associated with the development of science communication (with somewhat unlucky acronyms): 
	\begin{description}
		\item[SL] scientific literacy;
		\item[PUS] public understanding of science; and, 
		\item[PEST] public engagement with science and technology (the current challenge).
	\end{description}
	Similarly the recent \cite[Introduction, p. 1-4]{Illingworth-Allen:2016}.

	Another study \cite[pp. 106f]{Hartomo-Cribb:2002} of 2002 recalls the continuous flow of firm recommendations for public
	consultation to become an integral part of doing science -- not an optional
	add-on. They comment in rather sharp wording: ``This may seem a bit heretical in lands where science policy is still in the
	hands of the science mafia, and the game is how to limit and exclude
	rather than to engage, listen and learn. But there is more than a grain of
	commonsense in it." They quote from the British Council's useful report on the democratization of science a 
	list of essential preconditions, among them: \textit{independent advice and research}; and \textit{initiatives to forecast, recognize and resolve conflict}.
	Surely, that are valuable guidelines for communicating the emergence of multiple time scales in climate  change and transforming for sustainability, combined with the advice of \cite{Illingworth-Allen-ch5:2016}: There is an obligation for scientists to communicate their research to the rest of society, to inform people about scientific advances, and to ultimately engage them in a two-way dialog so that the general public does not just understand what science is doing, but that they also have a say in what is being done.
	
	\subsubsection{Urgent tasks for the IPCC} Clearly, one has to be grateful to the IPCC that it opts in The IPCC Special Report on 1.5$^\circ$C \cite{ClimateChange:2018} for a rigorous interpretation of the 1.5$^\circ$C  limit on global warming. We deplore though, that the report solely describes the means to reach that goal and the necessary adaptations but underplays the alarming fact that global warming is accelerating and that there are no signs for a decrease in global carbon emissions (see \cite{GlobalCarbonBudget:2018}). Based on our insight into the emergence of multiple time scales we fully agree with the comment \cite{Xu-et-al:2018} in \textit{Nature} that 
	\begin{quote}
		``Policymakers should ask the IPCC for another special report, this time on the rates of climate change over the next 25 years... 
		Researchers should improve climate models to describe the next 25 years in more detail, including the latest data on the state of the oceans and 
		atmosphere, as well as natural cycles. 
		They should do more to quantify the odds and impacts of extreme events. 
		The evidence will be hard to muster, but it will be more useful in 
		assessing real climate dangers and responses."
	\end{quote}

	In multiscale modelling and simulation, the general wisdom (quoted occasionally by \textsc{W. E}) is
	\begin{quote}
		``For most of the problems we are facing in science and engineering, the theoretical challenges lie in mathematics and algorithms.'' 
	\end{quote}
	Regarding the emergence of multiple time scales in climate change, an even better advice might be to recall what we have learned as children, namely to take care of our resources and not to throw our garbage in the environment, visible or invisible, --- and to follow that teaching rigorously.

\newcommand{\bysame}{}
\bibliography{Multiscale}

\def\cprime{$'$}
\renewcommand*{\bysame}[1]{---\thinspace}
\renewcommand*{\bfdefault}{b}
\providecommand{\MR}{\relax\ifhmode\unskip\space\fi MR }
\providecommand{\MRhref}[2]{%
  \href{http://www.ams.org/mathscinet-getitem?mr=#1}{#2}
}
\providecommand{\href}[2]{#2}
\begin{thebibliography}{100}

\bibitem{ParisAgreement:2015}
\emph{1/CP.21 Adoption of the Paris Agreement}, Dec. 13 2015, Report of the
  Conference of the Parties on its twenty-first session, held in Paris from 30
  November to 13 December 2015, Addendum Part two: Action taken by the
  Conference of the Parties at its twenty-first session.

\bibitem{GretaThunberg:2019}
\textsc{G.~Thunberg}, \emph{How dare you}, Sept. 24 2019, Full speech to world
  leaders at opening of UN Climate Action Summit on Sept. 24, 2019, video and
  transcription at
  https://www.ft.com/video/a851a48e-abab-4e1f-a2b7-c4b811bb74ae.

\bibitem{ClimateChange:2019}
\textsc{H.-O. P{\"o}rtner and {al.}} (eds.), \emph{IPCC Special Report on the
  Ocean and Cryosphere in a Changing Climate}. World Meteorological
  Organization, Geneva, Switzerland, 2019, In press,
  \url{https://www.ipcc.ch/srocc/home/}.

\bibitem{Cheng:2019}
\textsc{L.~Cheng, J.~Abraham, Z.~Hausfather and K.~E. Trenberth}, `How fast are
  the oceans warming?'. \emph{Science} \textbf{363}/6423 (2019), 128--129.

\bibitem{Resplandy:2018}
\textsc{L.~Resplandy, R.~F. Keeling, Y.~Eddebbar, M.~K. Brooks, R.~Wang,
  L.~Bopp, M.~C. Long, J.~P. Dunne, W.~Koeve and A.~Oschlies}, `Quantification
  of ocean heat uptake from changes in atmospheric O2 and CO2 composition'.
  \emph{Nature} \textbf{563}/7729 (2018), 105--108.

\bibitem{Behrens:2016}
\textsc{J.~Behrens}, `Numerical methods and scientific computing for climate
  and geosciences'. In: \emph{Mathematics and Society, edited by Wolfang
  K{\"o}nig}. European Math. Soc., Z{\"u}rich, 2016, pp.~281--293.

\bibitem{Hansen-et-al:2013}
\textsc{J.~Hansen, M.~Sato, G.~Russell and P.~Kharecha}, `Climate sensitivity,
  sea level and atmospheric carbon dioxide'. \emph{Philos. Trans. Royal Soc. A}
  \textbf{371}/2001 (28 October 2013), 20120294, One contribution of 11 to a
  Discussion Meeting Issue \textit{Warm climates of the past --? a lesson for
  the future?}, compiled and edited by Daniel J. Lunt, Harry Elderfield,
  Richard Pancost and Andy Ridgwell, 28 October 2013, PTRSA \textbf{371}/2001.

\bibitem{Horstemeyer:2012}
\textsc{M.~F. Horstemeyer}, \emph{Integrated Computational Materials
  Engineering (ICME) for Metals --- Using Multiscale Modeling to Invigorate
  Engineering Design with Science}. John Wiley \& Sons, Inc., Hoboken, New
  Jersey, 2012, \copyright\ by The Minerals, Metals \& Materials Society.

\bibitem{Burke-et-al:2018}
\textsc{K.~D. Burke, J.~W. Williams, M.~A. Chandler, A.~M. Haywood, D.~J. Lunt
  and B.~L. Otto-Bliesner}, `Pliocene and Eocene provide best analogs for
  near-future climates'. \emph{PNAS} (2018), published ahead of print December
  5, 2018.

\bibitem{Lehner:2018}
\textsc{F.~Lehner}, \emph{Model simulations of past climates},
  \url{https://edition.cnn.com/2018/12/10/world/climate-change-pliocene-study/},
  10 December 2018, Comments by FL to \cite{Burke-et-al:2018} cited in article
  ``In 200 years, humans reversed a climate trend lasting 50 million years,
  study says" by Susan Scutti, CNN.

\bibitem{Hirsch-Smale:1974}
\textsc{M.~W. Hirsch and S.~Smale}, \emph{Differential Equations, Dynamical
  Systems, and Linear Algebra}. Academic Press [A subsidiary of Harcourt Brace
  Jovanovich, Publishers], New York-London, 1974, Pure and Applied Mathematics,
  Vol. 60.

\bibitem{Rosling:2018}
\textsc{H.~Rosling, O.~Rosling and A.~Rosling~R{\"o}nnlund}, \emph{Factfulness:
  Ten Reasons were Wrong about}. Sceptre, 2018.

\bibitem{Steffen:2018}
\textsc{W.~Steffen, J.~Rockstr{\"o}m, K.~Richardson, T.~M. Lenton, C.~Folke,
  D.~Liverman, C.~P. Summerhayes, A.~D. Barnosky, S.~E. Cornell, M.~Crucifix,
  J.~F. Donges, I.~Fetzer, S.~J. Lade, M.~Scheffer, R.~Winkelmann and H.~J.
  Schellnhuber}, `Trajectories of the Earth System in the Anthropocene'.
  \emph{PNAS} \textbf{115} (August 14, 2018), 8252--8259, published ahead of
  print August 6, 2018.

\bibitem{Chandler:1987}
\textsc{D.~Chandler}, \emph{Introduction to Modern Statistical Mechanics}. The
  Clarendon Press, Oxford University Press, New York, 1987.

\bibitem{Archer:2009}
\textsc{D.~Archer, M.~Eby, V.~Brovkin, A.~Ridgwell, L.~Cao, U.~Mikolajewicz,
  K.~Caldeira, K.~Matsumoto, G.~Munhoven, A.~Montenegro and K.~Tokos},
  `Atmospheric lifetime of fossil fuel carbon dioxide'. \emph{Annu. Rev. Earth
  Pl. Sc.} \textbf{37}/1 (2009), 117--134.

\bibitem{Osegovic-Tatro-Holman:2006}
\textsc{J.~P. Osegovic, S.~R. Tatro and S.~A. Holman}, `Physical chemical
  characteristics of natural gas hydrate'. In: \emph{\textsc{Michael D. Max et
  al.} (eds.), Economic Geology of Natural Gas Hydrate}, Coastal Systems and
  Continental Margins, vol.~9. Springer, Dordrecht, 2006, pp.~45--104.

\bibitem{ClimateChange:2013}
\textsc{T.~Stocker, D.~Qin, G.-K. Plattner, M.~Tignor, S.~Allen, J.~Boschung,
  A.~Nauels, Y.~Xia, V.~Bex and P.~Midgley} (eds.), \emph{Climate Change 2013:
  The Physical Science Basis: Working Group I Contribution to the Fifth
  Assessment Report of the Intergovernmental Panel on Climate Change}.
  Cambridge University Press, Cambridge, United Kingdom and New York, NY, USA,
  2013.

\bibitem{Kirschvink:1992}
\textsc{J.~L. Kirschvink}, `Late Proterozoic low-latitude global glaciation:
  the snowball Earth'. In: \emph{The Proterozoic Biosphere. A Multidisciplinary
  Study} (J.~W. Schopf and C.~Klein, eds.). Cambridge University Press,
  Cambridge, GB, 1992, pp.~51--52.

\bibitem{Hoffman:2011}
\textsc{P.~F. Hoffman}, `Snowball Earth'. In: \emph{Encyclopedia of Geobiology}
  (J.~Reitner and V.~Thiel, eds.). Springer Netherlands, Dordrecht, 2011,
  pp.~814--824.

\bibitem{ClimateChange:2007}
\textsc{S.~Solomon, D.~Qin, M.~Manning, K.~Averyt and M.~Marquis},
  \emph{Climate Change 2007 - The Physical Science Basis: Working Group I
  Contribution to the Fourth Assessment Report of the IPCC}, Assessment report
  (Intergovernmental Panel on Climate Change). Cambridge University Press,
  2007.

\bibitem{Gao-House-Chapman:2005}
\textsc{S.~Gao, W.~House and W.~Chapman}, `NMR/MRI study of gas hydrate
  mechanisms'. \emph{J. Phys. Chem. B} \textbf{109}/41 (September 28, 2005),
  19090--19093, PMID 16853461.

\bibitem{Reay-et-al:2018}
\textsc{D.~S. Reay, P.~Smith, T.~R. Christensen, R.~H. James and H.~Clark},
  `Methane and global environmental change'. \emph{Annual Review of Environment
  and Resources} \textbf{43}/1 (2018), 165--192, First published as a Review in
  Advance on June 8, 2018.

\bibitem{Avery:2018}
\textsc{J.~S. Avery}, \emph{The Climate Emergency: Two Time Scales}, 1st ed..
  Danish Peace Academy, 2018, Draft of November 2017 free available at
  \url{http://www.fredsakademiet.dk/library/climate.pdf}.

\bibitem{Milkov:2004}
\textsc{A.~V. Milkov}, `Global estimates of hydrate-bound gas in marine
  sediments: how much is really out there?'. \emph{Earth-Science Reviews}
  \textbf{66}/3 (2004), 183--197.

\bibitem{Li-et-al:2018}
\textsc{J.-f. Li and e.~a. {}}, `The first offshore natural gas hydrate
  production test in South China Sea'. \emph{China Geology} \textbf{1}/1
  (2018), 5--16.

\bibitem{Zhang-et-al:2018}
\textsc{R.-w. Zhang and e.~a. {}}, `Distribution of gas hydrate reservoir in
  the first production test region of the Shenhu area, South China Sea'.
  \emph{China Geology} \textbf{4} (2018), 493--504.

\bibitem{Logan:2013}
\textsc{J.~D. Logan}, \emph{Applied Mathematics}, fourth ed.. John Wiley \&
  Sons, Inc., Hoboken, NJ, 2013.

\bibitem{Engquist:2015}
\textsc{B.~Engquist} (ed.), \emph{Encyclopedia of Applied and Computational
  Mathematics}. Springer, 2015.

\bibitem{E:2011}
\textsc{W.~E}, \emph{Principles of Multiscale Modeling}. Cambridge University
  Press, Cambridge, 2011.

\bibitem{Engquist-et-al:2005}
\textsc{B.~Engquist, P.~L\"{o}tstedt and O.~Runborg} (eds.), \emph{Multiscale
  Methods in Science and Engineering}, Lecture Notes in Computational Science
  and Engineering, vol.~44. Springer-Verlag, Berlin, 2005, Papers from the
  conference held in Uppsala, January 26--28, 2004.

\bibitem{Engquist-et-al:2009}
\textsc{B.~Engquist, P.~L\"{o}tstedt and O.~Runborg} (eds.), \emph{Multiscale
  Modeling and Simulation in Science}, Lecture Notes in Computational Science
  and Engineering, vol.~66. Springer-Verlag, Berlin, 2009, Papers from the
  Summer School held in Liding\"{o}, June 2007.

\bibitem{Majda:2016}
\textsc{A.~J. Majda}, \emph{Introduction to Turbulent Dynamical Systems in
  Complex Systems}, Frontiers in Applied Dynamical Systems: Reviews and
  Tutorials, vol.~5. Springer, 2016.

\bibitem{Klein:2015}
\textsc{R.~Klein}, `Multiscale numerical methods in atmospheric science'. In:
  \emph{Encyclopedia of Applied and Computational Mathematics, edited by B.
  Engquist, = \cite{Engquist:2015}}. Springer, Berlin-Heidelberg-New York,
  2015, pp.~1002--1006.

\bibitem{Madsen-et-al:2017}
\textsc{M.~S. Madsen, P.~L. Langen, F.~Boberg and J.~H. Christensen}, `Inflated
  uncertainty in multimodel--based regional climate projections'.
  \emph{Geophys. Res. Lett.} \textbf{44} (2017), 606--613, Published online 23
  Nov. 2017.

\bibitem{Commoner:1992}
\textsc{B.~Commoner}, \emph{Making Peace with the Planet}. New Press, 1992.

\bibitem{Commoner:1971}
\bysame, `The ecological crisis'. In: \emph{The Social Responsibility of the
  Scientist. Edited by Martin Brown}. The Free Press. A Division of The
  Macmillan Company, New York, N.Y., U.S.A., 1971, pp.~174--183.

\bibitem{E-Engquist:2003}
\textsc{W.~E and B.~Engquist}, `Multiscale modeling and computation'.
  \emph{Notices Amer. Math. Soc.} \textbf{50}/9 (October 2003), 1062--1070.

\bibitem{Donth:1982}
\textsc{E.~Donth}, `The size of cooperatively rearranging regions at the glass
  transition'. \emph{J. Non-Cryst. Solids} \textbf{53}/3 (1982), 325--330.

\bibitem{Sastry:1998}
\textsc{S.~Sastry, P.~G. Debenedetti and F.~H. Stillinger}, `Signatures of
  distinct dynamical regimes in the energy landscape of a glass-forming
  liquid'. \emph{Nature} \textbf{393}/6685 (1998), 554--557.

\bibitem{Garrahan:2002}
\textsc{J.~P~Garrahan and D.~Chandler}, `Geometrical explanation and scaling of
  dynamical heterogeneities in glass forming systems'. \emph{Phys. Rev. Lett.}
  \textbf{89} (2002), 035704.

\bibitem{Dyre:2006}
\textsc{J.~C. Dyre}, `The glass transition and elastic models of glass-forming
  liquids'. \emph{Reviews of Modern Physics} \textbf{78} (2006), 953--972.

\bibitem{Biroli:2013}
\textsc{G.~Biroli and J.~P. Garrahan}, `Perspective: The glass transition'.
  \emph{J. Chem. Phys.} \textbf{138}/12 (2013), 12A301, 1--13.

\bibitem{Gundermann:2011}
\textsc{D.~Gundermann, U.~R. Pedersen, T.~Hecksher, N.~P. Bailey, B.~Jakobsen,
  T.~Christensen, N.~B. Olsen, T.~B. Schr{\o}der, D.~Fragiadakis, R.~Casalini,
  C.~M. Roland, J.~C. Dyre and K.~Niss}, `Predicting the density-scaling
  exponent of a glass-forming liquid from Prigogine-Defay ratio measurements'.
  \emph{Nature Physics} \textbf{7} (2011), 816--821.

\bibitem{Frenkel:2002}
\textsc{D.~Frenkel and B.~Smit}, \emph{Understanding Molecular Simulation: From
  Algorithms to Applications}, 2nd edition ed.. Academic Press, 2002.

\bibitem{Pedersen:2010}
\textsc{U.~R. Pedersen, T.~B. Schr{\o}der and J.~C. Dyre}, `A repulsive
  reference potential reproducing the dynamics of a liquid with attractions'.
  \emph{Phys. Rev. Lett.} \textbf{105} (2010), 157801, 1--4.

\bibitem{rumd}
\textsc{N.~Bailey, T.~Ingebrigtsen, J.~S. Hansen, A.~Veldhorst, L.~B{\o}hling,
  C.~Lemarchand, A.~Olsen, A.~Bacher, L.~Costigliola, U.~Pedersen, H.~Larsen,
  J.~C. Dyre and T.~Schr{\o}der}, `{RUMD}: A general purpose molecular dynamics
  package optimized to utilize {GPU} hardware down to a few thousand
  particles'. \emph{{SciPost} Phys.} \textbf{3}/6 (2017), 038, 1--20.

\bibitem{Hecksher-et-al:2019}
\textsc{A.~Sanz, T.~Hecksher, H.~W. Hansen, J.~C. Dyre, K.~Niss and U.~R.
  Pedersen}, `Experimental evidence for a state-point-dependent density-scaling
  exponent of liquid dynamics'. \emph{Phys. Rev. Lett.} \textbf{122} (2019),
  055501.

\bibitem{Eigen:1975}
\textsc{M.~Eigen and R.~Winkler}, \emph{Das Spiel: Naturgesetze steuern den
  Zufall}, Serie Piper. Piper, 1975.

\bibitem{Thom:1989}
\textsc{R.~{Thom}}, \emph{{Structural Stability and Morphogenesis: An Outline
  of a General Theory of Models. Transl. from the French edition, as updated by
  the author, and by D. H. Fowler. Reprint from the 2nd Engl. ed.}}, reprint
  from the 2nd engl. ed. ed.. Redwood City, CA: Addison-Wesley Publishing
  Company, Inc., 1989 (English).

\bibitem{Schneppen:2017}
\textsc{S.~B. Nissen, M.~Perera, J.~M. Gonzalez, S.~M. Morgani, M.~H. Jensen,
  K.~Sneppen, J.~M. Brickman and A.~Trusina}, `Four simple rules that are
  sufficient to generate the mammalian blastocyst'. \emph{PLOS Biology}
  \textbf{15}/7 (2017), 1--30.

\bibitem{Puri&Hebrock:2007}
\textsc{S.~Puri and M.~Hebrok}, `Dynamics of embryonic pancreas development
  using real-time imaging'. \emph{Developmental Biology} \textbf{306}/1 (2007),
  82--93.

\bibitem{Nyeng-et-al:2019}
\textsc{P.~Nyeng, S.~Heilmann, Z.~L{\"o}f-{\"O}hlin, N.~Pettersson, F.~Hermann,
  A.~Reynolds and H.~Semb}, `p120ctn-Mediated organ patterning precedes and
  determines pancreatic progenitor fate'. \emph{Developmental Cell} \textbf{49}
  (2019), 31--47.

\bibitem{Simonsen-et-al:2019-2}
\textsc{K.~Shioda, C.~Schuck-Paim, R.~J. Taylor, R.~Lustig, L.~Simonsen, J.~L.
  Warren and D.~M. Weinberger}, `Challenges in estimating the impact of
  vaccination with sparse data'. \emph{Epidemiology} \textbf{30}/1 (2019).

\bibitem{Simonsen-et-al:2019}
\textsc{C.~Schuck-Paim, R.~J. Taylor, W.~J. Alonso, D.~M. Weinberger and
  L.~Simonsen}, `Effect of pneumococcal conjugate vaccine introduction on
  childhood pneumonia mortality in Brazil: a retrospective observational
  study'. \emph{The Lancet Global Health} \textbf{7} (2019), 249--256.

\bibitem{ViggoAndreasen-et-al:2015}
\textsc{H.~Heesterbeek, R.~Anderson, V.~Andreasen, S.~Bansal, D.~{De Angelis},
  C.~Dye, K.~Eames, W.~Edmunds, S.~Frost, S.~Funk, T.~Hollongsworth, T.~House,
  V.~Isham, P.~Klepac, J.~Lessler, J.~Lloyd-Smith, C.~Metcalf, D.~Mollison,
  L.~Pellis, J.~Pulliam, M.~Roberts, C.~Viboud and {Isaac Newton Institute IDD
  Collaboration}}, `Modeling infectious disease dynamics in the complex
  landscape of global health'. \emph{Science} \textbf{347}/6227 (2015)
  (English).

\bibitem{Boo:2012}
\textsc{B.~Boo{\ss}-Bavnbek}, `Towards a nano geometry? {G}eometry and dynamics
  on nano scale'. In: \emph{Analysis, Geometry and Quantum Field Theory},
  Contemp. Math., vol. 584. Amer. Math. Soc., Providence, RI,
  \auindex{Boo{\ss}--Bavnbek,\ B.|bind}2012, pp.~147--162. \url{arXiv:1202.5115
  [math-ph]}.

\bibitem{Sneyd-et-al:2016}
\textsc{G.~Dupont, M.~Falcke, V.~Kirk and J.~Sneyd}, \emph{Models of Calcium
  Signalling}, Interdisciplinary Applied Mathematics, vol.~43. Springer, 2016.

\bibitem{Andersen-et-al:2017}
\textsc{M.~Andersen, Z.~Sajid, R.~Pedersen, J.~Gudmand-Hoeyer, C.~Ellervik,
  V.~Skov and e.~{}}, `Mathematical modelling as a proof of concept for MPNs as
  a human inflammation model for cancer development'. \emph{PLoS ONE}
  \textbf{12} (2017), e0183620, 1--18.

\bibitem{Beck:2008}
\textsc{N.~Beck}, \emph{Diagnostic Hematology}. Springer London, 2008.

\bibitem{De:2014}
\textsc{S.~De, G.~J. Williams, A.~Hayen, P.~Macaskill, M.~McCaskill, D.~Isaacs
  and J.~C. Craig}, `Value of white cell count in predicting serious bacterial
  infection in febrile children under 5 years of age'. \emph{Archives of
  Disease in Childhood} \textbf{99}/6 (2014), 493--499.

\bibitem{Renstrom2011}
\textsc{E.~Renstr{\"o}m}, `Established facts and open questions of regulated
  exocytosis in $\beta$-cells - A background for a focused systems analysis
  approach'. In: \emph{Beta{S}ys --- {S}ystems Biology of Regulated Exocytosis
  in Pancreatic $\beta$-Cells = \cite[Ch. 2]{Beta:2011}} (B.~Boo{\ss}-Bavnbek,
  B.~Kl{\"o}sgen, J.~Larsen, F.~Pociot and E.~Renstr{\"o}m, eds.), Systems
  Biology, vol.~2. Springer Science+Business Media, New York, NY, 2011,
  pp.~25--52.

\bibitem{Grodsky-et-al:1968}
\textsc{D.~L. Curry, L.~L. Bennett and G.~M. Grodsky}, `Dynamics of insulin
  secretion by the perfused rat pancreas'. \emph{Endocrinology} \textbf{83}
  (September 1968), 572--584.

\bibitem{Rorsman-Renstrom:2003}
\textsc{P.~Rorsman and E.~Renstr{\"o}m}, `Insulin granule dynamics in
  pancreatic beta cells'. \emph{Diabetologia} \textbf{46} (2003), 1029--1045.

\bibitem{Beta:2011}
\textsc{B.~Boo{\ss}-Bavnbek, B.~Kl{\"o}sgen, J.~Larsen, F.~Pociot and
  E.~Renstr{\"o}m} (eds.), \emph{Beta{S}ys --- {S}ystems Biology of Regulated
  Exocytosis in Pancreatic $\beta$-Cells}, Systems Biology, vol.~2. Springer,
  Berlin-Heidelberg-New York, 2011, Comprehensive review in
  \textit{Diabetologia}, \url{DOI 10.1007/s00125-011-2269-3}.

\bibitem{Grodsky-et-al:1970}
\textsc{G.~M. Grodsky, H.~Landahl, D.~L. Curry and L.~L. Bennett}, `A
  two-compartmental model for insulin secretion'. \emph{Adv. Metab. Disord.}
  \textbf{1} (1970), 45--50, Suppl. 1, edited by Camerini-D{\'a}valos, Rafael
  A. and Cole, Harold S. under the title \textit{Early Diabetes}.

\bibitem{Sherman-et-al:2008}
\textsc{Y.~Chen, S.~Wang and A.~Sherman}, `Identifying the targets of the
  amplifying pathway for insulin secretion in pancreatic $\beta$--cells by
  kinetic modeling of granule exocytosis'. \emph{Biophysical Journal}
  \textbf{95} (September 2008), 2226--2241.

\bibitem{Pedersen:2009}
\textsc{M.~G. Pedersen}, `Contributions of mathematical modeling of beta cells
  to the understanding of beta-cell oscillations and insulin secretion'.
  \emph{Journal of Diabetes Science and Technology (JDST)} \textbf{3}/1 (2009),
  12--20, PMID: 20046647.

\bibitem{Pedersen-Sherman:2009}
\textsc{M.~G. Pedersen and A.~Sherman}, `Newcomer insulin secretory granules as
  a highly calcium-sensitive pool'. \emph{Proceedings of the National Academy
  of Sciences} \textbf{106}/18 (2009), 7432--7436.

\bibitem{Pedersen-et-al:2014}
\textsc{M.~Riz, M.~Braun and M.~G. Pedersen}, `Mathematical modeling of
  heterogeneous electrophysiological responses in human $\beta$-cells'.
  \emph{PLoS Computational Biology} \textbf{10}/1 (2014), e1003389, 1--14.

\bibitem{ApApBoKo:2011}
\textsc{D.~Apushkinskaya, E.~Apushkinsky, B.~Boo{\ss}-Bavnbek and M.~Koch},
  `Geometric and electromagnetic aspects of fusion pore making'. In:
  \emph{Beta{S}ys --- {S}ystems Biology of Regulated Exocytosis in Pancreatic
  $\beta$-Cells = \cite[Ch. 23]{Beta:2011}}, Systems Biology, vol.~2. Springer
  Science+Business Media, \auindex{Boo{\ss}--Bavnbek,\ B.|bind}2011,
  pp.~505--538. \url{arXiv:0912.3738 [math.AP]}.

\bibitem{Juselius:2006}
\textsc{K.~Juselius}, \emph{The Cointegrated VAR Model: Methodology and
  Applications}, Advanced texts in econometrics. Oxford University Press, 2006.

\bibitem{Peirce:2012}
\textsc{C.~Peirce}, \emph{Philosophical Writings of Peirce}. Dover
  Publications, 2012, Selected and edited by J. Buchler.

\bibitem{Spengler:1918}
\textsc{O.~Spengler}, \emph{Der Untergang des Abendlandes --- Umrisse einer
  Morphologie der Weltgeschichte, 2 vols.}. Braum{\"u}ller/C.H. Beck,
  Wien/M{\"u}nchen, 1918/1922, Numerous reprints and translations. English: The
  Decline of the West, C.F. Atkinson translator, Knopf, 1926.

\bibitem{Strauss-Howe:1997}
\textsc{W.~Strauss and N.~Howe}, \emph{The Fourth Turning: What the Cycles of
  History Tell Us About America's Next Rendezvous with Destiny}. Broadway
  Books, New York, 1997.

\bibitem{Marx:1887}
\textsc{K.~Marx}, \emph{Capital, Volume I: A Critique of Political Economy},
  The Works of Karl Marx and Friedrich Engels. Progress Publishers, Moscow,
  USSR; Dover Publications, 2011, First published in German in 1867; first
  English edition of 1887, translated by Samuel Moore and Edward Aveling,
  edited by Frederick Engels, free available at
  \url{https://www.marxists.org/archive/marx/works/1867-c1/index.htm}.

\bibitem{Keynes:1936}
\textsc{J.~M. Keynes}, \emph{The General Theory of Employment, Interest and
  Money}. Macmillan, London, 1936, Free available at
  \url{https://www.marxists.org/reference/subject/economics/keynes/general-theory/index.htm}.

\bibitem{Schumpeter:1939}
\textsc{J.~Schumpeter}, \emph{Business Cycles: A Theoretical, Historical, and
  Statistical Analysis of the Capitalist Process, Vols. I, II}. McGraw-Hill
  Book Company, Inc., New York and London, 1939, Vol. I of xvi+448 pp. was
  first published in 1923 and reprinted in 1939 jointly with Vol. II of ix+647
  pp. Both volumes were reprinted in 2005 by Martino Pub., Mansfield Centre,
  Connecticut.

\bibitem{Freeman:2009}
\textsc{C.~Freeman}, `Schumpeter's business cycles and techno--economic
  paradigms'. In: \emph{Techno--Economic Paradigms: Essays in Honour of Carlota
  Perez, edited by W. Drechsler, E. Reinert, R. Kattel}, Other Canon Economics.
  Anthem Press, London, 2009, pp.~125--144.

\bibitem{Freeman:1996}
\textsc{C.~Freeman} (ed.), \emph{Long Wave Theory}, International Library of
  Critical Writings in Economics. Edward Elgar, 1996.

\bibitem{Burns:1969}
\textsc{A.~F. Burns}, `The nature and causes of business cycles'. In: \emph{The
  Business Cycle in a Changing World, edited by A.F. Burns}. National Bureau of
  Economic Research (NBER), Cambridge, Mass., U.S.A., 1969, pp.~3--53,
  Originally published in \textit{International Encyclopedia of the Social
  Sciences}, Vol. 2, pp. 226--245, Sills (ed.), Crowell Collier and Macmillan,
  Inc.

\bibitem{Burns-Mitchell:1946}
\textsc{A.~F. Burns and W.~C. Mitchell}, \emph{Measuring Business Cycles},
  Studies in business cycles. National Bureau of Economic Research, 1946.

\bibitem{Goodwin:1951}
\textsc{R.~M. Goodwin}, `The non-linear accelerator and the persistence of
  business cycles'. \emph{Econometrica} \textbf{19} (1951), 1--17.

\bibitem{Goodwin:1989}
\bysame, \emph{Essays in Nonlinear Economic Dynamics}. Peter Lang, Bern, 1989.

\bibitem{Friedman:1968}
\textsc{M.~Friedman}, `The role of monetary policy'. \emph{American Economic
  Review} \textbf{58} (1968), 1--17.

\bibitem{Flaschel:2009}
\textsc{P.~Flaschel}, \emph{The Macrodynamics of Capitalism: Elements for a
  Synthesis of Marx, Keynes and Schumpeter}, 2 ed., Business and Economics.
  Springer Berlin Heidelberg, 2009, 1st edition 1993.

\bibitem{Flaschel-Kauermann-Teuber:2005}
\textsc{P.~Flaschel, G.~Kauermann and T.~Teuber}, `Long cycles in employment,
  inflation and real unit wage costs, qualitative analysis and quantitative
  assessment'. \emph{Am. J. Appl. Sci.} \textbf{2} (2005), 69--77.

\bibitem{Norlund:1924}
\textsc{N.~E. N{\"o}rlund}, \emph{Vorlesungen {\"u}ber Differenzenrechnung},
  Grundlehren der Mathematischen Wissenschaften, vol.~13. J. Springer, Berlin,
  1924, Reprinted 1954 by Chelsea Publishing Company, New York.

\bibitem{Kauermann-Teuber-Flaschel:2012}
\textsc{G.~Kauermann, T.~Teuber and P.~Flaschel}, `Exploring US business cycles
  with bivariate loops using penalized spline regression'. \emph{Computational
  Economics} \textbf{39} (2012), 409--427.

\bibitem{Groth-Madsen:2016}
\textsc{C.~Groth and J.~B. Madsen}, `Medium-term fluctuations and the ``Great
  Ratios'' of economic growth'. \emph{Journal of Macroeconomics} \textbf{49}
  (2016), 149--176.

\bibitem{Smithers:2019}
\textsc{A.~Smithers}, `Secular stagnation claims are stifling discussion'.
  \emph{Financial Times} (March 14, 2019), London based Andrew Smithers is the
  founder of Smithers \& Co., which provides economics-based asset allocation
  advice to over 100 fund management companies.

\bibitem{Summers:2015}
\textsc{L.~H. Summers}, `The age of secular stagnation.'. \emph{Foreign
  Affairs} \textbf{95}/2 (2016), 2--9.

\bibitem{Rachel+Summers:2019}
\textsc{{\L}.~Rachel and L.~H. Summers}, `On falling neutral real rates, fiscal
  policy, and the risk of secular stagnation'. \emph{Brookings Papers on
  Economic Activity (BPEA)} (BPEA Conference Draft March 7, 2019), 1--66.

\bibitem{Wolf:2019}
\textsc{M.~Wolf}, `Monetary policy has run its course. It has made secular
  stagnation worse. Fiscal alternatives look a safer bet'. \emph{Financial
  Times} (March 12, 2019), Martin Wolf is chief economics commentator at the
  Financial Times, London.

\bibitem{Liu-et-al:2018}
\textsc{J.~Han, W.-Q. Chen, L.~Zhang and G.~Liu}, `Uncovering the
  spatiotemporal dynamics of urban infrastructure development: A high spatial
  resolution material stock and flow analysis'. \emph{Environ. Sci. Technol.}
  \textbf{52} (2018), 12122--12132, Advance Access published October 2, 2018.

\bibitem{Puu:2003}
\textsc{T.~Puu}, \emph{Mathematical Location and Land Use Theory --- An
  Introduction}, second ed., Advances in Spatial Science. Springer-Verlag,
  Berlin, 2003.

\bibitem{Krugman:1993}
\textsc{P.~Krugman}, \emph{Geography and Trade}, Gaston Eyskens lecture series.
  MIT Press (MA), 1993.

\bibitem{Fujita:2001}
\textsc{M.~Fujita, P.~Krugman and A.~Venables}, \emph{The Spatial Economy:
  Cities, Regions, and International Trade}, Mit Press. MT Press, 2001.

\bibitem{EGR:2019}
\textsc{A.~Olhoff and J.~Christensen} (eds.), \emph{Emissions Gap Report 2019}.
  United Nations Environment Programme, Nairobi, Kenya, Nov. 2019,
  \url{https://newclimate.org/2019/11/26/emissions-gap-report-2019/}.

\bibitem{Majda:2018}
\textsc{A.~J. Majda}, \emph{Mathematical Strategies for Climate and Long Range
  Weather Forecasting in Hierarchy of Models}. Berlin: Springer, 2019,
  announced.

\bibitem{Konig:2016}
\textsc{W.~K{\"o}nig} (ed.), \emph{Mathematics and Society}. European
  Mathematical Society, Z{\"u}rich, 2016.

\bibitem{Santos-et-al:2016}
\textsc{P.~Santos, P.~Bacelar-Nicolau, M.~Pardal, L.~Bacelar-Nicolau and
  U.~Azeiteiro}, `Assessing student perceptions and comprehension of climate
  change in Portuguese higher education institutions'. In: \emph{Implementing
  Climate Change Adaptation in Cities and Communities: Integrating Strategies
  and Educational Approaches, W.L. Filho et al. (eds.),
  \cite{Illingworth-et-al:2016}}. Springer International Publishing, 2016,
  pp.~221--236.

\bibitem{Allen:2018}
\textsc{K.~Allen}, `Green bond market faces its first real test. Conditions for
  environmental finance tougher even as evidence of climate change grows'.
  \emph{Financial Times} (November 6, 2018), Kate Allen is a capital markets
  correspondent for the Financial Times.

\bibitem{Dale:2018}
\textsc{S.~Dale}, `Energy in 2017: two steps forward, one step back. Group
  chief economist's analysis'. In: \emph{BP Statistical Review of World Energy
  2018}, BP Statistical Review of World Energy, vol.~67. British Petrol,
  London, June 2018, pp.~3--7, You can order BP's printed publications, free of
  charge from bp.com/papercopies.

\bibitem{Neuhaus:2013}
\textsc{W.~Nordhaus}, \emph{The Climate Casino: Risk, Uncertainty, and
  Economics for a Warming World}. Yale University Press, 2013.

\bibitem{Wolf:2018}
\textsc{M.~Wolf}, `Inaction over climate change is shameful. We need to shift
  the world on to a different investment and growth path immediately'.
  \emph{Financial Times} (October 23, 2018), Martin Wolf is chief economics
  commentator at the Financial Times, London.

\bibitem{Bek-Thomsen-et-al:2017}
\textsc{J.~Bek-Thomsen, C.~O. Christiansen, S.~G. Jacobsen and M.~Thorup},
  \emph{History of Economic Rationalities: Economic Reasoning as Knowledge and
  Practice Authority}. Springer, March 2017.

\bibitem{Bullard-Muller:2012}
\textsc{N.~Bullard and T.~M{\"u}ller}, `Beyond the ``Green Economy": system
  change, not climate change?'. \emph{Development} \textbf{55}/1 (March 2012),
  54--62.

\bibitem{Jacobsen:2018}
\textsc{S.~G. Jacobsen} (ed.), \emph{Climate Justice and the Economy: Social
  Mobilization, Knowledge and the Political}, Routledge Advances in Climate
  Change Research. Taylor \& Francis, 2018.

\bibitem{Porter-Kramer:2011}
\textsc{M.~Porter and M.~Kramer}, `Creating shared value: how to reinvent
  capitalism --- and unleash a wave of innovation and growth'. \emph{Harvard
  Business Review} \textbf{89}/1/2 (2011), 62--77.

\bibitem{Bowater-Yeoman:2012}
\textsc{L.~Bowater and K.~Yeoman}, \emph{Science Communication: A Practical
  Guide for Scientists}. Wiley, 2012.

\bibitem{Illingworth-Allen:2016}
\textsc{S.~Illingworth and G.~Allen}, \emph{Effective Science Communication}.
  IOP Publishing, 2016.

\bibitem{Hartomo-Cribb:2002}
\textsc{T.~Hartomo and J.~Cribb}, \emph{Sharing Knowledge: A Guide to Effective
  Science Communication}, 2002.

\bibitem{Illingworth-Allen-ch5:2016}
\textsc{S.~Illingworth and G.~Allen}, `Outreach and public engagement'. In:
  \emph{Effective Science Communication, \cite[Ch. 5]{Illingworth-Allen:2016}}.
  Institute of Physics, IOP Publishing, Bristol, 2016, pp.~5--1 to 5--23.

\bibitem{ClimateChange:2018}
\textsc{V.~Masson-Delmotte and {al.}} (eds.), \emph{IPCC 2018. Global warming
  of $1.5^{\circ}$C. An IPCC Special Report on the impacts of global warming of
  $1.5^{\circ}$C above pre-industrial levels and related global greenhouse gas
  emission pathways, in the context of strengthening the global response to the
  threat of climate change, sustainable development, and efforts to eradicate
  poverty}. World Meteorological Organization, Geneva, Switzerland, 2018.

\bibitem{GlobalCarbonBudget:2018}
\textsc{C.~Le~Qu{\'e}r{\'e} and {al.}}, `Global carbon budget 2018'.
  \emph{Earth Syst. Sci. Data} \textbf{10} (2018), 2141--2194, Published also
  in \textit{Nature} and \textit{Environmental Research Letters} on December 5,
  2018.

\bibitem{Xu-et-al:2018}
\textsc{Y.~Xu, V.~Ramanathan and D.~G. Victor}, `Global warming will happen
  faster than we think. Three trends will combine to hasten it'. \emph{Nature}
  \textbf{564} (2018), 30--32, Comment.

\bibitem{Illingworth-et-al:2016}
\textsc{W.~Filho, K.~Adamson, R.~Dunk, U.~Azeiteiro, S.~Illingworth and
  F.~Alves} (eds.), \emph{Implementing Climate Change Adaptation in Cities and
  Communities: Integrating Strategies and Educational Approaches}, 1 ed.,
  Climate Change Management. Springer International Publishing, 2016.

\end{thebibliography}
\bibliographystyle{amsusrt-jl}

\end{document}